\documentclass{statsoc}
\usepackage[a4paper]{geometry}
\usepackage{epstopdf}
\usepackage{amsmath}
\usepackage{graphicx}
\usepackage[textwidth=8em,textsize=normal]{todonotes}

\usepackage{times}
\usepackage{amssymb}
\usepackage{bm}
\usepackage{natbib}
\usepackage{xcolor}
\usepackage{subfigure}
\usepackage{multirow}
\usepackage{url}
\usepackage{epsfig}
\usepackage{algorithmic}

\usepackage[plain,noend]{algorithm2e}
\newtheorem{theorem}{Theorem}
\newtheorem{remark}{Remark}  

\newtheorem{condition}{Condition}

\usepackage{siunitx}
\usepackage{booktabs}

\title[Temporal network analysis via a degree-corrected Cox model]{Temporal network analysis via a degree-corrected Cox model}
\author[Yuguo Chen {\it et al.}]{Yuguo Chen}
\address{Department of Statistics, University of Illinois at Urbana-Champaign, Champaign, Illinois 61820, U.S.A.}

\author{Lianqiang Qu}
\address{School of Mathematics and Statistics,
	Central China Normal University,
	Wuhan, Hubei 430079, China
}
\author{Jinfeng Xu}
\address{Department of Biostatistics, City University of Hong Kong, Tat Chee Avenue, Kowloon, Hong Kong
}
\author{Ting Yan}
\address{School of Mathematics and Statistics,
	Central China Normal University,
	Wuhan, Hubei 430079, China
}
\author[Chen et al.]{and Yunpeng Zhou}
\address{Department of Statistics and Actuarial Science,
	The University of Hong Kong, Hong Kong
}

\begin{document}

\begin{abstract}
Temporal dynamics, characterised by time-varying degree heterogeneity and homophily effects, are often exhibited in many real-world networks.
As observed in an MIT Social Evolution study, the in-degree and out-degree of the nodes show considerable heterogeneity that varies with time. Concurrently,
homophily effects, which explain why nodes with similar characteristics are more likely to connect  with each other, are also time-dependent.
To facilitate the exploration and understanding
of these dynamics, we propose a novel degree-corrected Cox  model for directed networks, where
the way for degree-heterogeneity or homophily effects to change with time is left completely unspecified.
Because each node has individual-specific in- and out-degree parameters that vary over time,
the number of unknown parameters grows with the number of nodes, leading to a high-dimensional estimation problem.
Therefore, it is highly nontrivial to make inference.
We develop a local estimating equations approach to estimate the unknown parameters
and establish the consistency and asymptotic normality of
the proposed estimators in the high-dimensional regime.
We further propose test statistics to check whether
temporal variation or degree heterogeneity is present in the network
and  develop a graphically diagnostic method to evaluate goodness-of-fit for
dynamic network models.
Simulation studies and two real data analyses
are provided to assess the finite sample performance of the proposed method and illustrate its practical utility.
\end{abstract}


\section{Introduction}
\label{section-introduction}

Temporal networks emerge in diverse domains, encompassing fields such as social sciences and biological sciences, as well as various communication networks,
including email networks, collaboration networks, and social media platforms.
Considering the temporal variation observed in the interactions among nodes within these networks,
a primary focus of network analysis is to accurately model and infer their dynamic structures and evolutionary patterns.
To achieve this objective, one approach involves transforming continuous network data into a sequence of static networks through aggregation within discrete time intervals.
The resulting discrete time network data can then be effectively analyzed using dynamic network models,
including temporal exponential random graph models \citep{hanneke2010discrete,krivitsky2014separable},
dynamic stochastic block models \citep{yang2011detecting},
and dynamic latent space models \citep{sewell2015analysis,sewell2015latent}.
However, in many scenarios, the interactions among nodes often occur irregularly and continuously \citep{perry2013point,DBS2013,MRV2018}. Inference based on the
aggregation of data is highly sensitive to the choice of time intervals and may lead to unreliable conclusions \citep{Choudhury2010}.

Alternatively, a more appealing approach is to directly model the temporal interactions using the counting process (e.g., \citealp{butts20084, Vu-hunter2011dynamic, DBS2013, perry2013point, MRV2018, Alexander2021, Sit2020}).  \cite{perry2013point} proposed a Cox-type regression model and utilized the conditional intensity function to capture the influence of nodes' past interaction history on both current and future interactions.
By assuming the constancy of covariate effects across time and node pairs, they developed asymptotic theory for partial likelihood estimation
when the time horizon $T$ goes to infinity.  
In a subsequent work, \cite{kreib2019} allowed
covariate effects to change over time and established asymptotic theory for nonparametric likelihood-based estimation via kernel smoothing
in an asymptotic framework that lets the number of nodes, rather than $T$,
diverge.
Most recently, \cite{Sit2020} used mean and rate functions in the Cox-type model to model
the dependence of interactions amongst nodes and obtained
asymptotic results by restricting the growing rate of the number of dependent nodes.

\begin{figure}[h]
	\centering
	\subfigure[The in-degree curves of some selected students.]{\includegraphics[width=0.36\textwidth]{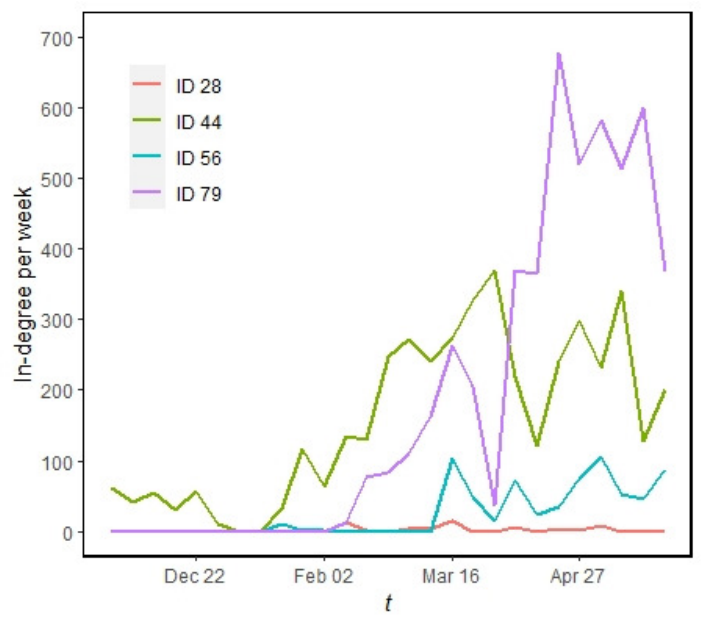}}
	\qquad
    \subfigure[The curves of the average number of interactions per pair of students
    based on whether they are from the same floor or different floors.]{\includegraphics[width=0.36\textwidth]{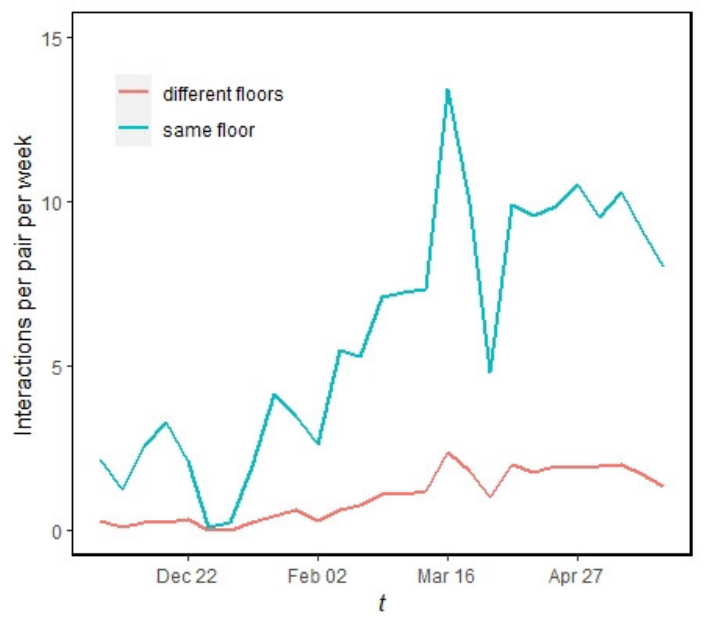}}
	\caption{Preliminary analysis of MIT Social Evolution data.
 }
	\label{fig:realHT2}
\end{figure}

Two prominent features, homophily and degree heterogeneity, are widely observed in network data.
Homophily refers to the tendency of individuals with similar covariate values to more easily establish connections with each other,
while degree heterogeneity indicates significant variation in individuals' interactions with others.
For example, in the  MIT Social Evolution study  \citep{madan2010} that motivated our research,
interactions were observed by monitoring cellular phones.
Specifically, Bluetooth signals were captured whenever a sender's phone connected to a receiver's phone within a 10-meter range.
This enabled the utilization of the frequency of Bluetooth communication as an indicator of daily social activities among students and tutors.
Fig. \ref{fig:realHT2} presents a preliminary analysis of the data,
showcasing the curves of the in-degree (defined in Section \ref{section:model}) for some selected students
and the average number of interactions per pair of students
based on whether they are from the same floor or different floors of the dormitory.
Notably, the plots reveal not only significant time-varying degree heterogeneity within the network (see Fig. \ref{fig:realHT2}(a)),
but also a clear presence of homophily,
as students from the same floor are more likely to form connections compared to those from different floors (see Fig. \ref{fig:realHT2}(b)).
In addition, the homophily effects also vary with time. 

However, the existing literature (e.g., \citealp{perry2013point, Sit2020,kreib2019}) primarily concentrates on modeling covariate effects,
which can be seen as slight generalizations of homophily effects,
and none of these studies directly addresses the modeling of degree heterogeneity.
When degree heterogeneity is present but not considered in the model,
the estimators of homophily parameters may have a large bias; see Fig. \ref{fig:chet}.
Therefore, there is a pressing need to develop novel models and methodologies
that can effectively capture both degree heterogeneity and homophily effects in a time-varying manner.
The main contributions of our paper are as follows.
\begin{itemize}
\item We propose a novel model, referred to as the degree-corrected Cox network model, for dynamic network data analysis.
Specifically, the proposed model contains a $p$-dimensional regression parameter for the time-varying homophily effect
and incorporates a set of $2n$ degree parameters measuring the time-varying degree heterogeneity,
where $n$ denotes the number of nodes in the network.
Therefore, it effectively captures the dynamic degree heterogeneity and homophily effects,
distinguishing it from the focus of previous studies such as \cite{perry2013point}, \cite{kreib2019}, and \cite{Sit2020}.
Additionally, this modeling strategy also offers several advantages.
For example, it enables us to develop test statistics to check whether a network exhibits time-varying degree heterogeneity
and provides a clear interpretation for node popularity \citep{Sengupta-Chen-2017}.
We shall elaborate a comprehensive comparison between the proposed model and existing models (e.g., \citealp{perry2013point,kreib2019})
in Section \ref{section:model}.

\item We develop a local estimating equations (LEE) method to estimate the $2n+p$ unknown time-varying parameters.
Subsequently, the resulting estimators are referred to as LEE estimators for convenience.
We propose an efficient iterative procedure for obtaining the LEE estimators.
Bandwidth selection is crucial in nonparametric regression methods;
however, there remains a lack of research on data-driven methods for bandwidth selection in the network literature.
To enhance the flexibility and adaptability of the proposed method,
we introduce a cross-validation method for selecting an appropriate bandwidth.
Additionally, we develop
a graphical diagnostic method to assess the goodness-of-fit for the degree-corrected Cox network model. This method is not limited to our model but is also applicable to general dynamic network models.

\item
We establish consistency and asymptotic normality of the LEE estimators
when the number of nodes goes to infinity.
Our theoretical setting 
differs from existing literature in several key aspects.
First, all the unknown parameters in our model are allowed to vary over time, while
the parameters are time-invariant in \cite{perry2013point} and \cite{Sit2020}.
Second, our model contains $2n$ time-varying degree heterogeneity parameters,
which results in statistical inference
in a high-dimensional setting.
This is different from \cite{perry2013point}, \cite{Sit2020}, \cite{kreib2019}, and \cite{Alexander2021}, where
the number of unknown parameters is fixed.
In addition,  \cite{kreib2019} and \cite{Alexander2021} did not investigate the time-varying degree effects
and made a boundedness assumption on the parameters, which is not required in our setting.
These distinctions render the methods in the aforementioned literature
unsuitable for our approach, thus requiring the development of new techniques for theoretical analysis.
To accomplish this, we derive concentration inequalities and combine them with theories of martingales and counting processes to establish the asymptotic properties of the estimators.
We also propose a sandwich formula for estimating the covariance of the estimators,
which is consistent and performs well in our simulation studies.

\end{itemize}

The rest of this paper is organized as follows.
Section \ref{section:model} introduces the degree-corrected Cox network model,
and also develops a local estimating equations approach to estimate the unknown parameters.
Section \ref{section:asymptotic} establishes
uniform consistency of the estimators and their point-wise central limit theorems.
Section \ref{section:test} develops test statistics to test for trend and degree heterogeneity, and introduces
a goodness-of-fit method.
Section \ref{section:BS} presents an iterative algorithm to obtain the estimators
and a data-driven procedure for bandwidth selection.
In Section \ref{section:simulation}, we evaluate finite
sample performance of the proposed method and inference procedures via simulation studies.
In Section \ref{section:realdata}, we analyze two real-world datasets: MIT Social Evolution dataset and Capital Bikeshare dataset.
Some concluding remarks are given in Section \ref{section:concluding}.
All the proofs are presented in the Supplementary Material.
The codes for simulation studies and real data analyses are publicly available at
{\it \url{https://github.com/MinZenggit/DCCOX}}.

\setlength{\abovedisplayskip}{6pt}
\setlength{\belowdisplayskip}{6pt}
\allowdisplaybreaks

\section{Model and estimation}
\label{section:model}

\subsection{A degree-corrected Cox network model}
We consider a network with $n$ nodes labelled as ``$1, \ldots, n$". 
We use $[n]$ to denote the integer sets $\{1,\dots,n\}$.
For any fixed $i$, let $\sum_{j\ne i}x_{ij}$ denote the shorthand for $\sum_{j=1,j\neq i}^nx_{ij}$.
For any two nodes $i\neq j\in[n]$, we define
$N_{ij}(s,t)$ as the number of directed interactions from the head node $i$ to the tail node $j$ in the time interval $(s,t].$
We write $N_{ij}(t)=N_{ij}(0,t)$. Without loss of generality, we assume that the interaction process
starts at $t = 0$ with $N_{ij}(0) = 0$ for $1\le i\not=j \le n.$
The counting process $\{N_{ij}(t): t\ge 0\}$ encodes occurrences of directed interactions  from the head node $i$ to the tail node $j$.
Let $N_{ij}(t^-)$ denote the number of interactions before time $t$.
If $N_{ij}((t+\Delta t)^-)-N_{ij}(t^-)=1$ as $\Delta t\rightarrow 0,$
then a directed interaction from head node $i$ to tail node $j$ occurs at time $t,$
indicating a direct edge from $i$ to $j$ at time $t$.
The out-degree and in-degree for node $i$ over the time interval $(s,t]$ are then defined by
$\sum_{j\neq i}N_{ij}(s,t)$ and $\sum_{j\neq i}N_{ji}(s,t),$ respectively.

Let $Z_{ij}(t)$ be a $p$-dimensional time-varying covariate for the node pair $(i,j).$
The covariate $Z_{ij}(t)$ can be used to
measure the similarity or dissimilarity between individual-level characteristics.
For example, if an individual $i$ has a $d$-dimensional characteristic $X_i(t)$,
the pairwise covariate $Z_{ij}(t)$ can be constructed by setting $Z_{ij}(t)=X_i(t)^\top X_j(t)$,
$Z_{ij}(t)=\|X_i(t)-X_j(t)\|_2$, or $Z_{ij}(t)=X_i(t)\otimes X_j(t)$.
Here, $\|x\|_2$ denotes the $\ell_2$-norm of any vector $x\in\mathbb{R}^d$ and
$x\otimes y$ denotes the Kronecker product of vectors $x$ and $y$.

Let $\tau$ denote the termination time for the observation period.
Then, the observations consist of $\{N_{ij}(t), Z_{ij}(t): i\neq j\in[n],~ t\in [0,\tau]\}$,
which are assumed to be independent across $(i,j).$
Let $\mathcal{F}_{t}$ denote the history that represents everything
occurred up to time $t$.
Define
\[\lambda_{ij}(t|\mathcal{F}_{t^-})=\lim_{\Delta t\rightarrow 0^+}P\big(N_{ij}((t+\Delta t)^-)-N_{ij}(t^-)=1|\mathcal{F}_{t^-}\big)/\Delta t\]
as the intensity function of $N_{ij}(t)$.
We propose the following model for $\lambda_{ij}(t|\mathcal{F}_{t^-})$:
\begin{align}
\label{eq2.1}
\lambda_{ij}(t|\mathcal{F}_{t^-})
&=\exp\{\alpha_i(t)+\beta_j(t)+Z_{ij}(t)^\top\gamma(t)\}, \quad 1\le i\not= j \le n,
\end{align}
where $\alpha_i(t)$ denotes the outgoingness of node $i$, $\beta_j(t)$ denotes the popularity of node $j$,
and $\gamma(t)$ is a common regression coefficient function.
As discussed in \cite{perry2013point} and \cite{yan2019},  $\gamma(t)$ captures the homophily effects of covariates.

\begin{remark}
Although we focus on the case where $\{N_{ij}(t), Z_{ij}(t): i \neq j \in [n], t \in [0, \tau]\}$ are independent across $(i,j)$, our theoretical analysis can be extended to a conditionally independent setting under additional concentration conditions on $Z_{ij}(t)$. In this setting, $N_{ij}(t)$ is conditionally independent given all covariates $Z_{ij}(t)$.
Due to space constraints, details are provided in Section C of the Supplementary Material.
\end{remark}

Similar to the explanation for the model parameters in the $p_1$ model for static networks \citep{Paul:Leinhardt1981},
the parameter functions $\alpha_i(t)$ and  $\beta_j(t)$ measure degree heterogeneity.
To clearly illustrate this in our model, we consider a Poisson process for $N_{ij}(t)$.
It is a specific case of counting processes, where
the intensity $\lambda_{ij}(t|\mathcal{F}_{t^-})$ is independent of the history $\mathcal{F}_{t^-}$ \cite[e.g.,][]{Cook2007}.
Note that the out-degree and in-degree for node $i$ over time interval $(s,t]$
are $\sum_{k\neq i}N_{ik}(s,t)$ and $\sum_{k\neq i}N_{ki}(s,t),$ respectively.
In the case of the Poisson process, we have
\begin{align}
\sum_{k\neq i}^n\mathbb{E}\{N_{ik}(s,t)\}=&\int_{s}^t\exp\{\alpha_i(u)\}\sum_{k\neq i}^n\exp\{\beta_k(u)
+Z_{ik}(u)^\top\gamma(u)\}du \label{eq2.2}\\
\text{and}~~~\sum_{k\neq i}^n\mathbb{E}\{N_{ki}(s,t)\}=&\int_{s}^t\exp\{\beta_i(u)\}\sum_{k\neq i}^n\exp\{\alpha_k(u)
+Z_{ki}(u)^\top\gamma(u)\}du. \label{eq2.3}
\end{align}
As we can see, the larger $\alpha_i(t)$ is, the more likely it is that node $i$ interacts with the others.
Therefore, $\alpha_i(t)~(i\in[n])$  accommodates the out-degree heterogeneity across different nodes.
However,  the larger $\beta_i(t)$ is, the more likely node $i$ will receive  interactions from other nodes,
that is, $\beta_i(t)$ reflects the in-degree heterogeneity.
Consequently, the effect of degree heterogeneity can be clearly delineated by estimating
the node-specific parameters $\alpha_i(t)$ and $\beta_j(t)$.

In model \eqref{eq2.1}, if we consider $\lambda_{0,ij}(t):=\exp\{\alpha_i(t)+\beta_j(t)\}$ as the baseline intensity function,
it reduces to a network version of the well-known Cox regression model.
Therefore, we call model \eqref{eq2.1} the \emph{degree-corrected Cox network model}.
Note that if one entry of $Z_{ij}(t)$ is equal to $1$ for all $(i, j)$,
then the regression coefficient of the covariate $Z_{ij}(t)$ contains an intercept $\gamma_0(t)$.
In this case, if one transforms $\alpha_i(t)+\beta_j(t)+\gamma_0(t)$ to $\{[\alpha_i(t)+c_1(t)]+[\beta_j(t)-c_1(t)]+c_2(t)\}+[\gamma_0(t)-c_2(t)]$,
where $c_1(t)$ and $c_2(t)$ denotes two arbitrary functions, model (1) is unchanged.
To ensure model identifiability, we exclude the intercept term in the assumptions
and set $\beta_n(t)\equiv0$.

We now discuss the differences between our model and those proposed by \cite{Alexander2021}, \cite{kreib2019}, and \cite{perry2013point}.
\begin{itemize}
\item
In \cite{kreib2019} and \cite{Alexander2021}, the network is assumed to be known a priori.
Specifically, they treated $C_n(t)=(C_{n,ij}(t): i,j\in [n])$ as a time-varying adjacency matrix,
where $C_{n,ij}(t)=1$ for $1\le i\neq j\le n$ if nodes $i$ and $j$ are connected by an edge at time $t$
and $C_{n,ij}(t)=0$ otherwise. Additionally, $C_{n,ii}(t)=0$ for $i\in[n]$.
Given $C_{n,ij}(t)=1$, \cite{kreib2019} and \cite{Alexander2021} assumed that
the intensity function of $N_{ij}(t)$ follows a Cox model.
In addition, the network process $C_{n,ij}(t)$ is treated as known a priori.
Therefore, degree heterogeneity in the network is part of their models.
Moreover, if we set the values of $C_{n,ij}(t)$ to 1 for all $1\le i\neq j\le n$,
their models do not characterize the effect of degree heterogeneity.
In contrast, our model directly parameterizes degree heterogeneity
using time-varying degree heterogeneity parameters $\alpha_i(t)$ and $\beta_j(t)$.
Therefore, it can handle unobserved heterogeneity and also estimate it.
In \citet{perry2013point}, out-degree heterogeneity was considered,
while in-degree heterogeneity was implicitly controlled through a predictable finite subset.
In contrast, model \eqref{eq2.1} explicitly characterises
both out-degree and in-degree heterogeneities using $\alpha_i(t)$ and $\beta_j(t)$,
rather than assuming these heterogeneities to be observed a priori.
This approach enables us to account for unobserved degree heterogeneity by estimating $\alpha_i(t)$ and $\beta_j(t)$
and to infer the presence of degree heterogeneity;
see Section \ref{section:test} for further details.

\item

Model \eqref{eq2.1} provides a clear interpretation for the node popularity, another important network feature \citep{Sengupta-Chen-2017}.
To determine how the model \eqref{eq2.1} captures this feature, we compute the log ratio of
$u_i^{(1)}(t)$ to $u_j^{(1)}(t)$, where $u_i^{(1)}(t)$ denotes the first derivative of
$u_i(t)=\sum_{k\neq i}^nE[N_{ki}(t)]$ with respect to $t$.
Then, we have
\begin{eqnarray*}
\log\bigg(\frac{u_{i}^{(1)}(t)}{u_{j}^{(1)}(t)}\bigg)
=\beta_i(t)-\beta_j(t)+\log\bigg(\frac{\sum_{k\neq i}\exp\{\alpha_k(t)+Z_{ki}(t)^\top\gamma(t)\}}
{\sum_{k\neq j}\exp\{\alpha_k(t)+Z_{kj}(t)^\top\gamma(t)\}}\bigg).
\end{eqnarray*}
As we can see, node $i$ tends to be more popular in a network
than node $j$ if $\beta_i(t)>\beta_j(t)$ at time $t$.
However, in the existing models \citep{perry2013point,kreib2019,Alexander2021},
we observe that the ratio does not contain the difference $\beta_i(t)-\beta_j(t)$.

\item The existing methods (e.g., \citealp{kreib2019,Alexander2021}) primarily focus on estimating the effects of covariates.
However, our goal is to estimate $\gamma(t)$ as well as the $2n$ node-specified parameters $\alpha_i(t)$ and $\beta_j(t)$.
Therefore, the number of unknown parameters increases with the number of nodes in model \eqref{eq2.1},
which necessitates the development of new statistical inference methods.
\end{itemize}

We next compare our model with frailty models (e.g., \citealp{A1993, Bianchi2024}).
First, frailty in frailty models \citep[e.g.,][]{A1993} is a latent random variable that is usually used to account for correlations in multivariate event history data. In our proposed model, if we treat the unknown parameters $\alpha_i(t)$ and $\beta_j(t)$ as latent random variables, then the proposed model is related to frailty models as in Chapter IX of \cite{A1993}. The frailty model approach typically requires specifying a parametric distribution for the frailties (i.e., latent random variables).
In this context, the counting processes $N_{i1}(t),\dots, N_{in}(t)$ for any fixed $i\in[n]$ are dependent,
as they share the same frailty $\alpha_i(t)$. Analogously, the counting processes $N_{1j}(t),\dots, N_{nj}(t)$ for any fixed $j\in [n]$ are also dependent due to a shared frailty $\beta_j(t)$. However, we need to know the correlation structure amongst $\alpha_i(t)$'s and $\beta_j(t)$'s for making inference in frailty models. This is usually unknown in practical applications. In addition, it is very difficult to draw inference in a high-dimensional problem with complex dependent structure.
To address this issue, we treat $\alpha_i(t)$ and $\beta_j(t)$ as unknown parameters in our proposed model, rather than latent random variables.  Second, in \cite{Bianchi2024}, frailties are considered as random effects,
accounting for the effects of the latent heterogeneity (e.g., receiver's popularity and sender's expansiveness). If we model degree heterogeneity in the framework of frailty, the degrees will be treated as covariates and the aim
will be to estimate
their effects
through
associated
coefficients, which are also treated as random variables (e.g., assumed to follow
a normal or Gamma distribution).
In contrast to this approach, our model directly accounts for unobserved heterogeneity by employing nonrandom parameters $\alpha_i(t)$ and $\beta_j(t)$. This facilitates straightforward interpretation of model parameters and enables the development of hypothesis testing methods to assess the presence of degree heterogeneity or temporal degree effects.

Finally, our model allows the network to have different edge densities. This enables the network to be sparse.
We call a dynamic network sparse if the ratio of the expected number of edges to $n^2$ over any time interval $(s,t]\subset(0,\tau)$
goes to zero as $n\rightarrow\infty.$
Specifically, consider the case in which $\sup_{t\in[0,\tau]}[\alpha_i(t)+\beta_j(t)+Z_{ij}(t)^\top\gamma(t)]=-q_n,$
where $q_n$ is a positive constant depending on $n.$
Then, by \eqref{eq2.2} and \eqref{eq2.3}, for any $0<s<t<\tau$,
the expected number of edges averaged over time interval $(s,t]$ satisfies
\begin{align}\label{eq2.4}
\sum_{i=1}^n\sum_{j\neq i}\frac{\mathbb{E}[N_{ij}(s,t)]}{t-s}
=&\sum_{i=1}^n\sum_{j\neq i}\frac{1}{t-s}\int_s^t\exp\{\alpha_i(u)+\beta_j(u)+Z_{ij}(u)^\top\gamma(u)\}du \nonumber \\
\leq & n^2e^{-q_n}.
\end{align}
The upper bound on the right-hand side of \eqref{eq2.4} resembles the definitions of $l_n$ in \cite{kreib2019} and equation (8) in \cite{kreib2023}.
It indicates that $q_n$ determines the sparsity level of the network;
specifically, as $q_n$ increases, the network becomes sparser.
If $q_n$ is sufficiently large, there are few interactions overall,
resulting in many pairs of nodes exhibiting no interactions at all.
For example, if we set $q_n=c\log(n)$, where $c\in (0,1]$ is a constant,
then the order of the right-hand side of \eqref{eq2.4} is $n^{2-c},$
which is less than $n^2$, indicating that the dynamic network is sparse.

\subsection{Local estimating equations approach}

We define $\theta(t)=(\alpha(t)^\top,$ $\beta(t)^\top,\gamma(t)^\top)^\top,$
where $\alpha(t)=(\alpha_1(t),\dots,\alpha_n(t))^\top$ and
$\beta(t)=(\beta_1(t),\dots,\beta_{n-1}(t))^\top$.
We use the superscript * to denote the true value (e.g., $\theta^*(t)$ is
the true value of $\theta(t)$).
Let $\mathcal{K}_h(u)=h^{-1}\mathcal{K}(u/h),$ where $\mathcal{K}(\cdot)$ is a kernel function and $h$ denotes the bandwidth.
Further, we define
\[
\mathcal{M}_{ij}(t; \alpha_i(t),\beta_j(t),\gamma(t))=N_{ij}(t)-\int_0^t\exp\{\alpha_i(s)+\beta_j(s)+Z_{ij}(s)^\top\gamma(s)\}ds.
\]
Under model \eqref{eq2.1}, $\mathcal{M}_{ij}(t)=\mathcal{M}_{ij}(t; \alpha_i^*(t),\beta_j^*(t),\gamma^*(t))$ is a zero-mean martingale process (see Lemma 2.3.2 in \citealp{FH1991}).
We next introduce a ``localness" assumption for the parameters:
$\alpha_i^*(t),$ $\beta_j^*(t),$ and $\gamma^*(t)$ are sufficiently smooth in the sense that as $s\to t$,
\begin{align*}
\max_{i\in[n]} |\alpha_i^*(s)-\alpha_i^*(t)| \rightarrow 0, \quad
\max_{j\in[n-1]} |\beta_j^*(s)-\beta_j^*(t)| \rightarrow 0, \quad
\max_{k\in[p]} |\gamma_k^*(s)-\gamma_k^*(t)| \rightarrow 0.
\end{align*}
When $s$ is close to $t,$ this leads to
\begin{align*}
d\mathcal{M}_{ij}(s)=&dN_{ij}(s)-\exp\{\alpha_i^*(s)+\beta_j^*(s)+Z_{ij}(s)^\top\gamma^*(s)\}ds\\
\approx & dN_{ij}(s)-\exp\{\alpha_i^*(t)+\beta_j^*(t)+Z_{ij}(s)^\top\gamma^*(t)\}ds,
\end{align*}
where $dN_{ij}(t)=N_{ij}((t+dt)^-)-N_{ij}(t^{-}).$
Therefore, this assumption allows us to leverage information
from nearby locations to estimate the unknown parameters.
For simplicity, we define $$d\mathcal{M}_{ij}(s,t;\alpha_i(t),\beta_j(t),\gamma(t))=dN_{ij}(s)-\exp\{\alpha_i(t)+\beta_j(t)+Z_{ij}(s)^\top\gamma(t)\}ds.$$

Based on the above, we propose the following local estimating equation:
\begin{align}\label{eq2.5}
(F(\theta(t))^\top, Q(\theta(t))^\top)^\top=0,
\end{align}
where $F(\theta(t))=(F_1(\alpha_1(t),\beta(t),\gamma(t)),\dots,F_{2n-1}(\alpha(t),\beta_{n-1}(t),\gamma(t)))^\top$
with
\begin{align*}
F_i(\alpha_i(t),\beta(t),\gamma(t))=&\frac{1}{n-1}\sum_{j=1,j\neq i}^n \int_{0}^{\tau}\mathcal{K}_{h_1}(s-t)d\mathcal{M}_{ij}(s,t;\alpha_i(t),\beta_j(t),\gamma(t)),~~i\in[n],
\\
F_{n+j}(\alpha(t),\beta_j(t),\gamma(t))=&\frac{1}{n-1}\sum_{j=1,j\neq i}^n \int_{0}^{\tau}\mathcal{K}_{h_1}(s-t)d\mathcal{M}_{ij}(s,t;\alpha_i(t),\beta_j(t),\gamma(t)),~~j\in[n-1],
\end{align*}
and
$$
Q(\alpha(t),\beta(t),\gamma(t))=\frac{1}{n(n-1)}\sum_{i=1}^n \sum_{j=1,j\neq i}^n
\int_{0}^{\tau}Z_{ij}(s)\mathcal{K}_{h_2}(s-t)d\mathcal{M}_{ij}(s,t;\alpha_i(t),\beta_j(t),\gamma(t)).
$$
We estimate $\theta^*(t)$ using the solution to equation \eqref{eq2.5}, denoted by $\widehat{\theta}(t)=(\widehat{\alpha}(t)^\top, \widehat{\beta}(t)^\top,\widehat{\gamma}(t)^\top)^{\top}$.
To obtain $\widehat\theta(t),$ we adopt a combination of the fixed point iteration and Newton-Raphson methods
by alternatively solving $F(\theta(t))=0$ and $Q(\theta(t))=0$.
This is implemented in Algorithm 1 in Section \ref{section:BS}.

\begin{remark}\label{MLE:Remark}
We examine the relationships and differences between the LEE method and the global or local maximum likelihood approaches.
Specifically, within our framework, the logarithm of the global log-likelihood can be written as
$$
\sum_{i=1}^n \sum_{j\neq i}\bigg[\int_0^\tau \big[\alpha_i(s)+\beta_j(s)+Z_{ij}(s)^\top\gamma(s)\big]dN_{ij}(s)
-\int_0^\tau \exp\{\alpha_i(s)+\beta_j(s)+Z_{ij}(s)^\top\gamma(s)\}dt\bigg].
$$
Since the global log-likelihood
involves integrals and the parameters $\alpha_i(s)$, $\beta_j(s)$, and $\gamma(s)$
are unknown functions, it is challenging to directly optimize the aforementioned log-likelihood.
To address this issue, we focus on estimating the unknown parameters at an arbitrarily given time point $t$.
Under the ``localness" assumption,
the local log-likelihood for the parameters at a given time point $t$ is
\begin{align*}
\sum_{i=1}^n \sum_{j\neq i}\bigg[\int_0^\tau\mathcal{K}_h(s-t)& \big[\alpha_i(t)+\beta_j(t)+Z_{ij}(s)^\top\gamma(t)\big]dN_{ij}(s)\\
&-\int_0^\tau\mathcal{K}_h(s-t)\exp\{\alpha_i(t)+\beta_j(t)+Z_{ij}(s)^\top\gamma(t)\}ds\bigg].
\end{align*}
If we set two different bandwidths, $h_1$ and $h_2$,
for estimating the degree parameter and the homophily parameter, respectively,
the method of local estimating equations and the local likelihood method produce exactly the same estimating equations.
That is, when estimating $\alpha_i(t)$ or $\beta_j(t)$, we use the local likelihood with the bandwidth $h_1$:
\begin{align*}
&\sum_{i=1}^n \sum_{j\neq i}\bigg[\int_0^\tau \mathcal{K}_{h_1}(s-t)(\alpha_i(t) + \beta_j(t) + Z_{ij}(s)^\top \gamma(t))dN_{ij}(s) \\
&\qquad\qquad\qquad- \int_0^\tau\mathcal{K}_{h_1}(s-t)
\exp(\alpha_i(t) + \beta_j(t) + Z_{ij}(s)^\top \gamma(t))ds\bigg],
\end{align*}
and when estimating $\gamma(t)$,  we use the local likelihood with the bandwidth $h_2$:
\begin{align*}
&\sum_{i=1}^n \sum_{j\neq i}\bigg[\int_0^\tau \mathcal{K}_{h_2}(s-t)(\alpha_i(t) + \beta_j(t) + Z_{ij}(s)^\top \gamma(t))dN_{ij}(s) \\
&\qquad\qquad\qquad- \int_0^\tau\mathcal{K}_{h_2}(s-t)
\exp(\alpha_i(t) + \beta_j(t) + Z_{ij}(s)^\top \gamma(t))ds\bigg].
\end{align*}
Taking derivatives with respect to $\alpha_{i}(t),\ \beta_j(t)$ and $\gamma(t)$,
we obtain the local estimating equations in \eqref{eq2.5}.
However, the bandwidths in the local likelihood usually share the same value when estimating different parameters.
In our context, the convergence rates of the resulting estimators for the homophily parameters and the degree parameters
are significantly different; see Theorems \ref{theorem-central-gamma} and \ref{theorem-central-degree}.
Therefore, we employ different bandwidths to balance the trade-off between the bias and variance of all estimators.
\end{remark}

\section{Theoretical properties}
\label{section:asymptotic}

In this section, we present the consistency and asymptotic normality of the estimator.
We first present some notations and assumptions.
Denote $\|a\|_{\infty}=\max_{1\le i\le n}|a_i|$ for a vector $a=(a_1,\dots,a_n)^\top$
and $\|A\|_{\max}=\max_{1\le i,j\le 2n-1} |a_{ij}|$ for an matrix $A=(a_{ij})\in\mathbb{R}^{(2n-1)\times(2n-1)}$.
Let $\eta(t)=(\alpha(t)^\top,\beta(t)^\top)^\top$ and $\pi_{ij,\gamma}^*(t)=\alpha_i^*(t)+\beta_j^*(t)+Z_{ij}(t)^\top\gamma(t)$.
Denote $V_{\gamma\gamma}(t)=[n(n-1)]^{-1}\sum_{i=1}^n\sum_{j\neq i}^n\mathbb{E}[Z_{ij}(t)^{\otimes 2}e^{\pi_{ij,\gamma}^*(t)}]$
and $V_{\gamma\eta^*}(t)=(\varpi_{1,\gamma},\dots,\varpi_{2n-1,\gamma})$,
where  $a^{\otimes2}=aa^\top$ for any vector $a,$ $\varpi_{i,\gamma}=(n-1)^{-1}\sum_{j\neq i}\mathbb{E}[Z_{ij}(t)e^{\pi_{ij,\gamma}^*(t)}]$ for $i\in[n]$
and $\varpi_{n+i,\gamma}=(n-1)^{-1}\sum_{j\neq i}\mathbb{E}[Z_{ji}(t)e^{\pi_{ij,\gamma}^*(t)}]$ for $i\in[n-1]$.
Define $v_{ij,\gamma}^*(t)$ as
\begin{align*}
v_{ii,\gamma}^*(t)=&(n-1)^{-1}\sum_{j\neq i}\mathbb{E}\{e^{\pi_{ij,\gamma}^*(t)}\},\quad~ i\in[n],\\
v_{(n+j)(n+j),\gamma}^*(t)=&(n-1)^{-1}\sum_{i\neq j}\mathbb{E}\{e^{\pi_{ij,\gamma}^*(t)}\},\quad~ j\in[n-1],\\
v_{i(n+j),\gamma}^*(t)=&v_{(n+j)i,\gamma}^*(t)=(n-1)^{-1}\mathbb{E}\{e^{\pi_{ij,\gamma}^*(t)}\},~i\in[n],~j\in[n-1],~i\neq j,
\end{align*}
and $v_{ij,\gamma}^*(t)=v_{ji,\gamma}^*(t)=0$ otherwise.
Let $\widehat V_{\eta\eta}(t,\gamma(t))$ denote the first-order derivative of $F(\theta(t))$ with respect to $\eta(t)$
and $V_{\eta^*\eta^*}(t,\gamma(t))=(v_{ij,\gamma}^*(t):1\le i,j\le 2n-1)$.
By Lemma B.4 in the Supplementary Material, we have
$$\|\mathbb{E}\{\widehat V_{\eta^*\eta^*}(t,\gamma(t))\}-V_{\eta^*\eta^*}(t,\gamma(t))\|_{\max}\rightarrow 0
~~\text{as}~~n\rightarrow\infty.$$
Thus, $v_{ij,\gamma}^*(t)$ can be viewed as the $(i,j)$th component of the expectation of the first-order derivative of $F(\theta(t))$ with respect to $\eta(t)$.
In addition, $V_{\eta^*\eta^*}(t,\gamma(t))$ is diagonally dominant and, therefore, positive definite.
This implies that its inverse exists.
For simplicity, let $v_{ij}^*(t)=v_{ij,\gamma^*}^*(t).$
Further, we define $S^*(t)=(s_{ij}^*(t): i, j\in [2n-1])$ as an approximation of the inverse of $V_{\eta^*\eta^*}(t,\gamma^*(t)),$
where $s_{ij}^*(t)$ is defined as
\begin{align*}
s_{ij}^*(t)=
\begin{cases}
\frac{\delta_{ij}}{v_{ii}^*(t)} + \frac{1}{v_{(2n)(2n)}^*(t)}, &~~ 1\le  i, j\le n~~
\text{or}~~ n+1\le i,j\le 2n-1,\\
\frac{1}{v_{(2n)(2n)}^*(t)}, &~~~\text{otherwise},
\end{cases}
\end{align*}
and $v_{(2n)(2n)}^*(t)=\sum_{i=1}^nv_{ii}^*(t)-\sum_{i=1}^n\sum_{j\neq i}^{2n-1}v_{ij}^*(t).$
Here, $\delta_{ij}=1$ if $i=j$ and $\delta_{ij}=0$ otherwise.
Let $f_{ij}(z|t)$ represent the density function of $Z_{ij}(t)$,
and $f_{ij}(z|s,t)$ denote the joint density function of $(Z_{ij}(s),Z_{ij}(t))^\top$.
We impose the following conditions.

\begin{condition}
\label{condition:CB1}
The covariates $Z_{ij}(t)~(1\le i\neq j\le n)$ are predictable
and satisfy $\|Z_{ij}(t) \|_\infty\le \kappa_n$ almost surely,
where $\kappa_n$ may diverge with $n$.
The functions $f_{ij}(z|t)$ and $f_{ij}(z|s,t)$ are twice continuously differentiable with respect to $t$ and $(s,t)$, respectively.
Additionally, $\mathcal{M}_{ij}(\cdot)~(1\le i\neq j \le n)$ have bounded total variations,
i.e., for all $1\le i\neq j \le n$, $\int_0^\tau|d\mathcal{M}_{ij}(t)|<\kappa_0$ almost surely,
where $\kappa_0$ is a constant free of $n$.
\end{condition}

\begin{condition}
\label{condition:PB2}
The functions $\alpha_i^*(t),~\beta_j^*(t)$, and $\gamma^*(t)$ are twice continuously differentiable.
In addition, each element of $\gamma^{(l)}(t)~(l=0,1,2)$ is a bounded function and
\begin{align*}
\theta^*(t)\in\mathcal{B}=\bigg\{\theta(t):
\sup_{t\in[a,b],|z|\le \kappa_n}|\alpha_i^{(l)}(t)+\beta_j^{(l)}(t)+z^\top\gamma^{(l)}(t)|\le q_{n}~(l=0,1,2)\bigg\},
\end{align*}
where $q_{n}$ is allowed to diverge with $n$, $[a,b]$ is a subset of $(0,\tau)$
and $g^{(l)}(t)$ denotes
the $l$th-order derivative of $g(t)$.
\end{condition}

\begin{condition}\label{condition:H3}
There exists some positive constant $\varsigma$ such that
\begin{align*}
\inf_{t\in[a,b]}\inf_{\gamma(t)\in \mathcal{J}}\rho_{\min}\big\{H_Q(\gamma(t))\big\}> \varsigma>0,
\end{align*}
where $H_Q(\gamma(t))=\lim_{n\rightarrow\infty}\mathbb{E}\{V_{\gamma\gamma}(t)-V_{\gamma\eta^*}(t)
V_{\eta^*\eta^*}(t,\gamma(t))^{-1}V_{\gamma\eta^*}(t)^\top\},$
$\rho_{\min}(M)$ denotes the smallest eigenvalue of the matrix $M,$
and $\mathcal{J}$ denotes the class of functions that take values from a compact set.
\end{condition}

\begin{condition}\label{condition:kernel4}
$\mathcal{K}(x)$ is a symmetric density function with support $[-1,1].$
Additionally, there exits a constant $\kappa_1$ such that $\sup_{x\in[-1,1]}|\mathcal{K}^{(1)}(x)|\le \kappa_1$.
\end{condition}

\begin{condition}\label{condition:bandwidth5}
$nh_1^2\rightarrow\infty$ and $nq_{n}^4e^{12q_{n}}h_1^5\rightarrow 0$ as $n\rightarrow \infty.$
Moreover, $h_2=O(h_1^2).$
\end{condition}

The predictability assumption in Condition \ref{condition:CB1} is mainly used to guarantee the martingale property of $\mathcal{M}_{ij}(t),$
which is widely used in various studies
such as survival data analyses and econometrics; see \cite{FH1991}.
If $Z_{ij}(t)$ is a fixed variable such as age or gender,
it is predictable with respect to $\{\mathcal{F}_t:t\ge 0\}$.
The boundedness of covariates is required to simplify the proof of the following theorems;
however, this can be relaxed to the sub-Gaussian case.
The twice differentiable assumption in Condition \ref{condition:CB1} is also adopted by \cite{kreib2019}.
The assumption of bounded total variation in Condition \ref{condition:CB1} is realistic for studies with a finite follow-up duration.
For instance, when $q_n$ is a constant, we can directly assume that $N_{ij}(\tau) < \infty$ almost surely,
which has been widely utilized in the context of recurrent event analysis (e.g., \citealp{AG1982,L2000}).
The uniform consistency of the estimators requires this condition since the number of unknown parameters diverges with $n$.
This is different from the existing literature (e.g., \citealp{perry2013point,kreib2019,Sit2020}),
where the parameter estimators are either not time-varying or pointwise consistent.

Condition \ref{condition:PB2} concerns
the network sparsity assumption,
which is akin to condition (A3) in \cite{kreib2019} and condition (A.2) in \cite{Alexander2021}.
This condition implies a ``localness" of the true parameter values.
It also requires that the unknown functions are sufficiently smooth
and the sum of the unknown functions and the sum of their first and second derivatives are bounded above by $q_n$.
In practical applications, one way to verify this assumption is based on \eqref{eq2.4}.
Specifically, we estimate the left-hand side of \eqref{eq2.4} using its observed value
$\sum_{i=1}^n\sum_{j\neq i}N_{ij}(s,t)/(t-s)$ for $0\le s<t\le \tau$.
If the value of $\sum_{i=1}^n\sum_{j\neq i}N_{ij}(s,t)/[n^2(t-s)]$ is quite small,
then equation \eqref{eq2.4} and Condition 2 approximately hold.
Condition \ref{condition:H3} implies that $H_Q(\gamma(t))$ is positive definite,
which ensures the identifiability of $\gamma^*(t)$.

Condition \ref{condition:kernel4} is a standard assumption in nonparametric statistics.
The kernel function affects the convergence rate of the estimators only
by multiplicative constants and thus has little impact on the rate of convergence (e.g. \citealp{FG1996}).
Condition \ref{condition:bandwidth5} determines the vanishing rate of bandwidths $h_1$ and $h_2$.
Further details on the selections of $h_1$ and $h_2$ are provided in Remark \ref{Rem2} and Section \ref{section:BS}.

\begin{theorem}\label{theorem:consistency}
Suppose  that Conditions \ref{condition:CB1}-\ref{condition:bandwidth5} hold.
If
\begin{align}\label{condition:TM1}
\kappa_n^2\sqrt{e^{37q_n}\log(nh_1)/(nh_1)}\rightarrow 0~~\text{as}~~n\rightarrow \infty,
\end{align}
then the solution $\widehat\theta(t)=(\widehat\eta(t)^\top,\widehat\gamma(t)^\top)^\top$ to equation \eqref{eq2.5} exists
and satisfies
\begin{align*}
\sup_{t\in[a,b]}\|\widehat{\eta}(t)-\eta^*(t)\|_\infty=o_p(1)~~\text{and}~~
\sup_{t\in[a,b]}\|\widehat{\gamma}(t)-\gamma^*(t)\|_\infty=o_p(1),
\end{align*}
where $[a,b]$ is a subset of $(0,\tau).$
\end{theorem}

\begin{remark}\label{Rem2}
Condition (6) in Theorem \ref{theorem:consistency} implies that
the sparser the networks are, the larger the bandwidth $h_1$ should be.
An intuitive explanation is that
we need to expand the bandwidth to include more data to accurately estimate time-varying parameters
when there are too few edges in a network at some time point.
At the same time, the bias increases with $h_1$.
The condition $nq_{n}^4e^{12q_n}h_1^5\rightarrow 0$ restricts the
growth rate of $h_1$ and thus balances bias and variance,
while also providing an upper bound for bandwidth selection in sparse networks.

Additionally, under the case that all $\pi_{ij}(t):=\alpha_i(t) + \beta_j(t) + Z_{ij}^\top \gamma(t)$ for $i\neq j$ are close to each other, we have
$q_n \approx -\log\{(n^2\tau)^{-1}\sum_{i=1}^n\sum_{j\neq i}\mathbb{E}[N_{ij}(\tau)] \},$
which is the logarithm of the expected number of edges averaged over the time interval $(0,\tau]$.
Here, $q_n$ determines the sparsity level of the network.
In this case, the statements on Condition 5, condition (6) and additional conditions in Theorem \ref{theorem:consistency}
are naturally formulated in terms of the expected number of edges via the quantity $q_n$.
In general cases, there is no precise formulation of the statements related to the expected number of edges in the conditions,
but the smallest
expected number of edges is imposed through the quantity $q_n$.

Moreover, condition (6) in Theorem \ref{theorem:consistency} implies that $q_n$ can be chosen as $c\log(nh_1)$, where $c$ is a positive constant.
According to Condition 5, we can choose $h_1=o(n^{-(1+12c)/(5+12c)})$,  ignoring the logarithmic factor.
If $q_n$ is a constant, then it corresponds to $c=0$, where the network is dense.
In this case, Condition 5 reduces to $nh_1^5\rightarrow0$ as $n\rightarrow \infty$
and the bandwidth $h_1$ can be chosen as $h_1=o(n^{-1/5})$, which agrees with
those conditions on the bandwidth of the classical kernel smoothing method (e.g., \citealp{FG1996}).
\end{remark}

We define $\mu_{0}=\int \mathcal{K}^2(u)du,$ and write $H_Q(t)=H_Q(\gamma^*(t))$.
Next, we present the asymptotic normality of $\widehat\gamma(t)$.

\begin{theorem}
	\label{theorem-central-gamma}
Suppose that Conditions \ref{condition:CB1}-\ref{condition:bandwidth5} hold.
If $\kappa_n^2 \sqrt{e^{37q_{n}}\log(nh_1)^{3}/(nh_1)}=o(1)$,
then $(Nh_2)^{1/2}\Psi(t)^{-1/2}\{\widehat{\gamma}(t)-\gamma^*(t)- [H_Q(t)]^{-1}b_*(t)\}$ converges in distribution to a $p$-dimensional standard normal distribution,
where $\Psi(t)=[H_Q(t)]^{-1}\Sigma(t) [H_Q(t)]^{-1}$.
Here, $\Sigma(t)$ and $b_*(t)$ are
\begin{align*}
\Sigma(t)=&\frac{\mu_0}{N}\sum_{i=1}^n\sum_{j=1,\ j\neq i}^n\mathbb{E}\bigg[\Big(Z_{ij}(t)
-V_{\gamma^*\eta^*}(t)S^*(t)\iota_{ij}\Big)^{\otimes 2}e^{\pi_{ij}^*(t)}\bigg],\\
b_*(t)=&\frac{\mu_{0}}{2Nh_1}\bigg[ \sum_{i=1}^n \frac{\sum_{j\neq i}\mathbb{E}(Z_{ij}(t) \exp\{\pi_{ij}^*(t)\})}
{\sum_{j\neq i} \mathbb{E}(\exp\{\pi_{ij}^*(t)\})}+\sum_{j=1}^n \frac{  \sum_{i\neq j}\mathbb{E}(Z_{ij}(t) \exp\{\pi_{ij}^*(t)\})}
{\sum_{i\neq j} \mathbb{E}(\exp\{\pi_{ij}^*(t)\})}\bigg],
\end{align*}
where $N=n(n-1)$ and $\iota_{ij}$ is a $(2n-1)$-dimensional vector with the $i$th and $(n+j)$th elements being one
and others being zero.
\end{theorem}

\begin{remark}
Theorem \ref{theorem-central-gamma} indicates that the estimator $\widehat\gamma(t)$ is biased,
with the bias expressed as $[H_Q(t)]^{-1}b_*(t).$
In the econometric literature, this phenomenon is referred to as the incidental parameter problem \citep{NS1948,FW2016,Dzemski2017}.
A similar issue exists in the static network literature \citep{Graham2017, yan2019}.
The bias is due to different convergence rates of $\widehat\eta(t)$ and $\widehat \gamma(t)$.
\end{remark}

The asymptotic normality of $\widehat\eta(t)$ is presented in the following theorem.

\begin{theorem}\label{theorem-central-degree}
Suppose that Conditions \ref{condition:CB1}-\ref{condition:bandwidth5} hold. If
$\kappa_n^4 e^{33q_{n}}\log nh_1/\sqrt{nh_1}=o(1)$,
then for any fixed positive integer $k$,
$(nh_1)^{1/2}\big\{(\mu_{0}S^*(t))^{-1/2}[\widehat{\eta}(t)- \eta^*(t)]\big\}_{1:k}$
converges in distribution to a $k$-dimensional standard normal distribution,
where $x_{1:k}=(x_1,\dots,x_k)^\top$ for any vector $x\in \mathbb{R}^{2n-1}$.
\end{theorem}

Next, we estimate the variances of $\widehat\eta_i(t)$ and $\widehat\gamma_j(t)$.
We consider a sandwich-type estimator for the variance of $\sqrt{nh_1}\{\widehat\eta_i(t)-\eta_i^*(t)\}.$
Let $\widehat\pi_{ij}(s,t)=\widehat\alpha_{i}+\widehat\beta_j(t)+Z_{ij}(s)^\top\widehat\gamma(t).$
We first estimate $[V_{\eta^*\eta^*}(t,\gamma^*(t))]^{-1}$ using $\widehat{S}(t)=(\widehat{s}_{ij}(t): i,j\in [2n-1])$,
where $\widehat{s}_{ij}(t)$ is defined as
\[
\widehat{s}_{ij}(t)=
\begin{cases}
\frac{\delta_{ij}}{\widehat v_{ii}(t)} + \frac{1}{\widehat v_{(2n)(2n)}(t)},  &~~~ 1\le i, j\le n~~
\text{or}~~ n+1\le i, j\le 2n-1, \\
-\frac{1}{\widehat v_{(2n)(2n)}(t)}, & ~~~\text{otherwise}.
\end{cases}
\]
Here, $\widehat{v}_{(2n)(2n)}=\sum_{i=1}^n \widehat{v}_{ii}(t)-\sum_{i=1}^n\sum_{j=1,\ j\neq i}^{2n-1} \widehat{v}_{ij}(t),$
\begin{align*}
\widehat v_{ii}(t)=&(n-1)^{-1}\sum_{j\neq i}\int_0^\tau\mathcal{K}_{h_1}(s-t)e^{\widehat\pi_{ij}(s,t)}ds,\quad~i\in[n],\\
\widehat v_{(n+j)(n+j)}(t)=&(n-1)^{-1}\sum_{i\neq j}\int_0^\tau\mathcal{K}_{h_1}(s-t)e^{\widehat\pi_{ij}(s,t)}ds,\quad~ j\in[n-1],\\
\widehat v_{i(n+j)}(t)=&\widehat v_{(n+j)i}(t)=(n-1)^{-1}\int_0^\tau\mathcal{K}_{h_1}(s-t)e^{\widehat\pi_{ij}(s,t)}ds,~i\in[n],~j\in[n-1],~i\neq j,
\end{align*}
and $\widehat v_{i,j}(t)=\widehat v_{j,i}(t)=0$ otherwise.

Second, we estimate the covariance $\Omega(t)$ of $\sqrt{h_1/n}\int_{0}^{\tau}\mathcal{K}_{h_1}(s-t)d\widetilde{\mathcal{M}}(s)$.
By Lemma B.4 in the Supplementary Material,
the distribution of $(nh_1)^{1/2}\big\{\widehat{\eta}(t)-\eta^*(t)\big\}$ is asymptotically  equivalent to
$S^*(t)(h_1/n)^{1/2} \int_{0}^{\tau}\mathcal{K}_{h_1}(s-t)d\widetilde{\mathcal{M}}(s),$
where $\widetilde{\mathcal{M}}(t)=(\widetilde{M}_{1}(s),\dots,\widetilde{M}_{2n-1}(t))^\top$ with
$$
\widetilde M_{i}(t)=\sum_{k\ne i} \mathcal{M}_{ik}(t)~ (i\in[n])~~~
 \text{and}~~~\widetilde M_{n+i}(t)=\sum_{k\ne i} \mathcal{M}_{ki}(t)~(i\in[n-1]).
 $$
Note that $\widetilde M_{i}(t) $ is a sum of local square-integrable martingales.
Therefore, by the martingale properties, we can estimate $\Omega(t)$ using $\widehat \Omega(t)=(\widehat\omega_{ij}(t): i,j\in[2n-1])$,
where $\widehat\omega_{ij}(t)$ is defined as
\begin{align*}
\widehat\omega_{ii}(t)=&\frac{h_1}{n}\sum_{j\neq i}\int_0^{\tau}\mathcal{K}_{h_1}^2(s-t)dN_{ij}(s),~i\in[n],\\
\widehat\omega_{n+j,n+j}(t)=&\frac{h_1}{n}\sum_{i\neq j}\int_0^{\tau}\mathcal{K}_{h_1}^2(s-t)dN_{ij}(s),~j\in[n-1],\\
\widehat\omega_{i,n+j}(t)=\widehat\omega_{n+j,i}(t)=&\frac{h_1}{n}\int_0^{\tau}\mathcal{K}_{h_1}^2(s-t)dN_{ij}(s), \qquad i\in[n],~j\in[n-1], ~i\neq j,\\
\widehat\omega_{ij}(t)=\widehat\omega_{ji}(t)=&0,\quad\quad\text{otherwise}.
\end{align*}
Finally, we estimate the variance of $\sqrt{nh_1}\{\widehat\eta_i(t)-\eta_i^*(t)\}$
by the $i$th diagonal element $\widehat\sigma_{ii}(t)$ of $\widehat S(t)\widehat \Omega(t)\widehat S(t)$.

Using arguments similar to those in the proof of Lemma B.4 in the Supplementary Material, we can show that
$\widehat S(t)\widehat \Omega(t)\widehat S(t)$ is a consistent estimator of the covariance of $\sqrt{nh_1}\{\widehat\eta(t)-\eta^*(t)\}$.
Let $z_{\alpha}$ be the $100(1-\alpha)$th percentile of the standard normal distribution.
Then, the $(1-\alpha)$-confidence interval  for $\eta_i^*(t)$ is given by
\begin{align}\label{confidence:eta}
\widehat \eta_i(t)\pm (nh_1)^{-1/2}z_{\alpha/2}\widehat\sigma_{ii}^{1/2}(t),~~i\in[2n-1].
\end{align}

Theorem \ref{theorem-central-gamma} indicates that $\widehat\gamma(t)$ is biased.
Therefore, bias correction is necessary for accurate statistical inference.
To do that, we define
\begin{align*}
\widehat b(t)=&\frac{h_1}{2N}\bigg[\sum_{i=1}^n\frac{\sum_{j\neq i}\int_0^\tau Z_{ij}(s)\mathcal{K}_{h_1}^2(s-t)dN_{ij}(s)}
{\sum_{j\neq i} \int_0^\tau\mathcal{K}_{h_1}(s-t)dN_{ij}(s)}+\sum_{j=1}^n\frac{\sum_{i\neq j}\int_0^\tau Z_{ij}(s)\mathcal{K}_{h_1}^2(s-t)dN_{ij}(s)}
{\sum_{i\neq j} \int_0^\tau\mathcal{K}_{h_1}(s-t)dN_{ij}(s)}\bigg],\\
\widehat H_Q(t)=&\frac{1}{N}\sum_{i=1}^n\sum_{j\neq i}\int_0^\tau Z_{ij}(s)^{\otimes2}\mathcal{K}_{h_2}(s-t)dN_{ij}(s)
-\widehat  V_{\hat\gamma,\hat\eta}(t)\widehat S(t)\widehat V_{\hat\eta,\hat\gamma}(t),
\end{align*}
where $\widehat V_{\hat \gamma,\hat\eta}(t)=N^{-1}(\widehat u_{1}(t),\dots,\widehat u_{2n-1}(t))$
and $\widehat  V_{\hat\eta,\hat\gamma}(t)=\widehat V_{\hat \gamma,\hat\eta}(t)^\top$
with
\begin{align*}
   &\widehat u_{i}(t)=\sum_{j\neq i}\int_0^\tau\mathcal{K}_{h_1}(s-t)Z_{ij}(s)dN_{ij}(s),~~~i\in[n],\\
\text{and}~~&\widehat u_{n+j}(t)=\sum_{i\neq j}\int_0^\tau\mathcal{K}_{h_1}(s-t)Z_{ij}(s)dN_{ij}(s),~~~j\in[n-1].
\end{align*}
In addition, by the martingale properties, $\Sigma(t)$ can be estimated  by
\begin{align*}
\widehat\Sigma(t)=&\frac{h_2}{N}\sum_{i=1}^n\sum_{j\neq i}\int_0^{\tau}\Big(Z_{ij}(u)-\widehat V_{\hat\gamma,\hat\eta}(u)\widehat S(u)\iota_{ij}\Big)^{\otimes 2}\mathcal{K}_{h_2}^2(u-t)dN_{ij}(u).
\end{align*}
Finally, we estimate the bias $[H_Q(t)]^{-1}b_*(t)$ using $[\widehat H_Q(t)]^{-1}\widehat b(t)$,
and estimate the covariance $\Psi(t)$ with $\widehat\Psi(t)=[\widehat H_Q(t)]^{-1}\widehat\Sigma(t)[\widehat H_Q(t)]^{-1}$.
By applying arguments similar to those in the proof of Lemma B.4 in the Supplementary Material,
we can show $\|[\widehat H_Q(t)]^{-1}\widehat b(t)-[H_Q(t)]^{-1} b_*(t)\|_{\infty}=o_p(1)$ and $\|\widehat \Psi(t)-\Psi(t)\|_{\max}=o_p(1).$
Let $\widetilde b_j(t)$ be the $j$th element of $[\widehat H_Q(t)]^{-1}\widehat b(t)$
and $\widehat \psi_{jj}(t)$ be the $j$th diagonal element of $\widehat \Psi(t).$
Then, the $(1-\alpha)$-confidence interval  for $\gamma_j^*(t)$ is given by
\begin{align}\label{confidence:gamma}
    \widehat \gamma_j(t)-\widetilde b_j(t)\pm (Nh_2)^{-1/2}z_{\alpha/2}\widehat \psi_{jj}^{1/2}(t),~~j\in[p].
\end{align}

\section{Hypothesis testing}
\label{section:test}

In this section, we develop hypothesis testing methods for the time-varying parameters and introduce a method for assessing model
goodness-of-fit. Specifically, in Section \ref{sec:Trt},  we propose a max-norm test statistic to test whether the in-degree and out-degree parameters exhibit temporal variation. In Section \ref{sec:dht}, we test whether there is degree heterogeneity and
develop a resampling approach to obtain the critical values for the proposed tests.
In Section \ref{sec:GOF}, we introduce a graphical diagnostic method for assessing goodness-of-fit of the degree-corrected Cox network model.

\subsection{Testing for temporal variation}\label{sec:Trt}

If both $\eta^*(t)$ and $\gamma^*(t)$ remain constant over time,  the network is  considered static.
Therefore, it is of interest to examine whether $\eta^*(t)$ and $\gamma^*(t)$ exhibit temporal variation.
Our problem can be formulated as follows:
\begin{align*}
& H_{0\eta}: \eta^*(t)=\eta^*~~ \text{for~all}~~t\in [a,b]~~
\text{versus }~~ H_{1\eta}: \eta^*(t)\neq \eta^*~~\text{for some} \ t\in [a,b]\\
\text{and}~~& H_{0\gamma}: \gamma^*(t)=\gamma^*~~ \text{for~all}~~ t\in [a,b]~
\text{ versus }~~ H_{1\gamma}: \gamma^*(t)\neq \gamma^*~~\text{for some}\ t\in [a,b],
\end{align*}
where $\eta^*$ and $\gamma^*$ are some unspecified vectors.

We see that if either $H_{0\eta}$ or $H_{0\gamma}$ holds, then model \eqref{eq2.1} is a semi-parametric model.
If both $H_{0\eta}$ and $H_{0\gamma}$ hold, then model \eqref{eq2.1} becomes a completely parametric model.
\citet{Alexander2021} proposed a test statistic that compares the completely parametric
and non-parametric estimator using the $\ell_2$-distance
to test $H_{0\eta}$ and $H_{0\gamma}.$
However, because the dimension of the parameters grows with the number of nodes under model \eqref{eq2.1},
the method developed for a fixed dimension
in \cite{Alexander2021} cannot be applied directly.
We consider the following test statistics to test $H_{0\eta}$ and $H_{0\gamma},$ respectively:
\begin{align*}
\mathcal{T}_{\eta}&=\max_{i\in[2n-1]}\sup_{a\le t_1<t_2\le b}\sqrt{nh_1}\big|\widehat{\eta}_i(t_1)-\widehat{\eta}_i(t_2)\big|/\widehat\vartheta_{i,\eta}^{1/2}(t_1,t_2),\\
\mathcal{T}_{\gamma}&=\max_{j\in[p]}\sup_{a\le t_1<t_2\le b}\sqrt{Nh_2}\big|\widehat{\gamma}_j(t_1)-\widetilde{b}_j(t_1)-\widehat{\gamma}_j(t_2)+\widetilde b_{j}(t_2)\big|/\widehat\vartheta_{j,\gamma}^{1/2}(t_1,t_2),
\end{align*}
where $\widehat\vartheta_{i,\eta}(t_1,t_2)=\widehat\sigma_{ii}(t_1)+\widehat\sigma_{ii}(t_2)$
and $\widehat\vartheta_{j,\gamma}(t_1,t_2)=\widehat \psi_{jj}(t_1)+\widehat \psi_{jj}(t_2).$

The test statistics $\mathcal{T}_{\eta}$ and $\mathcal{T}_\gamma$ are close to zero under the nulls
$H_{0\eta}$ and $H_{0\gamma},$ respectively.
Therefore, we reject $H_{0\eta}$ if $\mathcal{T}_\eta>c_\eta(\nu),$
and reject $H_{0\gamma}$ if $\mathcal{T}_\gamma>c_\gamma(\nu),$
where $c_\eta(\nu)$ and $c_\gamma(\nu)$ are the critical values.
In practice, the critical values $c_\eta(\nu)$ and $c_\gamma(\nu)$ are unknown.
To obtain these critical values, we consider a resampling approach.
Using arguments similar to those in the proof of Theorem \ref{theorem-central-degree},
we can show that under the null $H_{0\eta},$
the distribution of $\sqrt{nh_1}[\widehat{\eta}(t_1)-\widehat{\eta}(t_2)]$ is asymptotically equivalent to
$\sqrt{h_1/n}[\widehat S(t_1)\int_0^{\tau} \mathcal{K}_{h_1}(u-t_1)d\widetilde{\mathcal{M}}(u)
-\widehat S(t_2)\int_0^{\tau} \mathcal{K}_{h_1}(u-t_2)d\widetilde{\mathcal{M}}(u)],$
and  $\sqrt{nh_1}[\widehat{\eta}(t_1)-\eta^*(t_1)] $ and $\sqrt{nh_1}[\widehat{\eta}(t_2)-\eta^*(t_2)]$
are asymptotically independent with $t_1\neq t_2.$
In addition, using arguments similar to those in the proof of Theorem \ref{theorem-central-gamma},
we find that under the null $H_{0\gamma},$
the distribution of $\sqrt{Nh_2}[\widehat{\gamma}(t_1)-\widehat{b}(t_1)-\widehat{\gamma}(t_2)+\widehat b(t_2)]$ is asymptotically equivalent to
\begin{align*}
&\sqrt{\frac{h_2}{N}}\sum_{i=1}^n\sum_{j\neq i} \bigg[\int_0^{\tau}\mathcal{K}_{h_2}(u-t_1)\Big(Z_{ij}(u)-\widehat V_{\hat\gamma,\hat\eta}(t_1)\widehat S(t_1)\iota_{ij}\Big)d\mathcal{M}_{ij}(u)\\
& \hspace{1in}-\int_0^{\tau}\mathcal{K}_{h_2}(u-t_2)\Big(Z_{ij}(u)-\widehat V_{\hat\gamma,\hat\eta}(t_2)\widehat S(t_2)\iota_{ij}\Big)d\mathcal{M}_{ij}(u)\bigg],
\end{align*}
and $\sqrt{Nh_2}[\widehat{\gamma}(t_1)-\widehat{b}(t_1)-\gamma^*(t_1)]$ and $\sqrt{Nh_2}[\widehat{\gamma}(t_2)-\widehat b(t_2)-\gamma^*(t_2)]$ are asymptotically independent with $t_1\neq t_2.$
A direct calculation shows that  the variance function of $\mathcal{M}_{ij}(u)$ is $E\{N_{ij}(u)\}$
(see Theorem 2.5.3 in \citealp{FH1991}).
Following \cite{LFW1994},
we replace $\mathcal{M}_{ij}(u)$ with $N_{ij}(u)G_{ij},$ that is,
\begin{align*}
\widetilde{T}_{\eta}(t_1,t_2)=&\sqrt{\frac{h_1}{n}}\bigg[\widehat S(t_1)\int_0^{\tau} \mathcal{K}_{h_1}(u-t_1)d\widetilde{\mathcal{N}}(u)
-\widehat S(t_2)\int_0^{\tau} \mathcal{K}_{h_1}(u-t_2)d\widetilde{\mathcal{N}}(u)\bigg],\\
\widetilde{T}_{\gamma}(t_1,t_2)=&\sqrt{\frac{h_2}{N}}\sum_{i=1}^n\sum_{j\neq i}\bigg\{\widehat H_Q^{-1}(t_1)\int_0^{\tau}\mathcal{K}_{h_2}(u-t_1)\Big(Z_{ij}(u)-\widehat V_{\hat\gamma,\hat\eta}(t_1)\widehat S(t_1)\iota_{ij}\Big)dN_{ij}(u)G_{ij}\\
&\hspace{0.65in}-\widehat H_Q^{-1}(t_2)\int_0^{\tau}\mathcal{K}_{h_2}(u-t_2)\Big(Z_{ij}(u)-\widehat V_{\hat\gamma,\hat\eta}(t_2)\widehat S(t_2)\iota_{ij}\Big)dN_{ij}(u)G_{ij}\bigg\},
\end{align*}
where $\widetilde{\mathcal{N}}(u)=(\widetilde N_1(u),\dots,\widetilde N_{2n-1}(u))^\top$ with
\begin{align*}
\widetilde N_i(u)=\sum_{k\ne i} N_{ik}(u)G_{ik}~ (i\in[n])~~~
\text{and}~~~\widetilde N_{n+i}(u)=\sum_{k\ne i} N_{ki}(u)G_{ki}~(i=[n-1]).
\end{align*}
Here, $G_{ii}=0$ and $G_{ij}\ (i\neq j\in[n])$ are independent standard normal variables which are independent of the observed data.
We define
\begin{align*}
&\widehat\vartheta_{\eta}(t_1,t_2)=\text{diag}\{\widehat\vartheta_{1,\eta}(t_1,t_2),\dots,\widehat\vartheta_{2n-1,\eta}(t_1,t_2)\}\\
\text{and}~~&\widehat\vartheta_{\gamma}(t_1,t_2)=\text{diag}\{\widehat\vartheta_{1,\gamma}(t_1,t_2),\dots,\widehat\vartheta_{p,\gamma}(t_1,t_2)\},
\end{align*}
where $\text{diag}\{a_1,\dots,a_n\}$ denotes a diagonal matrix with $a_i$ as its $i$th diagonal element.
By repeatedly generating normal random samples $G_{ij},$
the distributions of $\mathcal{T}_{\eta}$ and $\mathcal{T}_{\gamma}$
can be respectively approximated by the conditional distributions of $\widetilde{\mathcal{T}}_{\eta}$ and $\widetilde{\mathcal{T}}_{\gamma}$ given the observed data, where $\widetilde{\mathcal{T}}_{\eta}=\sup_{a\le t_1<t_2\le b} \|\widehat\vartheta_{\eta}^{-1/2}(t_1,t_2)\widetilde{T}_{\eta}(t_1,t_2)\|_{\infty}$
and $\widetilde{\mathcal{T}}_{\gamma}=\sup_{a\le t_1<t_2\le b} \|\widehat\vartheta_{\gamma}^{-1/2}(t_1,t_2)\widetilde{T}_{\gamma}(t_1,t_2)\|_{\infty}.$
Then, the critical values $c_\eta(\nu)$ and $c_\gamma(\nu)$ can be obtained from the upper $(1-\nu)$-percentile of the conditional distributions of $\widetilde{\mathcal{T}}_{\eta}$ and $\widetilde{\mathcal{T}}_{\gamma},$ respectively.

\subsection{Tests for degree heterogeneity}\label{sec:dht}

Degree heterogeneity is a notable characteristic frequently observed in real-world networks.
Therefore, it is essential to perform tests to identify the presence of degree heterogeneity in practical situations.
Consider the following hypotheses:
\begin{align*}
& H_{01}: \alpha_{i}^*(t)=\alpha^*(t)~~ \text{for~all}~~ i\in[n]\qquad
\text{ versus }~H_{11}: \alpha_{i}^*(t)\neq \alpha^*(t)~~\text{for some}\ i\in[n],\\
\text{and}~& H_{02}: \beta_{i}^*(t)=\beta^*(t)~~ \text{for~all}~~  i\in[n-1]~
\text{ versus }~ H_{12}: \beta_{i}^*(t)\neq \beta^*(t) \text{for some}\ i\in[n-1],
\end{align*}
where $\alpha^*(t)$ and $\beta^*(t)$ are unspecified functions.
We see that
\begin{itemize}
    \item If $H_{01}$ holds but $H_{02}$ does not, then the network has only in-degree heterogeneity.
    \item If $H_{02}$ holds but $H_{01}$ does not, then the network has only out-degree heterogeneity.
    \item If both $H_{01}$ and $H_{02}$ hold, then the network has no degree heterogeneity.
\end{itemize}
Let $e_{i,j}$ be a $(2n-1)$-dimensional vector in which the $i$th element is $1,$
$j$th element is $-1$, and other elements are zeros.
We consider the following test statistics for $H_{01}$ and $H_{02},$ respectively:
\begin{align*}
\mathcal{D}_\alpha=&\max_{1\le i<j\le n}\sup_{t\in[a,b]}
\sqrt{nh_1}|\widehat \alpha_{i}(t)-\widehat \alpha_{j}(t)|/\widehat\zeta_{ij,\alpha}^{1/2}(t)\\
\text{and}~~~\mathcal{D}_\beta=&\max_{1\le i<j\le n-1}\sup_{t\in[a,b]}
\sqrt{nh_1}|\widehat \beta_{i}(t)-\widehat \beta_{j}(t)|/\widehat\zeta_{ij,\beta}^{1/2}(t),
\end{align*}
where $\widehat\zeta_{ij,\alpha}(t)=e_{i,j}^\top\widehat S(t)\widehat\Omega(t)\widehat S(t) e_{i,j}$
and $\widehat\zeta_{ij,\beta}(t)=e_{n+i,n+j}^\top\widehat S(t)\widehat\Omega(t)\widehat S(t)e_{n+i,n+j}.$

The test statistics $\mathcal{D}_{\alpha}$ and $\mathcal{D}_\beta$ are close to zero under the nulls
$H_{01}$ and $H_{02},$ respectively.
Therefore, we reject $H_{01}$ if $\mathcal{D}_\alpha>c_1(\nu),$
and reject $H_{02}$ if $\mathcal{D}_\beta>c_2(\nu),$
where $c_1(\nu)$ and $c_2(\nu)$ are the critical values.
We consider a resampling approach to obtain these critical values.
Here, we focus only on obtaining $c_1(\nu)$, while
$c_2(\nu)$ can be obtained similarly.
Using arguments similar to those in the proof of Theorem \ref{theorem-central-degree},
we can show that under the null $H_{01},$
the distribution of $\sqrt{nh_1}\{\widehat \alpha_{i}(t)-\widehat \alpha_{j}(t)\}~(i\neq j\in[n])$ is asymptotically equivalent to
$\sqrt{h_1/n}e_{i,j}^\top \widehat S(t)\int_{0}^\tau \mathcal{K}_{h_1}(u-t) d\widetilde{\mathcal{M}}(u).$
As in Section \ref{sec:Trt}, we replace $\mathcal{M}_{is}(u)$ with $N_{is}(u)G_{is}$ in $\widetilde{\mathcal{M}}(u),$
that is, $\widetilde{\mathcal{D}}_{ij,\alpha}(t)=\sqrt{h_1/n}e_{i,j}^\top\widehat S(t)\int_0^{\tau} \mathcal{K}_{h_1}(u-t)d\widetilde N(u).$
By repeatedly generating a normal random sample $G_{ij},$
the distribution of $\mathcal{D}_{\alpha}$
can be approximated by the conditional distribution of $\widetilde{\mathcal{D}}_{\alpha}$ given the observed data, where
\begin{align*}
\widetilde{\mathcal{D}}_{\alpha}=&\max_{1\le i<j\le n}\sup_{t\in[a,b]} |\widetilde{\mathcal{D}}_{ij,\alpha}(t)|/\widehat\zeta_{ij,\alpha}^{1/2}(t).
\end{align*}
Then, the critical value $c_1(\nu)$ can be obtained from the upper $(1-\nu)$-percentile of the conditional distribution of $\widetilde{\mathcal{D}}_{\alpha}.$

\subsection{Goodness-of-fit}
\label{sec:GOF}

We develop a graphical diagnostic method to evaluate goodness-of-fit for dynamic network models.
Specifically, for each given $t\in(0,\tau),$ we first calculate the cumulative number $N_{ij}(t)$ of interactions
observed from node $i$ to node $j$. For each $i\in[n]$, define
$$
\bar N_i^O(t)=\sum_{j\neq i}N_{ij}(t)~~~~\text{and}~~~~\bar N_i^I(t)=\sum_{j\neq i}N_{ji}(t).
$$
We then estimate them respectively as follows:
$$
\widehat N_i^O(t)=\sum_{j\neq i}\widehat N_{ij}(t)~~~~\text{and}~~~~\widehat N_i^I(t)=\sum_{j\neq i}\widehat N_{ji}(t),
$$
where $\widehat N_{ij}(t)=\int_0^t \exp\{\widehat\alpha_i(s)+\widehat\beta_i(s)+Z_{ij}(s)^\top\widehat\gamma(s)\}ds$  is obtained under the proposed model.
We now plot the curves of $(\bar N_i^O(t),\widehat N_i^O(t))$ and $(\bar N_i^I(t),\widehat N_i^I(t))$, respectively.
This produces the so-called Arjas plot \citep{Arjas1988},
which compares the observed and expected occurrences.
If the proposed model fits the data well, the plots will be approximately
linear with a slope close to one.
Additionally, various forms of model misspecification often yield distinct patterns in the graphs, providing a valuable diagnostic tool. For numerical analysis, see Section \ref{sec:SGOF}.

\section{Implementation}
\label{section:BS}

\subsection{Computational Algorithm}
The estimators of $\alpha_i^*(t),$ $\beta_j^*(t)$ and $\gamma^*(t)$ are obtained through
solving the local estimating equations given in (5).
The classical Newton-Raphson method for our case may be time-consuming and may not work well,
due to the fact that the dimension of the unknown parameters grows with $n.$
To obtain $\widehat\theta(t),$ we adopt a combination of the fixed point iterative method and the Newton-Raphson method
by alternatively solving $F(\theta(t))=0$ and $Q(\theta(t))=0$.
This is implemented in Algorithm 1,
in which  {\bf Step 1} is about
solving $F(\theta(t))=0$ with a given $\gamma(t)$ via the fixed point iterative method,
and {\bf Step 2} is about solving $Q(\theta(t))=0$ with given $\alpha(t)$ and $\beta(t)$
via the Newton-Raphson method.
It has good performances in simulation studies.

\begin{algorithm}[htpb]
\caption{An iterative algorithm for solving local estimating equations}\label{algorithm-a}
\begin{algorithmic}
\STATE {{\bf Input}: $k=0,~\alpha_i^{[0]}(t)=0,~\beta_j^{[0]}(t)=0$, and $\gamma^{(0)}(t)=0$.}
\STATE {{\bf Step 1.} Calculate $\alpha_i^{[k+1]}(t)$ and $\beta_j^{[k+1]}(t)$ by
\begin{align*}
\alpha_i^{[k+1]}(t)=&\log\Bigg\{\frac{\sum_{j\neq i}\int_0^{\tau}\mathcal{K}_{h_1}(s-t)dN_{ij}(t)}
{\sum_{j\neq i}\int_0^{\tau}\mathcal{K}_{h_1}(s-t)\exp\{\beta_j^{[k]}(t)+Z_{ij}(s)^\top\gamma^{[k]}(t)\}ds}\Bigg\},\\
\beta_j^{[k+1]}(t)=&\log\Bigg\{\frac{\sum_{i\neq j}\int_0^{\tau}\mathcal{K}_{h_1}(s-t)dN_{ij}(t)}
{\sum_{i\neq j}\int_0^{\tau}\mathcal{K}_{h_1}(s-t)\exp\{\alpha_i^{[k]}(t)+Z_{ij}(s)^\top\gamma^{[k]}(t)\}ds}\Bigg\}.
\end{align*}}

\STATE {{\bf Step 2.} Update $\gamma^{[k+1]}(t)$ by the solution to
\[
\frac{1}{N}\sum_{i=1}^n \sum_{j\neq i}
\int_{0}^{\tau}Z_{ij}(s)\mathcal{K}_{h_2}(s-t)\big[dN_{ij}(s)- \exp\big\{\alpha_i^{[k]}(t)+\beta_j^{[k]}(t)+Z_{ij}(s)^\top\gamma(t)\big\}ds\big]=0.
\]
}
\STATE Stop the algorithm if
$$\max_{i\in[n]}|\alpha_i^{[k+1]}(t)-\alpha_i^{[k]}(t)|
+\max_{i\in[n-1]}|\beta_i^{[k+1]}(t)-\beta_i^{[k]}(t)|+\max_{j\in[p]}|\gamma_j^{[k+1]}-\gamma_j^{[k]}|\le 10^{-3}.$$
Otherwise, set $k=k+1$ and return to {\bf Step 1}.

\STATE{{\bf Output}: $\widehat\alpha_i(t)$, $\widehat\beta_j(t)$ and $\widehat\gamma(t).$}
\end{algorithmic}
\end{algorithm}

\begin{remark}\label{Rem4}
In Algorithm \ref{algorithm-a}, we set the initial values of the $2n+p$ unknown parameters to zero.
Although selecting data-driven initial values is a common practice,
it is challenging in the high-dimension setting.
As mentioned in Remark \ref{MLE:Remark} and Section \ref{section:asymptotic},
our method can be viewed as a localized version of the maximum likelihood approach.
Let $F_\gamma(\eta(t))$ be the version of $F(\eta(t), \gamma(t))$ with $\eta(t)$ as its argument and $\gamma(t)$ treated as a fixed value, and $Q_\eta(\gamma(t))$ can be defined similarly.
Because the negative Jacobian matrix of $F_\gamma(\eta(t))$ with respect to $\eta(t)$ for any given $\gamma(t)$
is diagonally dominant and its elements are nonnegative, it is positive definite.
Therefore, the solution to $F_\gamma(\eta(t))=0$ is  unique if it exists; see Theorem \ref{theorem:consistency}.
For the function $Q_\eta(\gamma(t))$, its Jacobian matrix is also positive definite under Condition 3,
so that the solution to $Q_\eta(\gamma(t))=0$ is unique for any given $\eta(t)$.
Additionally, the local estimating equations are continuous and monotonic with respect to the parameters.
As a result, our method does not suffer from the multiple-solution problem under the conditions of Theorem \ref{theorem:consistency}.
This also indicates that the choice of initial values primarily affects the computational time of Algorithm \ref{algorithm-a}, but
not the final optimal solution. For numerical analysis,
see Section \ref{section:5.1}.
On the other hand, if these conditions do not hold, there may exist multiple solutions.
This is an important issue; however, it is beyond the scope of this paper.
We intend to investigate this problem in future work.
\end{remark}

\subsection{Bandwidth selection}
\begin{figure}[!htpb]
\centering
\includegraphics[width=0.75\textwidth]{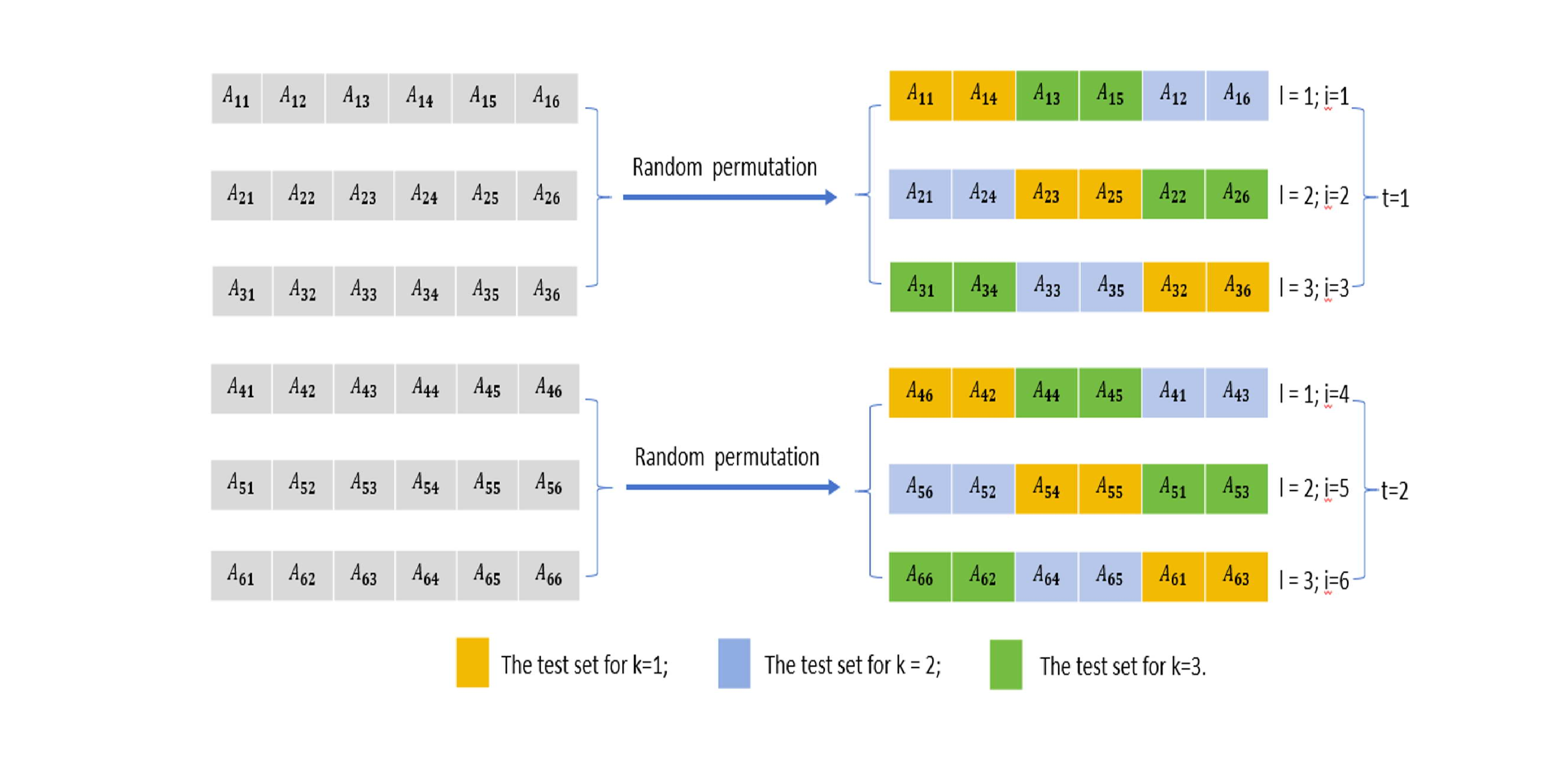}
	\caption{An example of the partition of a dataset with $n=6$ and $K=3$. Here, $A_{ij}$ denotes the observed data $\{(N_{ij}(t),Z_{ij}(t): t\in[0,\tau]\}$ for the pair $(i,j)$.}
	\label{fig:CV}
\end{figure}

Bandwidth selection is often a critical part of nonparametric regression.
Here we consider a $K$-fold cross-validation method to choose the bandwidth parameters $h_1$ and $h_2.$
Let $A_{ij}=\{(N_{ij}(t),Z_{ij}(t)): t\in[0,\tau]\}$ represent the observed data for the pair $(i,j)$.
We first partition randomly the dataset $\{A_{ij}: 1\le i\neq j\le n\}$ into $K$ groups of equal size.
Collect $A_{ij}$ into a matrix $A$, where the $(i,j)$-element is $A_{ij}$, and set $A_{ii}=0$.
For simplicity, we assume that $n/K$ is an integer, denoted by $K_0$.
Let $R_s=(A_{ij}: (s-1)K+1\le i\le sK, j\in[n])\in \mathbb{R}^{K_0\times n}$.
Thus, we partition $A$ into $K_0$ submarices by rows,
and each submatrix $R_s$ contains approximately the same number of rows.
To attain a random splitting, we consider a random permutation of the  columns of $R_s$.
Let $R_s=(R_{1s},\dots,R_{ns})$,
where $R_{js}=(A_{ij}:(s-1)K+1\le i\le sK)^\top$ denotes the $j$th column vector of $R_s$.
For each $s\in[K_0]$, let $\pi(R_s)=(R_{\pi_{1s}},\dots,R_{\pi_{ns}})$,
where $(\pi_{1s},\dots,\pi_{ns})$ is a random permutation of $(1,\dots,n)$.
The test set in the $k$th group is given by
\begin{align*}
\mathcal{N}_k=\bigcup_{s=1}^{K_0}\bigcup_{i=(s-1)K+1}^{sK}\Big\{A_{i\pi_j}: j=1+t~\text{mod}~n~~\text{for}~~(k+l-2)K_0&\le t\le (k+l-1)K_0-1,\\
&~~l=1+(i-1)~\text{mod}~K\Big\},
\end{align*}
where ``$x~\text{mod}~y$" represents the remainder of the division of $x$ by $y$.
An example with $n=6$ and $K=3$ is illustrated in Fig. \ref{fig:CV}.

Let $\widehat\alpha_i^{(-k)}(t),$ $\widehat\beta_i^{(-k)}(t)$ and $\widehat\gamma^{(-k)}(t)$ be an estimate of
$\alpha_i^*(t),$ $\beta_i^*(t)$ and $\gamma^*(t)$
using the data from all subgroups other than $\mathcal{N}_k,$ that is,
we obtain $\widehat\alpha_i^{(-k)}(t),$ $\widehat\beta_i^{(-k)}(t)$ and $\widehat\gamma^{(-k)}(t)$
by the solution to the following local estimating equation:
\begin{align*}
(F^{(-k)}(\theta(t))^\top, Q^{(-k)}(\theta(t))^\top)^\top=0,
\end{align*}
where $F^{(-k)}(\theta(t))=(F_1^{-k}(\theta(t)),\dots,F_{2n-1}^{(-k)}(\theta(t)))^\top$ with
\begin{align*}
F_i^{(-k)}(\theta(t))=&\frac{1}{n-1}\sum_{(i,j)\not\in \mathcal{N}_k}\int_{0}^{\tau}\mathcal{K}_{h_1}(s-t)d\mathcal{M}_{ij}(s,t),~~i\in[n],
\\
F_{n+j}^{(-k)}(\theta(t))=&\frac{1}{n-1}\sum_{(i,j)\not\in \mathcal{N}_k}\int_{0}^{\tau}\mathcal{K}_{h_1}(s-t)d\mathcal{M}_{ij}(s,t),~~j\in [n-1],
\end{align*}
and
$$Q^{(-k)}(\theta(t))=\frac{1}{N}\sum_{(i,j)\not\in \mathcal{N}_k}
\int_{0}^{\tau}Z_{ij}(s)\mathcal{K}_{h_2}(s-t)d\mathcal{M}_{ij}(s,t).
$$
Note that under model (1),  we have $\mathcal{M}_{ij}(t)$ is a zero-mean martingale process.
Thus, the $k$th prediction error can be obtained by
\[
\text{PE}_k(h_1,h_2)=\sum_{(i, j)\in \mathcal{N}_k}\int_{0}^{\tau}
\bigg[N_{ij}(t)-\int_{0}^t\exp\{\widehat\alpha_i^{(-k)}(s)+\widehat\beta_j^{(-k)}(s)+Z_{ij}(s)^\top\widehat\gamma^{(-k)}(s)\}ds\bigg]^2dt.
\]
The optimal bandwidth $(h_1,h_2)$ are obtained by
\[
(\widehat h_{1},\widehat h_{2})=\text{arg}\min_{(h_1, h_2)} \sum_{k=1}^K \text{PE}_k(h_1,h_2).
\]

\section{Numerical studies}
\label{section:simulation}
\vspace{-10pt}
\subsection{Consistency and normality}
\label{section:5.1}
In this section, we describe the simulation studies conducted to evaluate the finite sample performance of the proposed method.
We set $\alpha_i^*(t)=-c_0\log(n)+(2.5+\sin(2\pi t))$ if $i<n/2$
and $\alpha_i^*(t)=-c_0\log(n)+(1.5+t/2)$ otherwise.
In addition, we set $\beta_i^*(t)=-c_0\log(n)+(2.5+\cos(2\pi t))$ if $i<n/2$
and $\beta_i^*(t)=-c_0\log(n)+(1.5+t/2)$ otherwise.
Here, $c_0$ is used to specify sparse regimes.
We consider $c_0=0.5$; hence, $q_n\approx\log(n).$
The network sparsity level defined by
$\tau n^2e^{-q_n}$ is $O(n),$
which is less than $n^2$ and a moderately sparse network is generated.
In addition, we set $\gamma^*(t)=(\gamma_1^*(t),\gamma_2^*(t))^\top$ with $\gamma_1^*(t)=\gamma_2^*(t)=\sin(2\pi t)/3$.
The covariates $Z_{ij}$ are generated independently from the standard normal distribution.
We set $\tau=1$ and the number of nodes as $n=60, 100, 200$, and $500.$
The kernel is chosen as $\mathcal{K}(x)=\exp(-x^2/2)/(2\pi)^{1/2}$.
All of the results are based on $1000$ replications.
For the bandwidth, we consider a 5-fold cross validation method.
Specifically, we set $h_1\in\{0.05, 0.1, \dots, 0.5\}$ and $h_2\in\{0.002, 0.006, \dots, 0.03, 0.04, 0.08\}$.
We first partition the dataset $\{(N_{ij}(t),Z_{ij}(t)): 1\le i\neq j\le n\}$ into five subgroups using the procedure described in Section \ref{section:BS}.
For each pair $(h_1,h_2)$, we estimate the parameters using the training data,
and calculate the prediction error using the test data.
We then identify the optimal bandwidth that minimizes the prediction error.
Fig. S3 in the Supplementary Material shows the prediction errors for different $(h_1,h_2)$ values, confirming the effectiveness of the 5-fold cross-validation method.

To measure the error of the estimators, we use the mean integrated squared error (MISE), defined as
$\text{MISE}=1000^{-1}\sum_{k=1}^{1000}\int_{0}^{\tau}[\hat{f}_{k}(t)-f^*(t)]^2dt.$
Here, $f^*(t)$ represents the true value
while $\hat f_{k}(t)$ denotes its estimate at the $k$th replication.
The MISE for the estimators of $\alpha_1^*(t)$, $\alpha_{n/2+1}^*(t)$, $\beta_1^*(t)$, $\beta_{n/2+1}^*(t)$,
and $\gamma_1^*(t)$ are presented in Table \ref{MISEnew},
in which the estimators are evaluated at the time points $t_i=i\tau/100~(i=1,\dots,99).$
The results for other parameters are similar and omitted here.
From Table \ref{MISEnew}, we can see that all MISEs are small.
As expected, the MISE decreases as the sample size $n$ increases.
The MISE for $\gamma_1^*(t)$ is much smaller (by up to two orders of magnitude) than that for the degree parameters.
This is due to the fact that the dimension of regression coefficients is fixed while the number of degree parameters is of order $n$.
The averages of the 1000 estimated coefficient curves for $\alpha_1^*(t),~\beta_1^*(t)$ and $\gamma_1^*(t),$
and their pointwise $95\%$ confidence intervals are shown in Figures \ref{fig:sendnew} and S4 in the Supplementary Material.
The confidence intervals for $\alpha_1^*(t),~\beta_1^*(t)$, and $\gamma_1^*(t)$ are calculated as follows.
Let $t_i=i\tau/100$ for $i=1,\dots,99.$
We first estimate the parameters at each time point $t_i$.
Then, for each given $1\le i\le 99$, we calculate the confidence intervals for $\alpha_1^*(t_i),~\beta_1^*(t_i)$ and $\gamma_1^*(t_i)$, respectively, using \eqref{confidence:eta} and \eqref{confidence:gamma}.
As the number of nodes increases,
the estimated curves are closer to the true curves.

\begin{table}
\caption{\label{MISEnew}{\small The MISEs for $\alpha_1^*(t)$, $\alpha_{n/2+1}^*(t)$, $\beta_1^*(t)$, $\beta_{n/2+1}^*(t)$ and	$\gamma_1^*(t)$.}}
{\small\begin{tabular}{ccccccccccc}\hline
$n$&&  $\alpha_1^*(t)$ && $\alpha_{\frac{n}{2}+1}^*(t)$ & & $\beta_1^*(t)$ & &$\beta_{\frac{n}{2}+1}^*(t)$ &&$\gamma_1^*(t)$ \\
\hline
$60$  && 0.157   && 0.211   && 0.193   && 0.175 && 0.014\\
$100$ && 0.129   && 0.190   && 0.157   && 0.169 && 0.008\\
$200$ && 0.111   && 0.173   && 0.162   && 0.153 && 0.004\\ 	
$500$ && 0.104   && 0.169   && 0.149   && 0.149 && 0.002\\
\hline 			
\end{tabular}}
\end{table}

\begin{table}
\caption{\label{estimates}{\small The coverage probability ($\times 100$) for $\alpha_1^*(t)$, $\alpha_{n/2+1}^*(t)$, $\beta_1^*(t)$, $\beta_{n/2+1}^*(t)$ and $\gamma_1^*(t)$ at $t=0.4, 0.6$ and $0.8.$ The average length of 95\% confidence intervals among 1000 replications
is given in parentheses.}}
{\small\begin{tabular}{ccccccccccc} \hline
$n$&&&&  \multicolumn{1}{c}{$t=0.4$} &&& \multicolumn{1}{c}{$t=0.6$} & && \multicolumn{1}{c}{$t=0.8$}      \\
\hline
                    $60$	 &$\alpha_1^*(t)$          &&&	 95.3 (1.06)   &&& 93.7 (1.61)  && & 94.0 (1.50)\\
                        	 &$\alpha_{\frac{n}{2}+1}^*(t)$    &&&	94.0 (1.62)   &&& 93.8 (1.71) & & & 93.0 (1.44)\\
                        	 &$\beta_1^*(t)$           &&&	92.7 (1.17)   &&& 94.2 (1.63)  && & 94.0 (1.36)\\
                        	 &$\beta_{\frac{n}{2}+1}^*(t)$     &&&	94.3 (1.22)  & && 95.6 (1.65)  & && 93.5 (1.68)\\
                        	 &$\gamma_1^*(t)$          &&&	92.1 (0.36)   &&& 91.8 (0.48) & & & 90.3 (0.36)\\				
                    $100$	 &$\alpha_1^*(t)$          &&&	 95.5 (1.00)   &&& 92.3 (1.61)  && & 95.9 (1.45)\\
                        	 &$\alpha_{\frac{n}{2}+1}^*(t)$    &&&	93.7 (1.64)   &&& 93.8 (1.69) & & & 95.4 (1.39)\\
                        	 &$\beta_1^*(t)$           &&&	92.8 (1.12)   &&& 95.8 (1.60)  && & 92.7 (1.30)\\
                        	 &$\beta_{\frac{n}{2}+1}^*(t)$     &&&	92.6 (1.18)  & && 94.7 (1.65)  & && 93.9 (1.66)\\
                        	 &$\gamma_1^*(t)$          &&&	93.8 (0.31)   &&& 95.8 (0.44) & & & 94.7 (0.37)\\ 	
                    $200$	  &$\alpha_1^*(t)$         && &	 96.5 (0.95)  && & 97.8 (1.54) & & & 91.1 (1.36)\\
                        	 &$\alpha_{\frac{n}{2}+1}^*(t)$    &&&	95.8 (1.65)  & && 95.3 (1.63)  & && 95.9 (1.29)\\
                        	 &$\beta_1^*(t)$           &&&	90.7 (1.08)  && & 92.3 (1.57)  && & 94.4 (1.20)\\
                        	 &$\beta_{\frac{n}{2}+1}^*(t)$     &&&	96.1 (1.15) & & & 96.3 (1.61)  && & 95.5 (1.60)\\
                        	 &$\gamma_1^*(t)$         && &	95.3 (0.23)  && & 95.3 (0.33)  && & 95.3 (0.27)\\
                    $500$	 &$\alpha_1^*(t)$         && &	 95.5 (0.92) &&  & 97.1 (1.53) &&  & 94.8 (1.33)\\
                        	 &$\alpha_{\frac{n}{2}+1}^*(t)$    &&&	92.9 (1.73)  && & 94.5 (1.65)  && & 93.7 (1.24)\\
                        	 &$\beta_1^*(t)$           &&&	93.7 (1.08)  && & 94.9 (1.60)   &&& 93.8 (1.14)\\
                        	 &$\beta_{\frac{n}{2}+1}^*(t)$     &&&	95.5 (1.15) & & & 97.5 (1.54)  & && 95.4 (1.59)\\
                        	 &$\gamma_1^*(t)$          &&&	94.6 (0.16)  & && 95.3 (0.23)   &&& 95.2 (0.19)\\ 	
				\hline
			\end{tabular}
}
\end{table}

Table \ref{estimates} presents the coverage probabilities of the pointwise $95\%$ confidence intervals
and average lengths of the confidence intervals for  $\alpha_1^*(t),~\beta_1^*(t)$, and $\gamma_1^*(t).$
We observe that the coverage probabilities are close to the nominal level
and lengths of the confidence intervals decrease as $n$ increases.
In addition, Fig. \ref{fig:sendnew} shows that the confidence bands tend to cover the entire true curves.
Fig. S5 in the Supplementary Material presents the asymptotic distributions
of standardised $\widehat\alpha_1(t)$, $\widehat\beta_1(t)$, and $\widehat\gamma_1(t)~(t=0.6~\text{and}~0.8)$ with $n=500,$
which can be approximated well by the standard normal distribution.
This confirms the theoretical results of Theorems \ref{theorem-central-gamma} and \ref{theorem-central-degree}.

\begin{figure}[h]
	\centering
	\subfigure{
		\includegraphics[width=0.28\textwidth]{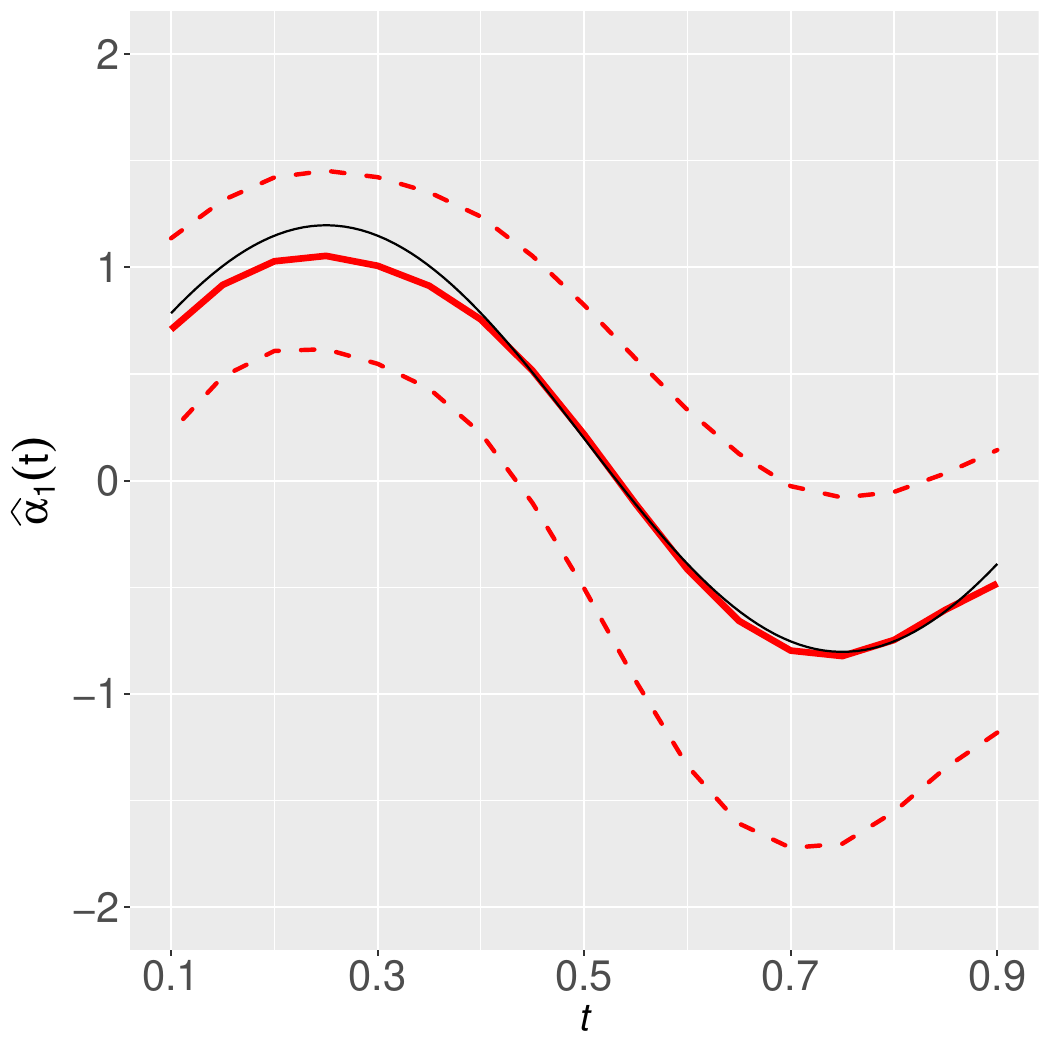}}
	\subfigure{
		\includegraphics[width=0.28\textwidth]{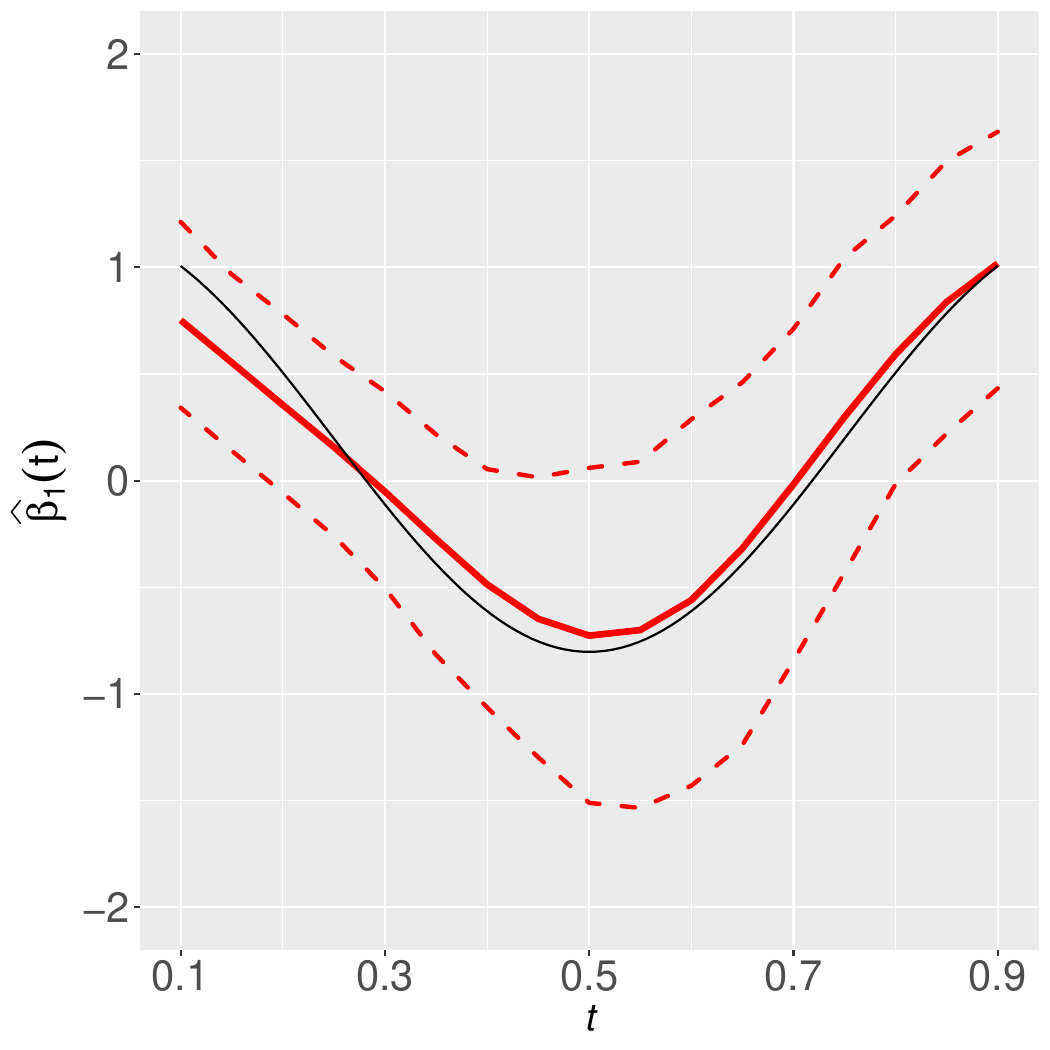}}
	\subfigure{
		\includegraphics[width=0.28\textwidth]{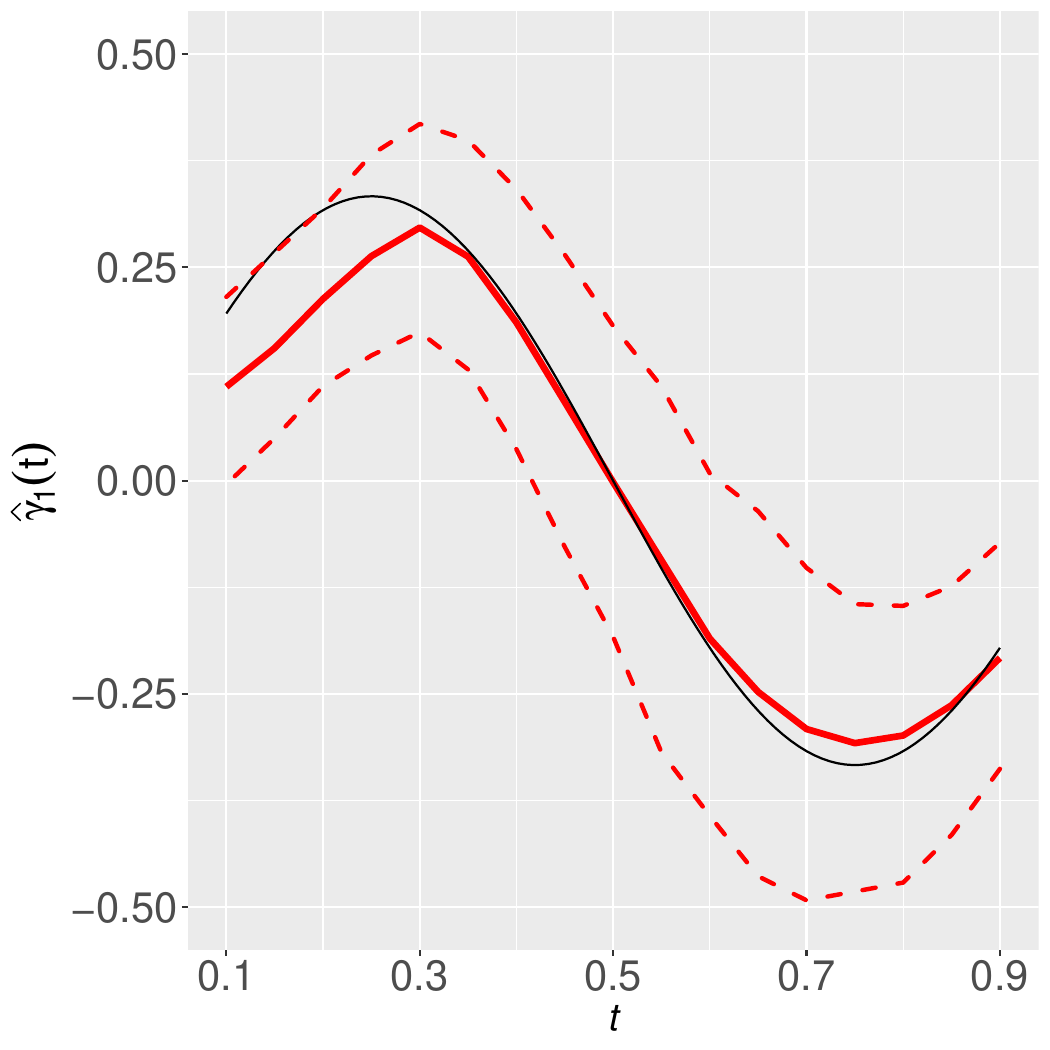}}
	\caption{The estimated curves for $\alpha_1^*(t),~\beta_1^*(t)$ and $\gamma_1^*(t)$ with $n=100$.
The black solid lines denote the true curves of $\alpha_1^*(t), ~\beta_1^*(t)$ and $\gamma_1^*(t).$
The red solid lines are the averages (over 1000 replications) of the proposed estimators
$\widehat\alpha_1(t),~\widehat\beta_1(t)$ and $\widehat\gamma_1(t)$,
while the red dashed lines represent the pointwise $95\%$ confidence intervals.}
	\label{fig:sendnew}
\end{figure}

We next conduct simulation studies to evaluate the stability of the proposed method
when a range of different initial values are chosen.
Specifically, for each $t\in\{i\tau/100: 1\le i\le 99\}$ with $\tau=1$,
we set the initial value of $\alpha_i^*(t)$ as $\alpha_i^*(t)+a$,
where $a$ is selected to be $0$, $-1$, $2$, and $5$, respectively.
The initial values for $\beta_{j}^*(t)$ and $\gamma^*(t)$ are determined in the same manner.
When $a=0$, the initial values are exactly the true values of the parameters.
After obtaining the estimators, we calculate the MISE and the coverage frequencies of the estimators.
The results are presented in Tables \ref{MISE:Initial} and \ref{CP:Initial}.
We observe that the performance of the proposed method is comparable across different choices of initial values.
Therefore, the proposed method is stable with respect to the choice of initial values.
In addition, we record the average computational time (in seconds) for the proposed method with $a=0,~-1,~2$, and $5$,
which are 1.04, 1.06, 1.87, and 3.77 seconds, respectively.
This indicates that the choice of initial values primarily affects the computational time of the proposed method
but has very little influence on the optimal solution.
Additionally, the findings indicate, to some extent,
that our method may not suffer from the issue of multiple solutions.

\begin{table}
\caption{\label{MISE:Initial}{\small The MISEs for $\alpha_1^*(t)$, $\alpha_{n/2+1}^*(t)$, $\beta_1^*(t)$, $\beta_{n/2+1}^*(t)$ and	$\gamma_1^*(t)$ with different initial values. The number of nodes is $n=100$.}}
{\small\begin{tabular}{rccccccccc} \hline
$a$&$\alpha_1^*(t)$&&$\alpha_{\frac{n}{2}+1}^*(t)$&&$\beta_1^*(t)$&&$\beta_{\frac{n}{2}+1}^*(t)$&&$\gamma_1^*(t)$\\
\hline
			  $0$ & 0.133 && 0.186 && 0.125 && 0.172 && 0.008\\
              $-1$&0.123 && 0.180 && 0.129 && 0.168 && 0.008\\
              2  &0.131 &&  0.179 &&  0.129 &&  0.172 &&  0.008\\
              5  &0.128 && 0.180 && 0.131 && 0.165 && 0.008\\
\hline
\end{tabular}}
\end{table}

\begin{table}
\caption{\label{CP:Initial}{\small The coverage probability ($\times 100$) for $\alpha_1^*(t)$, $\alpha_{n/2+1}^*(t)$, $\beta_1^*(t)$, $\beta_{n/2+1}^*(t)$ and $\gamma_1^*(t)$ at $t=0.4, 0.6$ and $0.8$ with different initial values.
The number of nodes is $n=100$.
 The average length of 95\% confidence intervals among 1000 replications is given in parentheses.}}
{\small\begin{tabular}{ccccccccccc} \hline
$a$&&&&  \multicolumn{1}{c}{$t=0.4$} &&& \multicolumn{1}{c}{$t=0.6$} & && \multicolumn{1}{c}{$t=0.8$}      \\
\hline
                    $0$	 &$\alpha_1^*(t)$          &&&	 94.4 (1.00)   &&& 92.9 (1.58)  && & 92.9 (1.44)\\
                        	 &$\alpha_{\frac{n}{2}+1}^*(t)$    &&&	94.0 (1.63)   &&& 92.7 (1.67) & & & 93.6 (1.38)\\
                        	 &$\beta_1^*(t)$           &&&	92.7 (1.12)   &&& 94.0 (1.62)  && & 94.8 (1.29)\\
                        	 &$\beta_{\frac{n}{2}+1}^*(t)$     &&&	94.5 (1.18)  & && 93.0 (1.62)  & && 95.3 (1.66)\\
                        	 &$\gamma_1^*(t)$          &&&	92.1 (0.29)   &&& 93.2 (0.40) & & & 91.6 (0.30)\\
                    $-1$	 &$\alpha_1^*(t)$          &&&	 94.4 (1.00)   &&& 93.6 (1.59)  && & 94.4 (1.44)\\
                        	 &$\alpha_{\frac{n}{2}+1}^*(t)$    &&&	94.5 (1.65)   &&& 94.7 (1.68) & & & 95.6 (1.37)\\
                        	 &$\beta_1^*(t)$           &&&	91.6 (1.11)   &&& 93.6 (1.62)  && & 93.8 (1.29)\\
                        	 &$\beta_{\frac{n}{2}+1}^*(t)$     &&&	94.7 (1.18)  & && 94.9 (1.63)  & && 94.8 (1.66)\\
                        	 &$\gamma_1^*(t)$          &&&	93.0 (0.29)   &&& 92.3 (0.40) & & & 91.1 (0.29)\\
                    $2$	  &$\alpha_1^*(t)$         && &	 95.7 (0.95)  && & 93.9 (1.58) & & & 93.6 (1.45)\\
                        	 &$\alpha_{\frac{n}{2}+1}^*(t)$    &&&	94.6 (1.63)  & && 94.0 (1.68)  & && 92.9 (1.38)\\
                        	 &$\beta_1^*(t)$           &&&	91.4 (1.11)  && & 95.3 (1.62)  && & 93.0 (1.30)\\
                        	 &$\beta_{\frac{n}{2}+1}^*(t)$     &&&	94.8 (1.19) & & & 95.1 (1.63)  && & 94.1 (1.66)\\
                        	 &$\gamma_1^*(t)$         && &	94.0 (0.29)  && & 93.8 (0.40)  && & 92.1 (0.30)\\
                    $5$	  &$\alpha_1^*(t)$         && &	 94.8 (1.00) &&  & 93.0 (1.57) &&  & 93.2 (1.44)\\
                        	 &$\alpha_{\frac{n}{2}+1}^*(t)$    &&&	94.9 (1.64)  && & 93.4 (1.67)  && & 94.1 (1.37)\\
                        	 &$\beta_1^*(t)$           &&&	91.5 (1.11)  && & 93.9 (1.62)   &&& 93.9 (1.29)\\
                        	 &$\beta_{\frac{n}{2}+1}^*(t)$     &&&	94.7 (1.19) & & & 95.8 (1.62)  & && 95.3 (1.65)\\
                        	 &$\gamma_1^*(t)$          &&&	92.0 (0.29)  & && 91.9 (0.40)   &&& 91.2 (0.30)\\
				\hline
\end{tabular}}
\end{table}

\subsection{Comparison with other methods}
We now compare our method with the method of \cite{kreib2019} on the performance of estimating homophily parameters.
In this simulation, the homophily parameter is set as $\gamma^*(t)=\sin(2\pi t)/3$ and the covariates $Z_{ij}$
are set to be 1 if $i\leq 4$ and $j\leq n/3$, and $0$ otherwise.
We set $\alpha_i^*(t)$ and $\beta_i^*(t)$ as $\alpha_i^{*}(t)=\beta_i^{*}(t)=b[-0.5\log(n)+(3+t/2)]$ if $i<n/2$,
and $\alpha_i^{*}(t)=\beta_i^{*}(t)=0$ otherwise.
When $b=0,$ the simulated network does not have degree heterogeneity,
and hence both methods yield consistent estimators.
As $b$ increases, the method of \cite{kreib2019} may give
biased estimates for homophily parameters,
due to the presence of degree heterogeneity.
We choose $b$ to be $0$, $1/3$, $1/2$ and $1$, and set $n=200.$

\begin{figure}[h]
\centering
	\subfigure[$b=0$]{
		\includegraphics[width=0.29\textwidth]{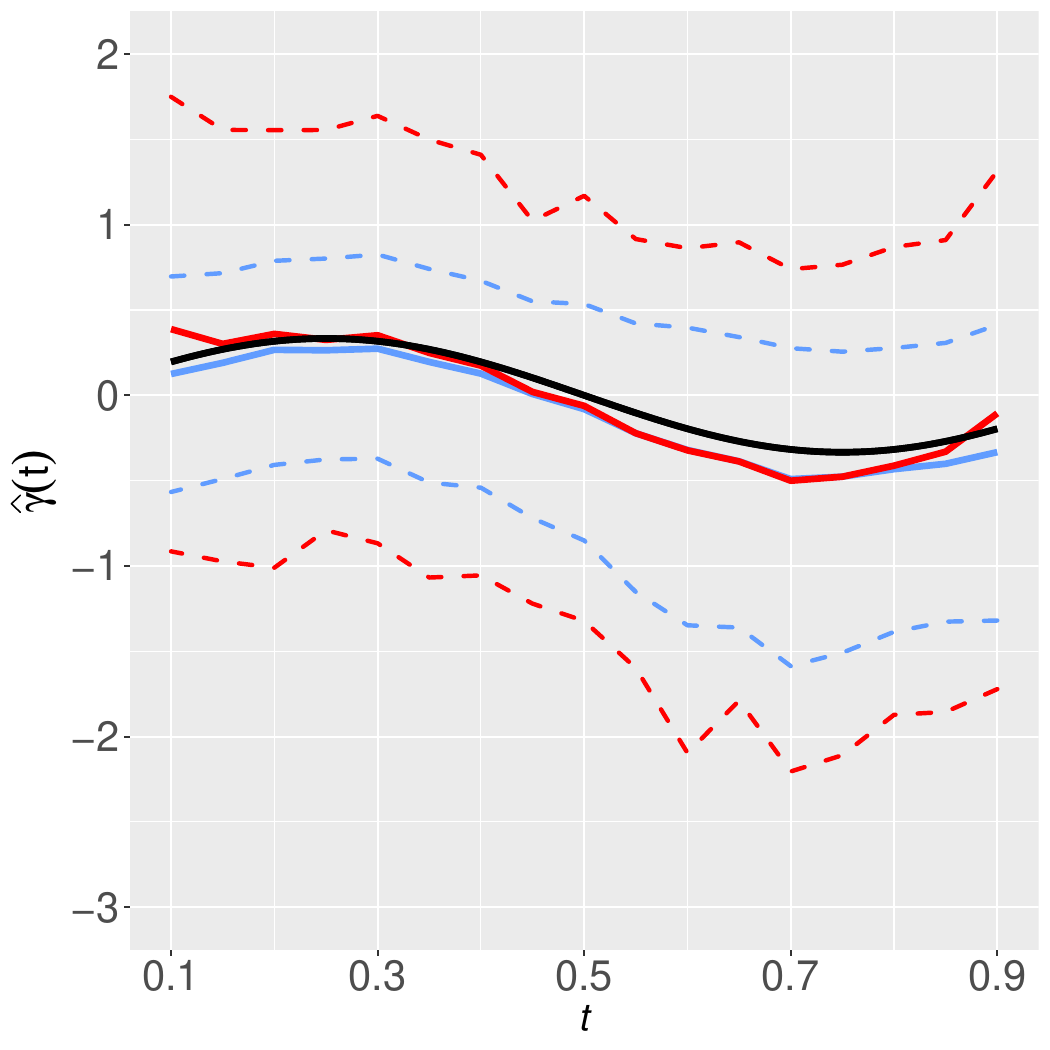}}
      \subfigure[$b=1/3$]{
		\includegraphics[width=0.29\textwidth]{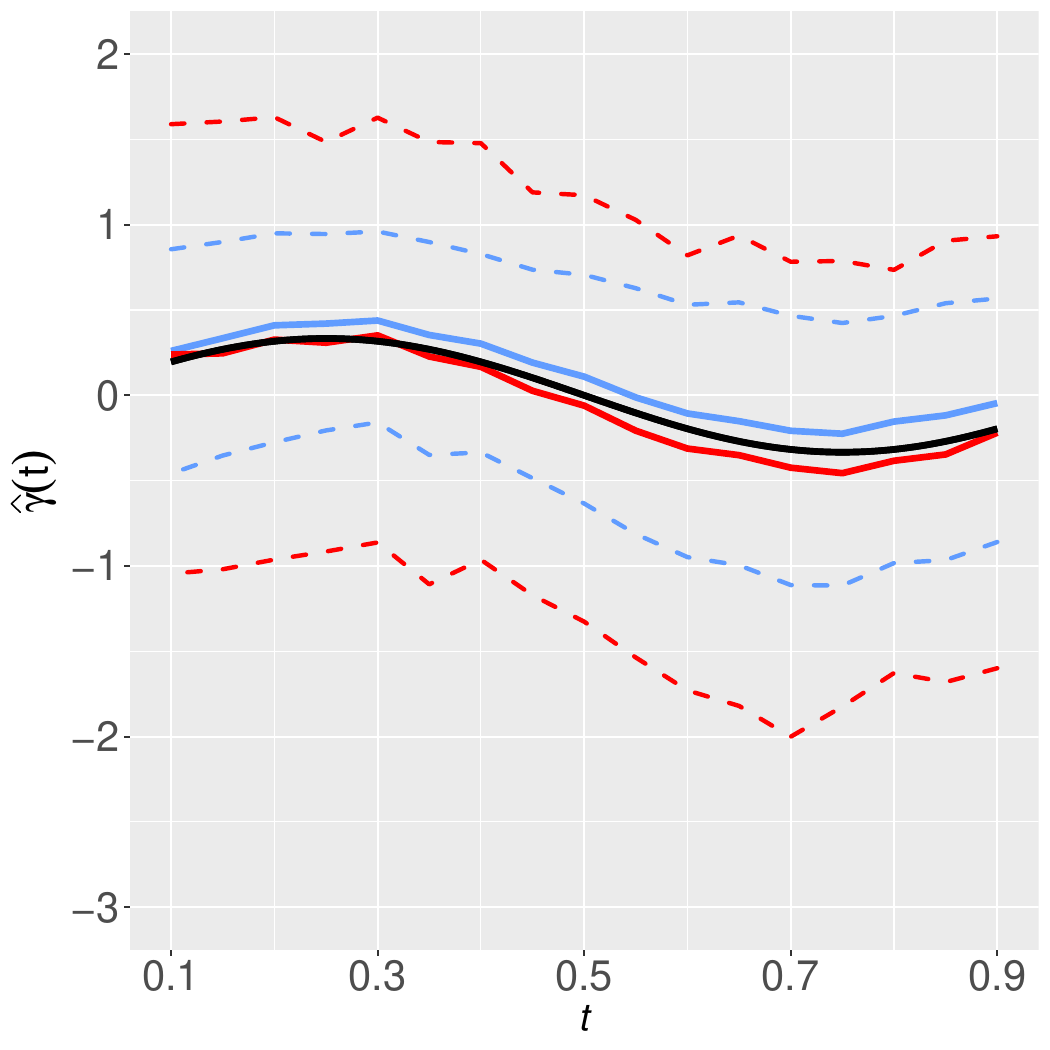}}\\
	\subfigure[$b=1/2$]{
		\includegraphics[width=0.29\textwidth]{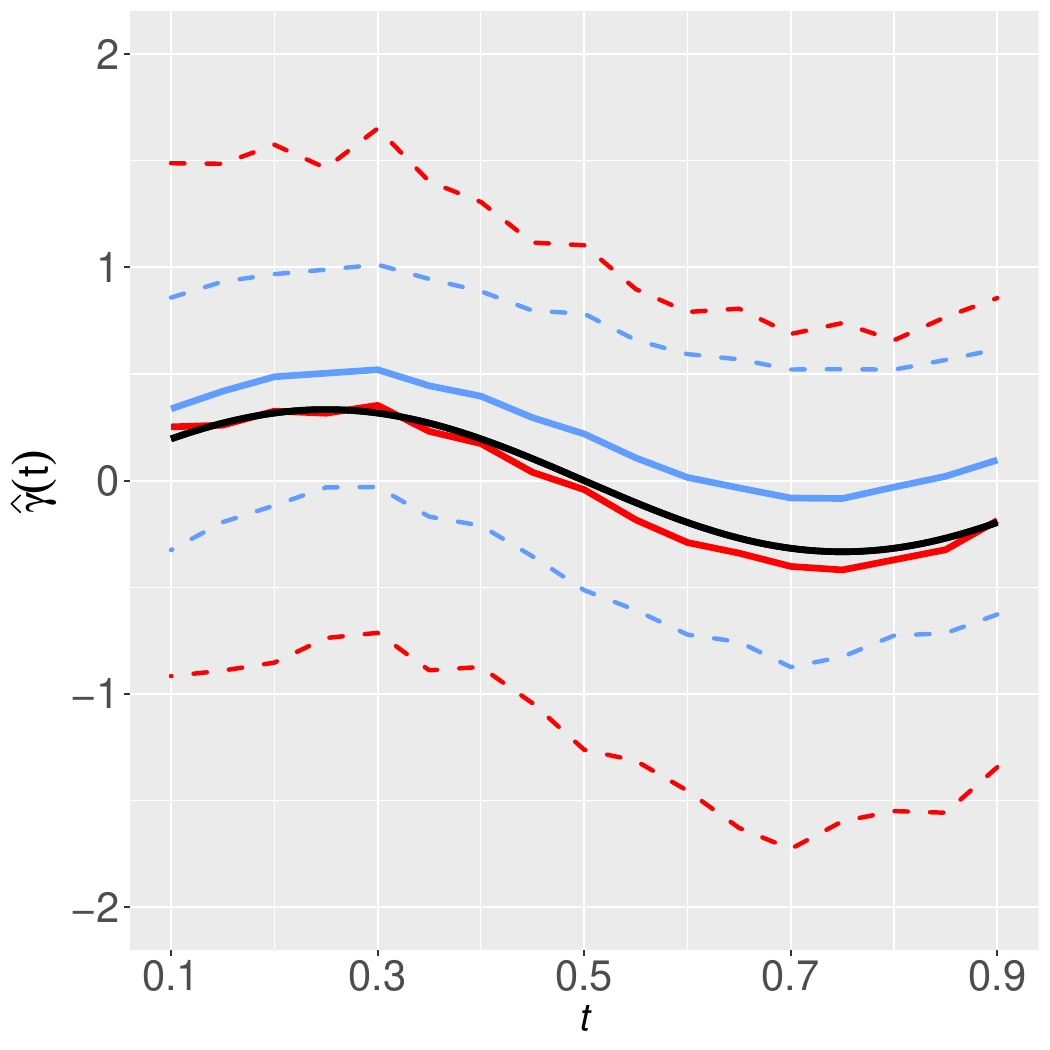}}
        \subfigure[$b=1$]{
            \includegraphics[width=0.29\textwidth]{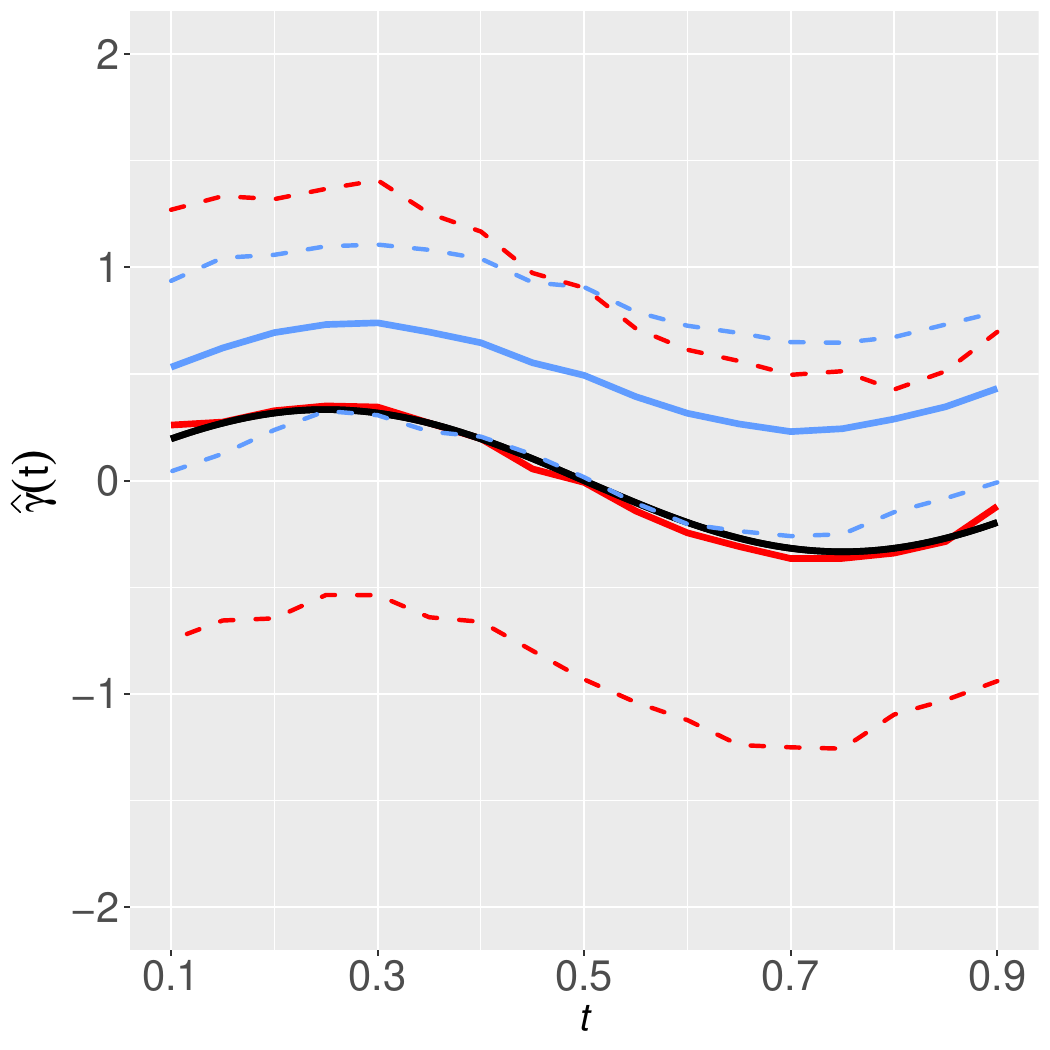}}
	\caption{The red solid lines are the average of the proposed estimators  of $\gamma^*(t)$ over 1000 replications,
	and the red dashed lines are the pointwise $95\%$ confidence intervals. The blue solid lines are the average of the estimators obtained by \cite{kreib2019} over 1000 replications, and the blue dashed lines are the pointwise $95\%$ confidence intervals. The black solid lines represent the true curves of $\gamma^*(t).$}
	\label{fig:chet}
\end{figure}

The results based on $1000$ replications are shown in Fig. \ref{fig:chet}.
We see that when $b=0$, the performance of the two methods are comparable in terms of bias.
The confidence band for our method is evaluated by (8).
Our method leads to a wider confidence band, which is not surprising
because there are $2n+p-1$ unknown parameters
in our model, while there are only $p$ unknown parameters in \cite{kreib2019}.
On the other hand, our model still performs well in estimating $\gamma^*(t)$ with $b=1/3$, $1/2$  and $1$,
but the method of \cite{kreib2019} yields a biased estimate for $\gamma^*(t)$.
The bias increases as $b$ varies from $1/3$ to $1.$
When $b=1$, the $95\%$ pointwise confidence band of \cite{kreib2019} even fails to cover the entire true curve  of $\gamma^*(t)$.
This indicates that when degree heterogeneity exists in a network,
neglecting this feature may result in a biased estimate for homophily effects.

\subsection{Size and power of tests}

We first examine the performance of the tests developed in Section \ref{sec:Trt} for time-varying trends.
For this purpose, we set $\alpha_{i}^*(t)=\beta^*_j(t)=-0.5\log(n)+(2.5+\tilde c_1\sin(2\pi t))$ for all $i\in[n]$ and $j\in[n-1]$,
and $\gamma^*(t)=\tilde c_2\sin(2\pi t)/3$.
The covariates $Z_{ij}$ are generated independently from the standard normal distribution.
The parameters $\tilde c_1$ and $ \tilde c_2$ indicate the trend level.
We observe that $H_{0\eta}$ holds if $\tilde c_1=0$,
and the departure from $H_{0\eta}$ increases as $\tilde c_1$ increases.
Similarly, $H_{0\gamma}$ holds if $\tilde c_2=0,$, and
the departure from $H_{0\gamma}$ increases as $\tilde c_2$ increases.
For simplicity, we choose $t_1$ and $t_2$ as $0.1, 0.2, \ldots, 0.9$ to calculate $\mathcal{T}_{\eta}$ and $\mathcal{T}_{\gamma}.$
The kernel function and bandwidth are chosen to be the same as those used previously.
The sample size is set to $n=100$ and $200,$ and the level $\nu$ is chosen as $0.05$.
The critical values are calculated using the resampling method with 1000 simulated realizations.

\begin{figure}[h]
\centering
	\subfigure[Size and power of $\mathcal{T}_{\eta}$]{
		\includegraphics[width=0.3\textwidth]{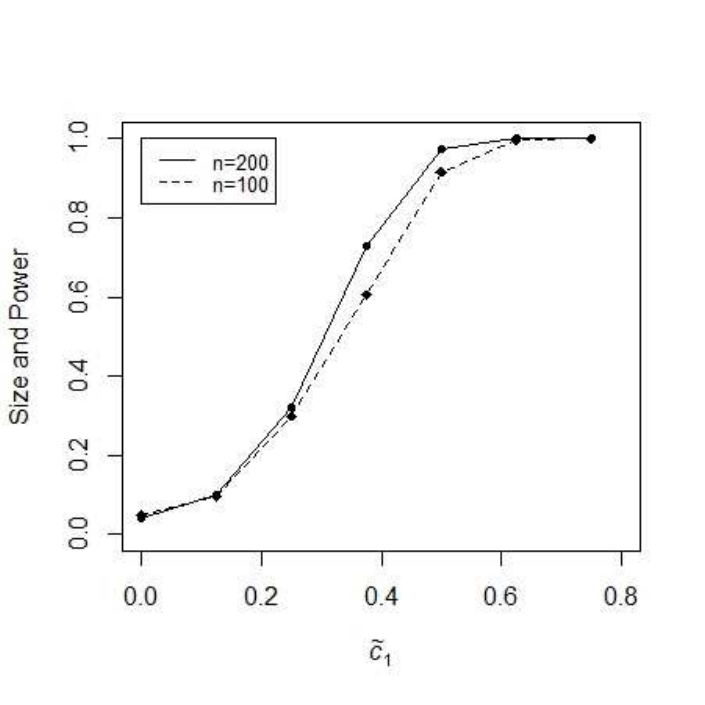}}
	\subfigure[Size and power of $\mathcal{T}_{\gamma}$]{
		\includegraphics[width=0.3\textwidth]{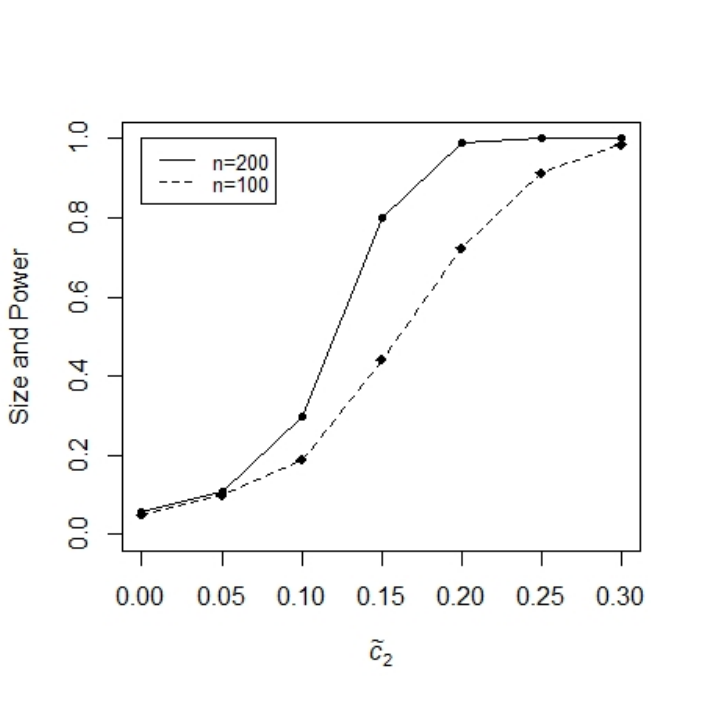}}
  \subfigure[Size and power of $\mathcal{D}_{\alpha}$]{
		\includegraphics[width=0.3\textwidth]{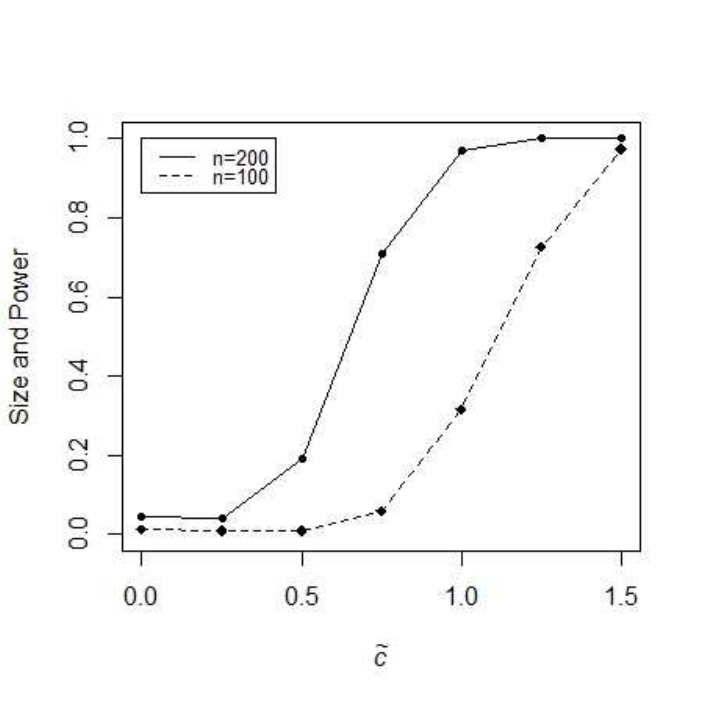}}
	\caption{The size and power of the test statistics $\mathcal{T}_{\eta},$ $\mathcal{T}_{\gamma}$ and $\mathcal{D}_{\alpha}.$ The level $\nu$ is $0.05.$}
	\label{fig:con}
\end{figure}

Fig. \ref{fig:con} depicts the size and power of the statistics $\mathcal{T}_\eta$ and $\mathcal{T}_{\gamma}.$
The estimated sizes of $\mathcal{T}_\eta$ and $\mathcal{T}_{\gamma}$ are approximately 0.05,
and the empirical powers of both test statistics increase
as $\tilde c_1$ and $\tilde c_2$ increase.
The powers also increase with the sample size.
The results show that the statistics ${\mathcal{T}}_{\eta}$ and $\mathcal{T}_{\gamma}$ perform well under the null hypothesis
and can also successfully detect time-varying trends in the parameters under the alternative hypothesis.

We now examine the performance of the test proposed in Section \ref{sec:dht} to test degree heterogeneity.
For this purpose, we set $\gamma^*(t)=\sin(2\pi t)/3$.
The covariates $Z_{ij}$ are generated independently from the standard normal distribution.
Let $\alpha_{i}^*(t)=\beta^*_i(t)=t/2~\text{for~all}~ i\in[n-1],$ $\alpha_n^*(t)=t/2 + \tilde c$,
and $\beta_n^*(t)=0$, where $\tilde c$ indicates the level of degree heterogeneity.
When $\tilde c=0,$ the null hypothesis $H_{01}$ holds, and
the departure from  the null $H_{01}$ increases as $\tilde c$ increases.
We choose $t$ as $0.1,0.2,\dots,0.9$ to calculate $\mathcal{D}_{\alpha}.$
The kernel function and bandwidth are chosen to be the same as those used previously.
The sample size is set to $n=100$ and $200,$ and the level $\nu$ is chosen as $0.05$.
The critical values are calculated using the resampling method with 1000 simulated realizations.

The estimated size and power of ${\mathcal{D}}_{\alpha}$ are presented in Fig. \ref{fig:con}. 
We observe that the estimated size is approximately 0.05 when $\tilde c=0,$
and the power of the proposed test increases as $\tilde c$ tends to 1.5.
In addition, the power increases when the sample size increases from $100$ to $200$.
The results show that the statistic ${\mathcal{D}}_{\alpha}$
performs well under the null hypothesis and can also successfully detect the existence of degree heterogeneity under the alternative hypothesis. The performance of ${\mathcal{D}}_{\beta}$ is similar to that of ${\mathcal{D}}_{\alpha}$ and omitted here.

\subsection{Goodness-of-fit comparison}
\label{sec:SGOF}

\begin{figure}[h]
\centering
	\subfigure[Our method for Case I]{
		\includegraphics[width=0.30\textwidth]{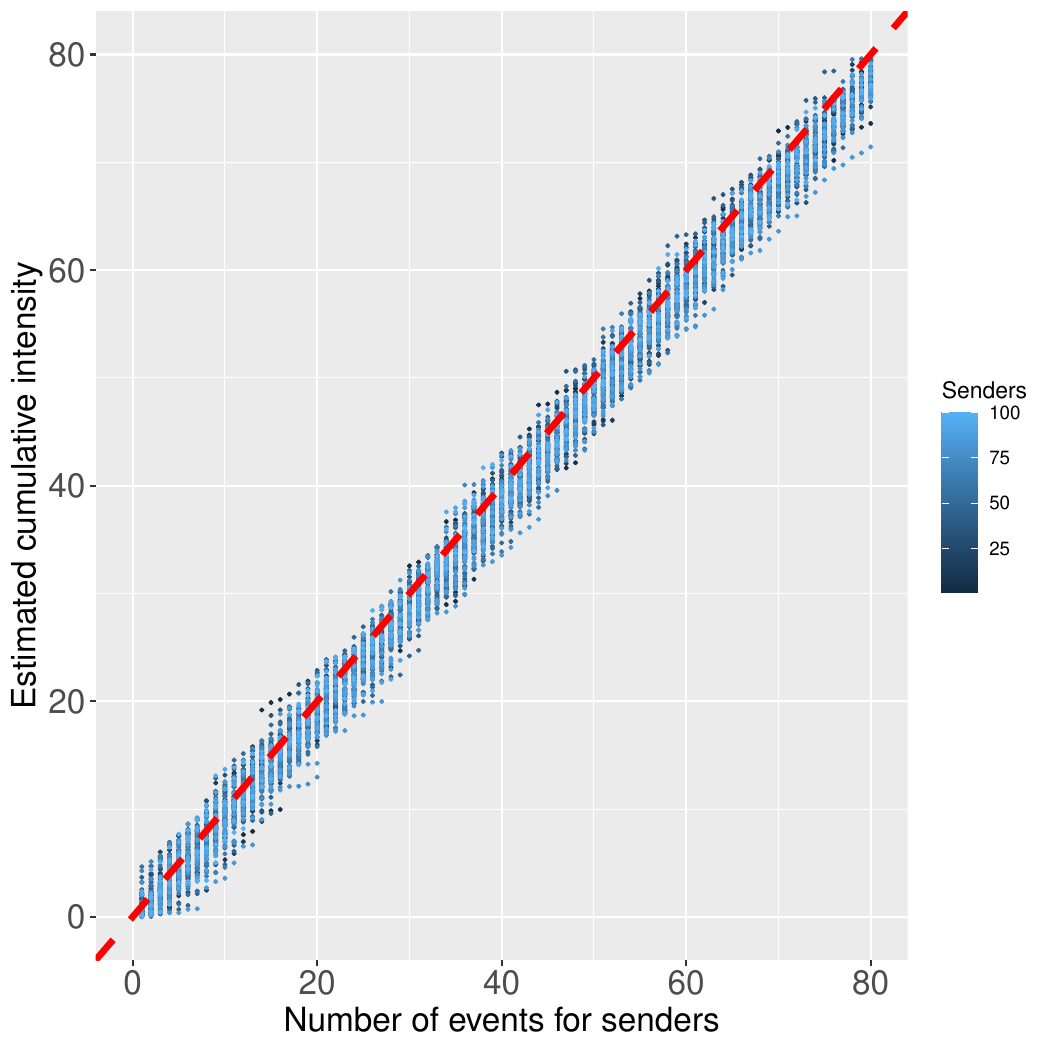}}
	\subfigure[Our method for Case II]{
		\includegraphics[width=0.30\textwidth]{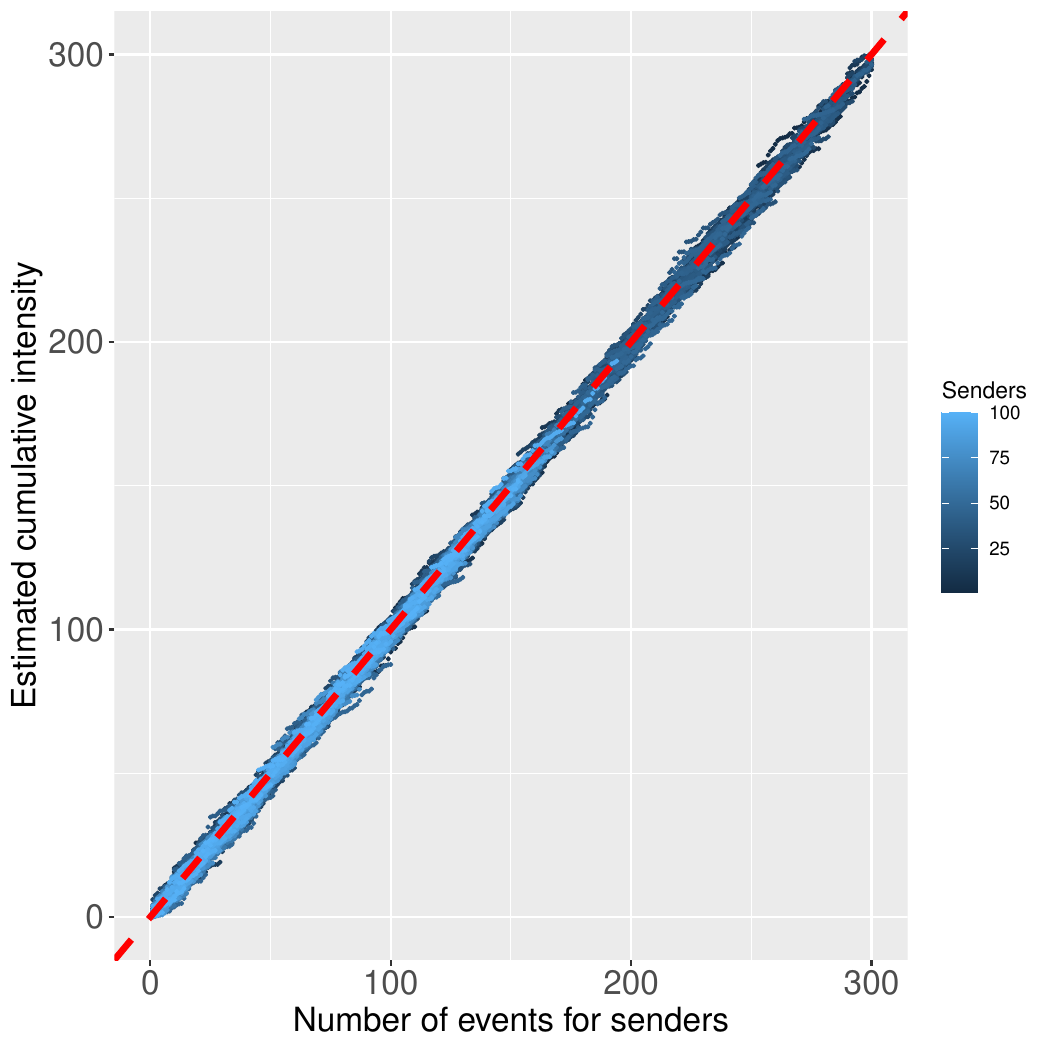}}
  \subfigure[Our method for Case III]{
		\includegraphics[width=0.30\textwidth]{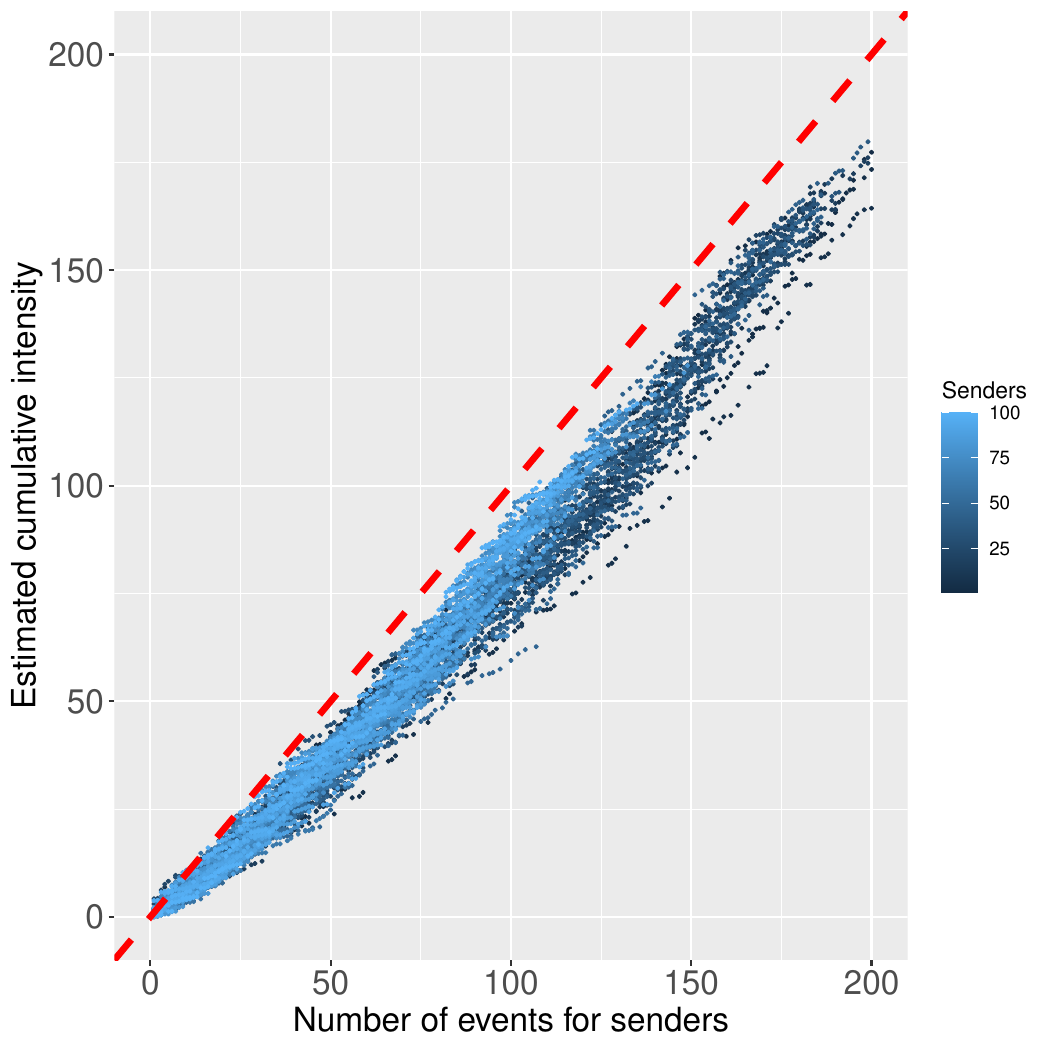}}
	\subfigure[Krei\ss's method for Case I]{
		\includegraphics[width=0.30\textwidth]{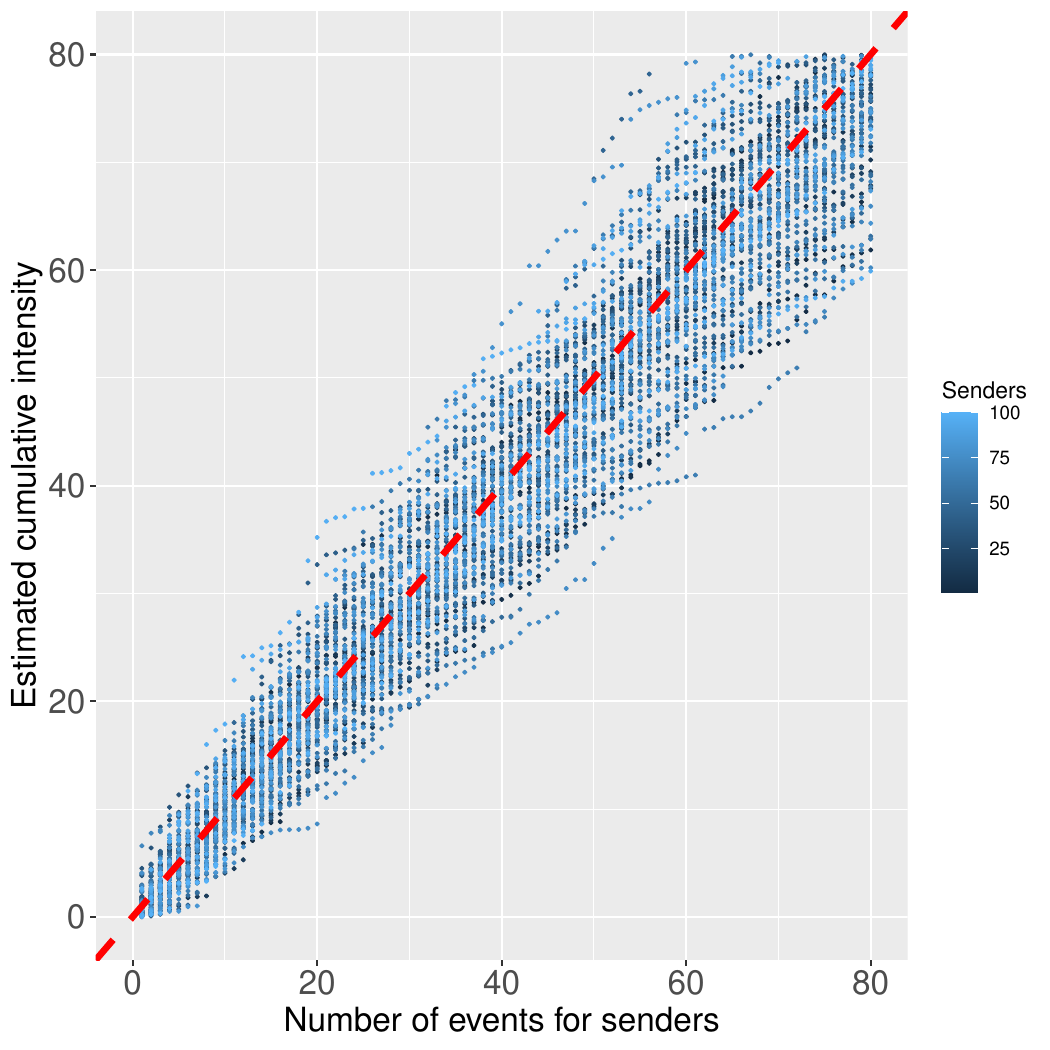}}
	\subfigure[Krei\ss's method for Case II]{
		\includegraphics[width=0.30\textwidth]{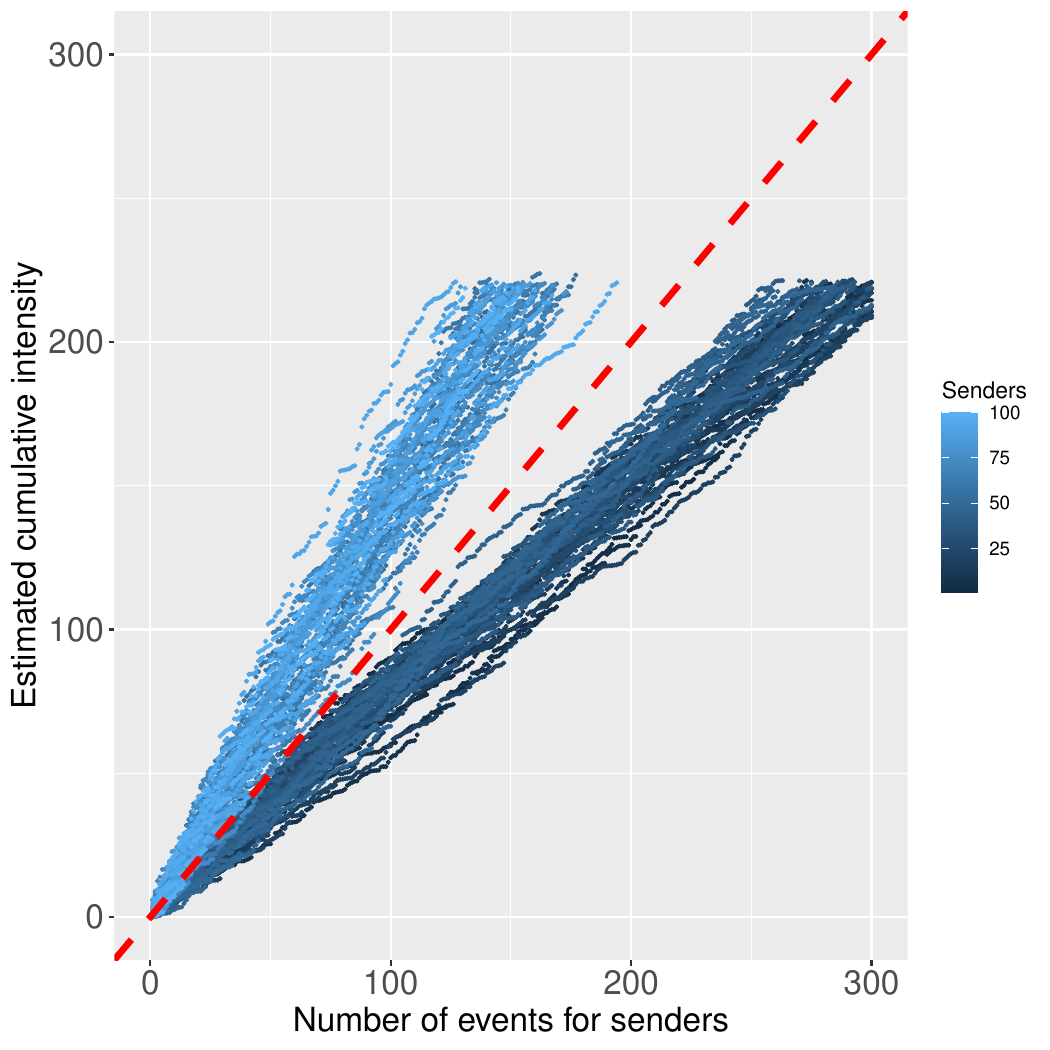}}
  \subfigure[Krei\ss's method for Case III]{
		\includegraphics[width=0.30\textwidth]{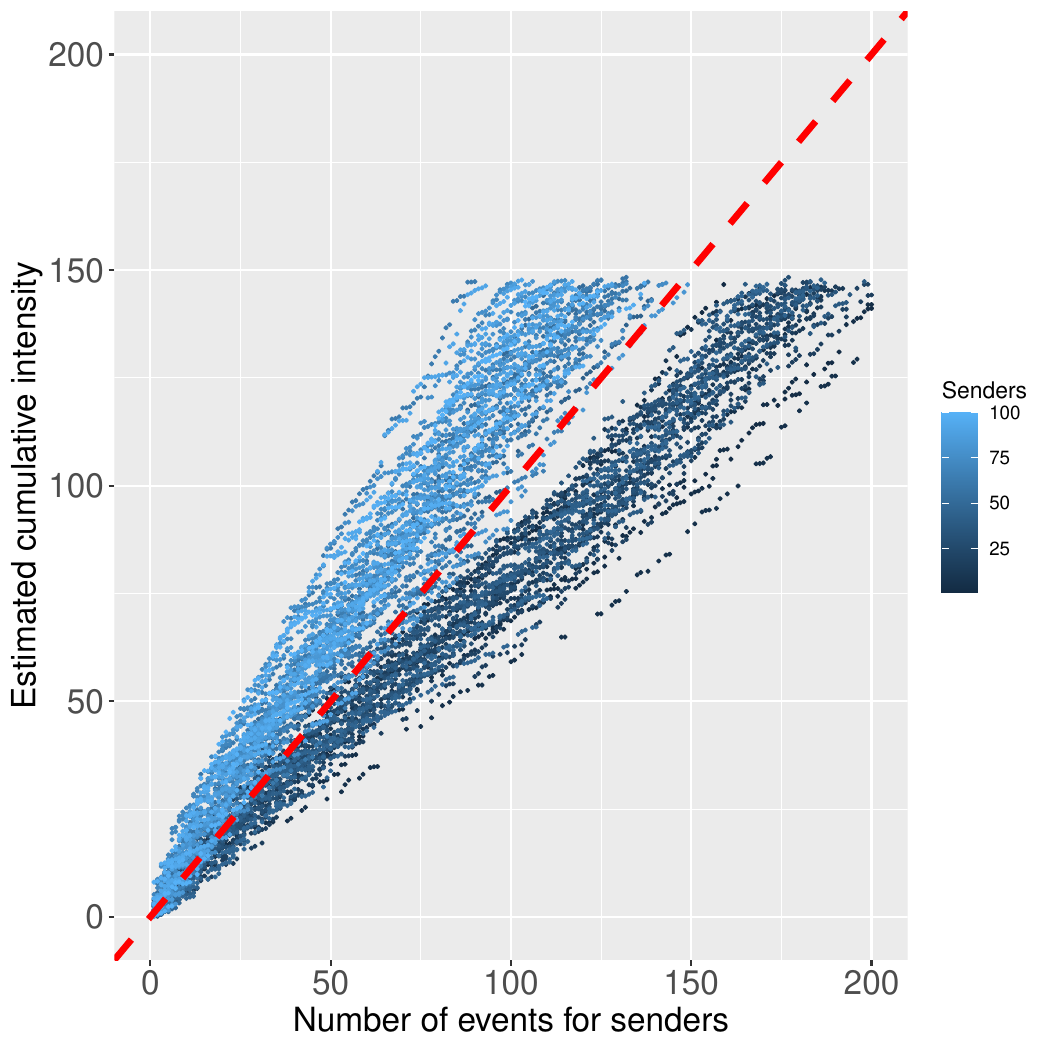}}
	\caption{Simulation comparison for goodness-of-fit. The red dashed line represents the identity line.}
	\label{fig:simulation:GOF}
\end{figure}

We use the graphical diagnostic method to assess the goodness-of-fit
in a comparison between our model and \citeauthor{kreib2019}'s model.
Let the covariate $Z_{ij}=(Z_{ij1},Z_{ij2},Z_{ij3})^\top$,
where $Z_{ijk}$ is generated from the standard normal distribution.
Let $\gamma_j^*(t)=\sin(2\pi t)/3$ for $j\in[3]$,
and we set $\gamma^*(t)=(\gamma_1^*(t),\gamma_2^*(t),\gamma_3^*(t))^\top$.
We consider the following three cases for the degree parameters:
\begin{itemize}
\item[]Case I :  $\alpha_i^*(t)=\beta_i^*(t)\equiv0$ for $i\in[n]$.
\item[]Case II:
$\alpha_i^{*}(t)=\beta_i^{*}(t)=-(0.5\log(n)+3+t/2)$ if $i<n/2$,
and $\alpha_i^{*}(t)=\beta_i^{*}(t)=0$ otherwise.
\item[]Case III: For $i\in[n/2]$,  we define $\alpha_i^*(t)$ as follows:  $\alpha_i^*(t)=-(0.5\log(n)+3+t/2)$ if $t\in[0,1/4]$, $\alpha_i^*(t)=0.5\log(n)+3+t/2$ if $t\in(1/4,1/2]$; $\alpha_i^*(t)=0.2$ if $t\in(1/2,3/4]$ and $\alpha_i^*(t)=0$ otherwise. For $i>n/2$, we set $\alpha^*(t)\equiv0$. The parameters $\beta_i^*(t)~(i\in[n])$ are defined as $\beta_{i}^*(t)=\alpha_{i}^*(t)$ for
    $i\in [n]$.
    The parameter $\gamma_{3}^*(t)$ is omitted, and only $\gamma_{1}^*(t)$ and $\gamma_2^*(t)$ are estimated when carrying out the estimation methods.
\end{itemize}
The
setting in Case I is correctly specified for our model and that of \cite{kreib2019},
while the
setting in Case II is correctly specified for our model but not for the model of \cite{kreib2019}.
In Case III, the specified
setting is incorrect for both models.
Here, we set $n=100$. The results are presented in Fig. \ref{fig:simulation:GOF}.
As expected, when the
setting is correctly specified for both models,
the curves closely align with the identity line (slope one), indicating a good model fit.
Conversely, when the
setting is correctly specified for our model but not for the model of \cite{kreib2019},
the curves for \cite{kreib2019}
show significant deviations
from the identity line,
whereas ours remain well-aligned.
However, if the
setting is incorrectly specified for both models,
the deviations of the curves become substantial, indicating poor model fit.
These results demonstrate that the Arjas plot is effective in assessing the goodness-of-fit
for both the proposed model and that of \cite{kreib2019}.

\section{Uncovering Temporal Patterns in Real-World Networks}
\label{section:realdata}

Dynamic network analysis has emerged as a powerful framework for uncovering temporal patterns in real-world networks. To demonstrate  our statistical method for analyzing continuous-time network interactions, we apply it to two real-world datasets: the MIT Social Evolution dataset, tracking students' social interactions from September 2008 to July 2009, and Washington, D.C.'s bike-sharing dataset, recording bicycle rentals and returns across stations from April to December 2018. Our approach elucidates evolving dynamics in social and bike-sharing networks by:
	\begin{enumerate}
		\item analyzing time-varying in- and out-degree heterogeneity to reveal distinct behavioral profiles of nodes (students or stations);
		\item estimating time-varying covariate effects to identify drivers of interaction or trip frequencies;
		\item classifying nodes into roles based on interaction intensity and tracking their evolution.
	\end{enumerate}
	These examples highlight the method's ability to uncover dynamic behavioral patterns across diverse time-evolving networks.

\subsection{MIT Social Evolution data}\label{Analysis:MIT}

The MIT Social Evolution dataset captures social interactions among students in an MIT dormitory over an academic year (September 1, 2008, to July 16, 2009). This dataset has been used to study infection spread \citep{Dong2012} and the diffusion of health-related behaviors among undergraduates \citep{madan2010}. Interactions were tracked via cellular phones, with Bluetooth signals recorded when a sender's phone connected to a receiver's phone within a 10-meter range, serving as a proxy for daily social activity. Through rigorous data cleaning \citep{lee2018inferring}, we constructed a network of 63 students, most of whom resided in the dormitory. We analyzed categorical covariates influencing pairwise communication, including same-floor residence and absolute differences in admission years (categorized as 0, 1, 2, or 3 years). We adopt the Gaussian kernel, $\mathcal{K}(x) = \exp(-x^2/2)/\sqrt{2\pi}$, with bandwidths selected via the proposed 5-fold cross-validation method.

\begin{figure}[h]
	\centering
	\subfigure[Subfigure 1 list of figures text][$\widehat\alpha_1'(t)$]{
		\includegraphics[width=0.32\textwidth]{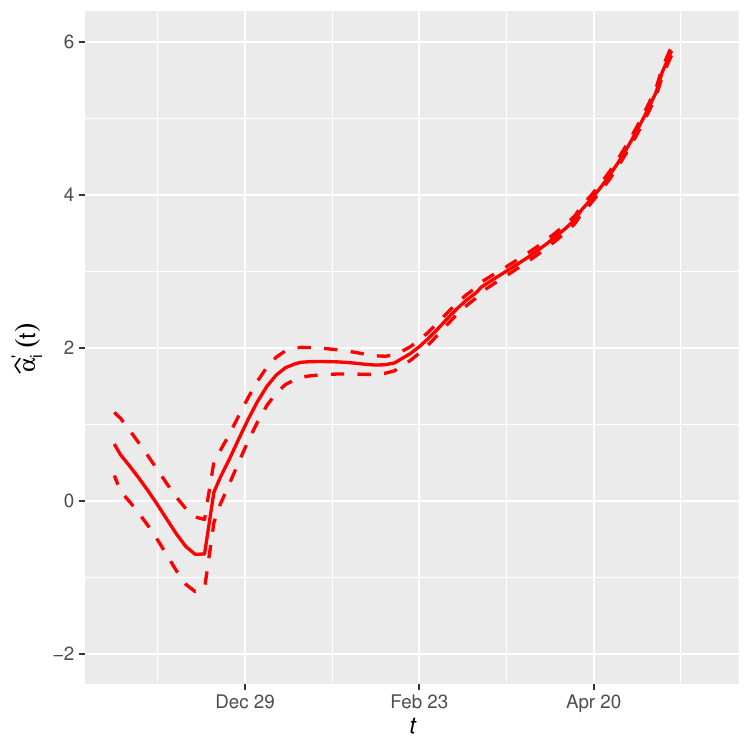}}
	\subfigure[Subfigure 2 list of figures text][$\widehat\alpha_{32}'(t)$]{
		\includegraphics[width=0.32\textwidth]{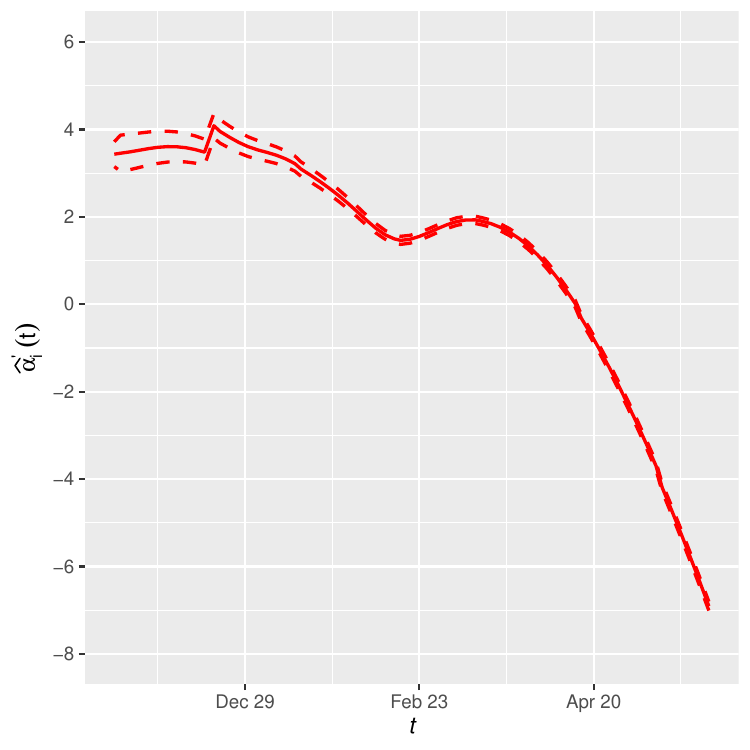}}
	\qquad
	\subfigure[Subfigure 1 list of figures text][$\widehat\beta_{1}'(t)$]{
		\includegraphics[width=0.32\textwidth]{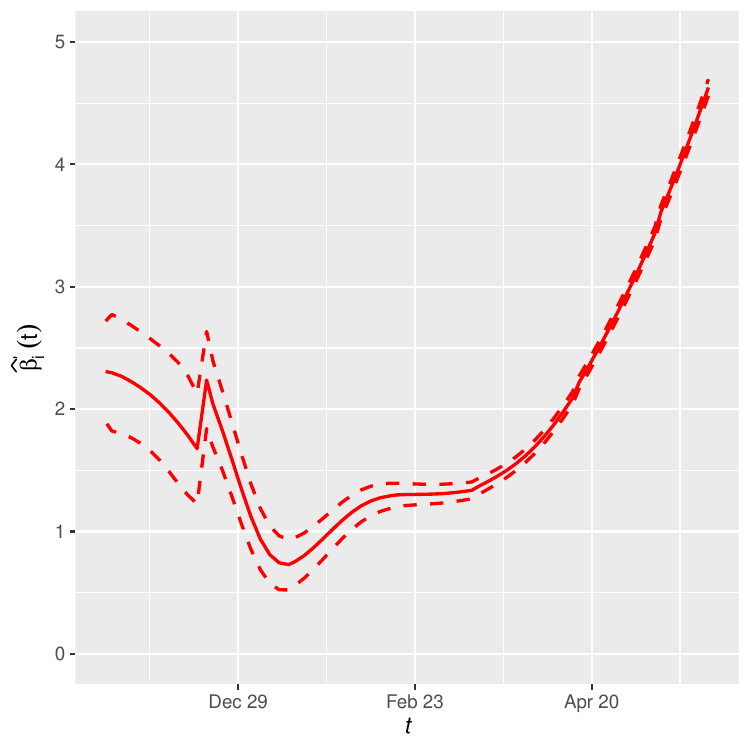}}
	\subfigure[Subfigure 2 list of figures text][$\widehat\beta_{32}'(t)$]{
		\includegraphics[width=0.32\textwidth]{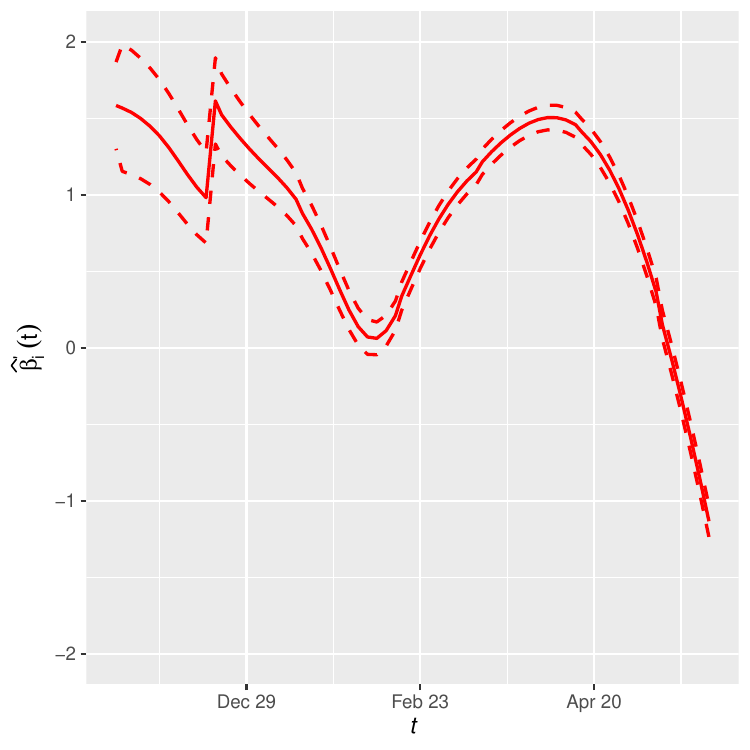}}
	\caption{MIT Social Evolution data: The estimated curves (solid lines) of $\alpha_i^*(t)$ and $\beta_i^*(t)$ for students with IDs $1$ and $32$. The dashed lines represent the pointwise $95\%$ confidence intervals.
 }
	\label{fig:realID}
\end{figure}

First, clear time-varying patterns were observed in the estimated in- and out-degree parameters of students. For example, Fig.~\ref{fig:realID} presents the estimated curves of $\alpha_i^*(t)$ and $\beta_i^*(t)$ for students with IDs 1 and 32. To avoid arbitrary baseline selection, Fig.~\ref{fig:realID} shows the adjusted curves $\widehat{\alpha}_i'(t) = \widehat{\alpha}_i(t) - \overline{\alpha}(t)$ and $\widehat{\beta}_i'(t) = \widehat{\beta}_i(t) - \overline{\beta}(t)$, where $\overline{\alpha}(t) = n^{-1}\sum_{i=1}^n \widehat{\alpha}_i(t)$ and $\overline{\beta}(t) = n^{-1}\sum_{i=1}^n \widehat{\beta}_i(t)$. These curves reveal that both students display time-varying communication patterns as senders and receivers of Bluetooth signals throughout the study period. We conducted the testing procedures detailed in Section~\ref{sec:Trt} to evaluate whether the degree parameters vary over time. The analysis produced a $p$-value below 0.001, confirming that the degree parameters are indeed time-varying.

Second, we implemented the testing procedure described in Section~\ref{sec:dht} to determine whether the degree parameters vary across individuals, thereby assessing the presence of degree heterogeneity. The $p$-values for testing $\alpha_i^*(t)$ and $\beta_i^*(t)$ are both below 0.001, providing strong evidence of significant degree heterogeneity within the network. This result is further supported by Fig.~\ref{fig:realID}, which illustrates distinct communication behaviors between students 1 and 32 throughout the study period.
\begin{figure}[h]
	\centering
	\subfigure[Subfigure 1 list of figures text][$\hat{\gamma}_1(t)$: Same floor]{
		\includegraphics[width=0.33\textwidth]{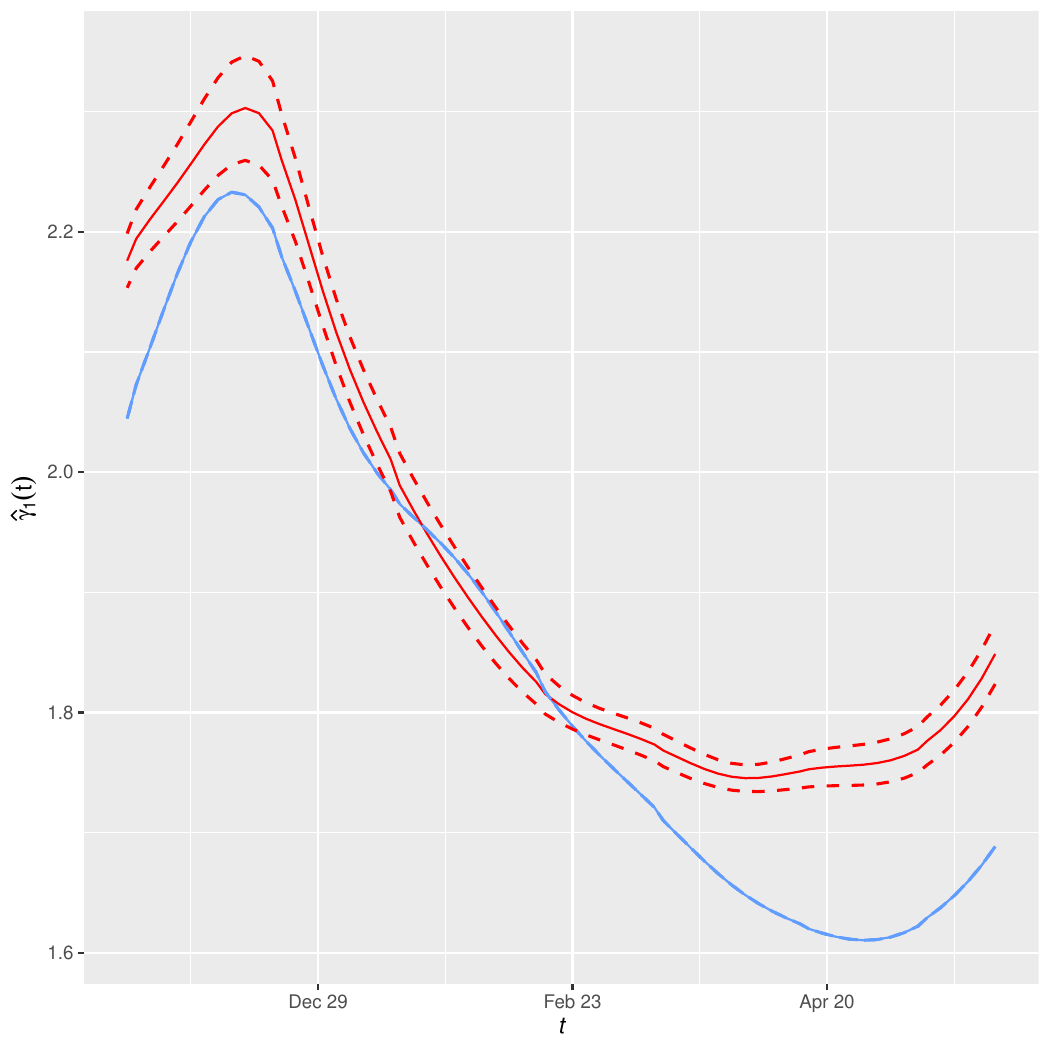}}
	\subfigure[Subfigure 2 list of figures text][$\hat{\gamma}_2(t)$: Same admission year]{
		\includegraphics[width=0.33\textwidth]{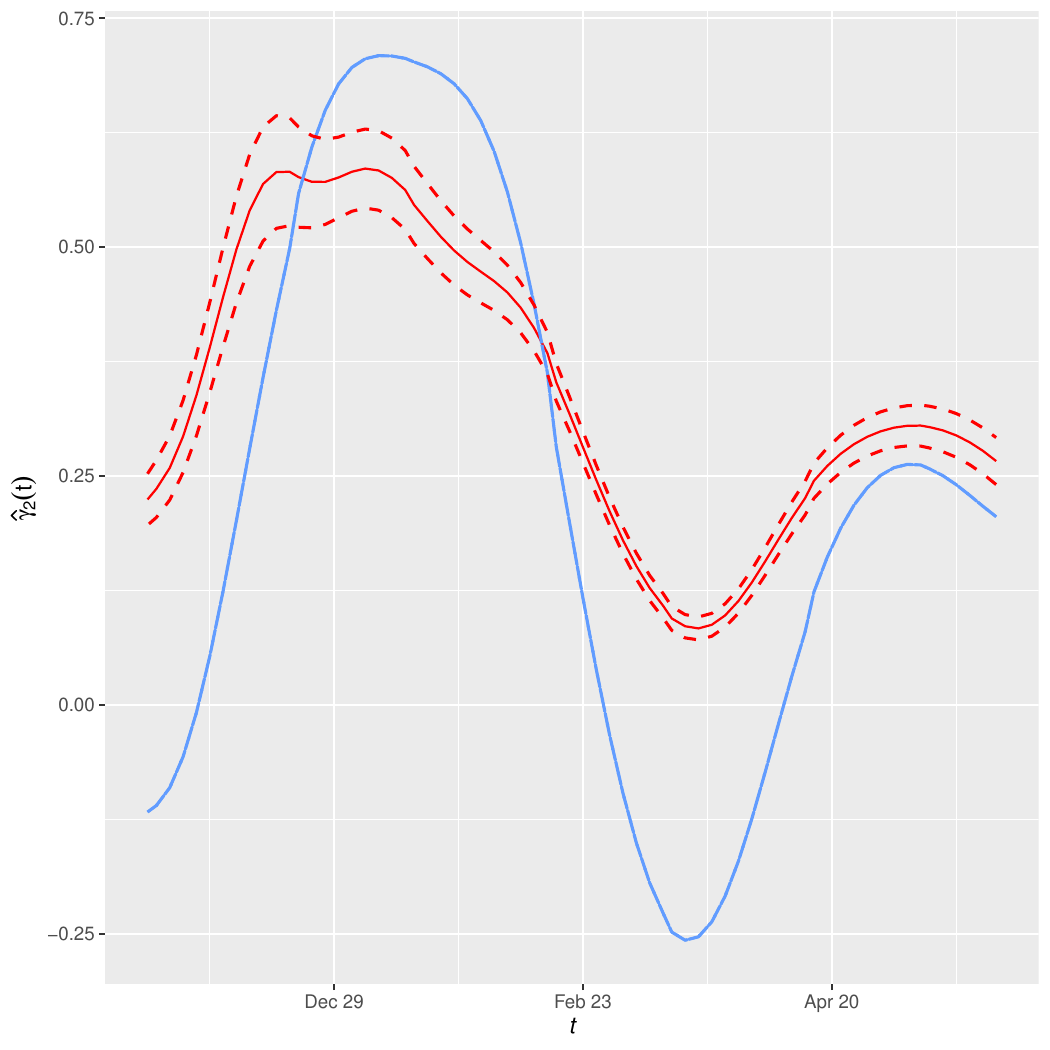}}
	\qquad
	\subfigure[Subfigure 1 list of figures text][$\hat{\gamma}_3(t)$: 1 year difference on admission]{
		\includegraphics[width=0.33\textwidth]{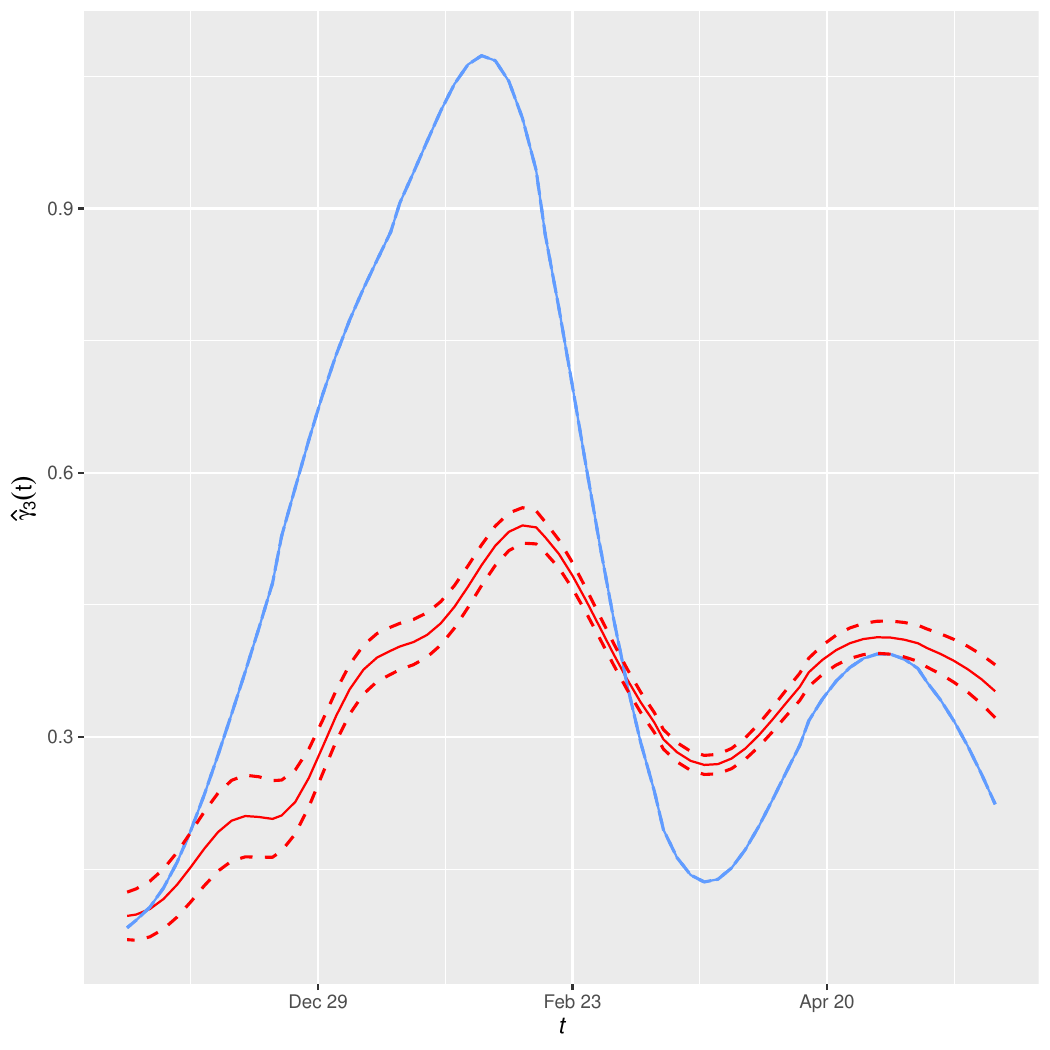}}
	\subfigure[Subfigure 2 list of figures text][$\hat{\gamma}_4(t)$: 2 years difference on admission]{
		\includegraphics[width=0.33\textwidth]{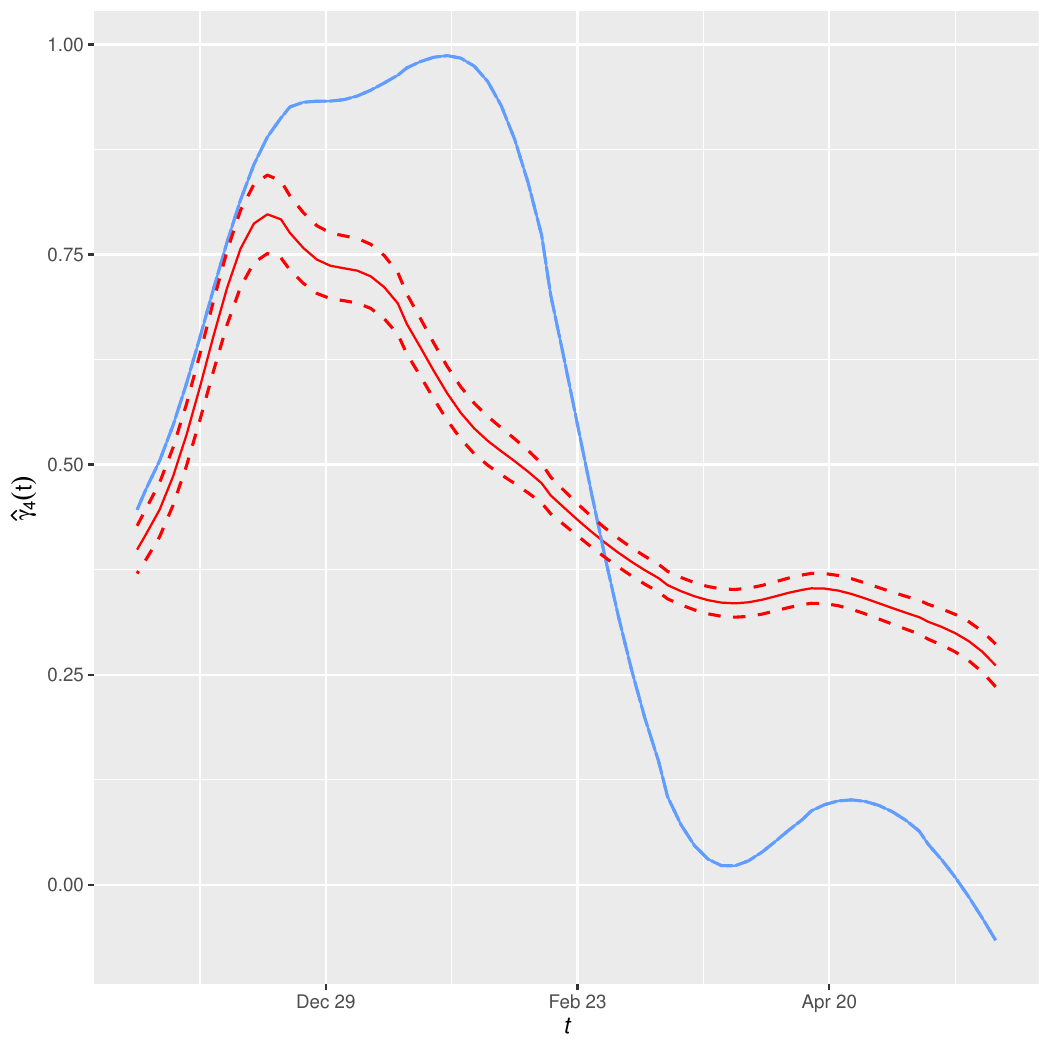}}
	\caption{MIT Social Evolution data: The red solid curves represent the estimates of covariate effects using our model,
while the blue curves represent the estimates obtained by the method in \cite{kreib2019}.
The red dashed lines represent the pointwise $95\%$ confidence intervals.
}
	\label{fig:realHT}
\end{figure}

\begin{figure}[h]
	\centering
		\subfigure[Subfigure 1 list of figures text][Our method for out-degrees]{
		\includegraphics[width=0.33\textwidth]{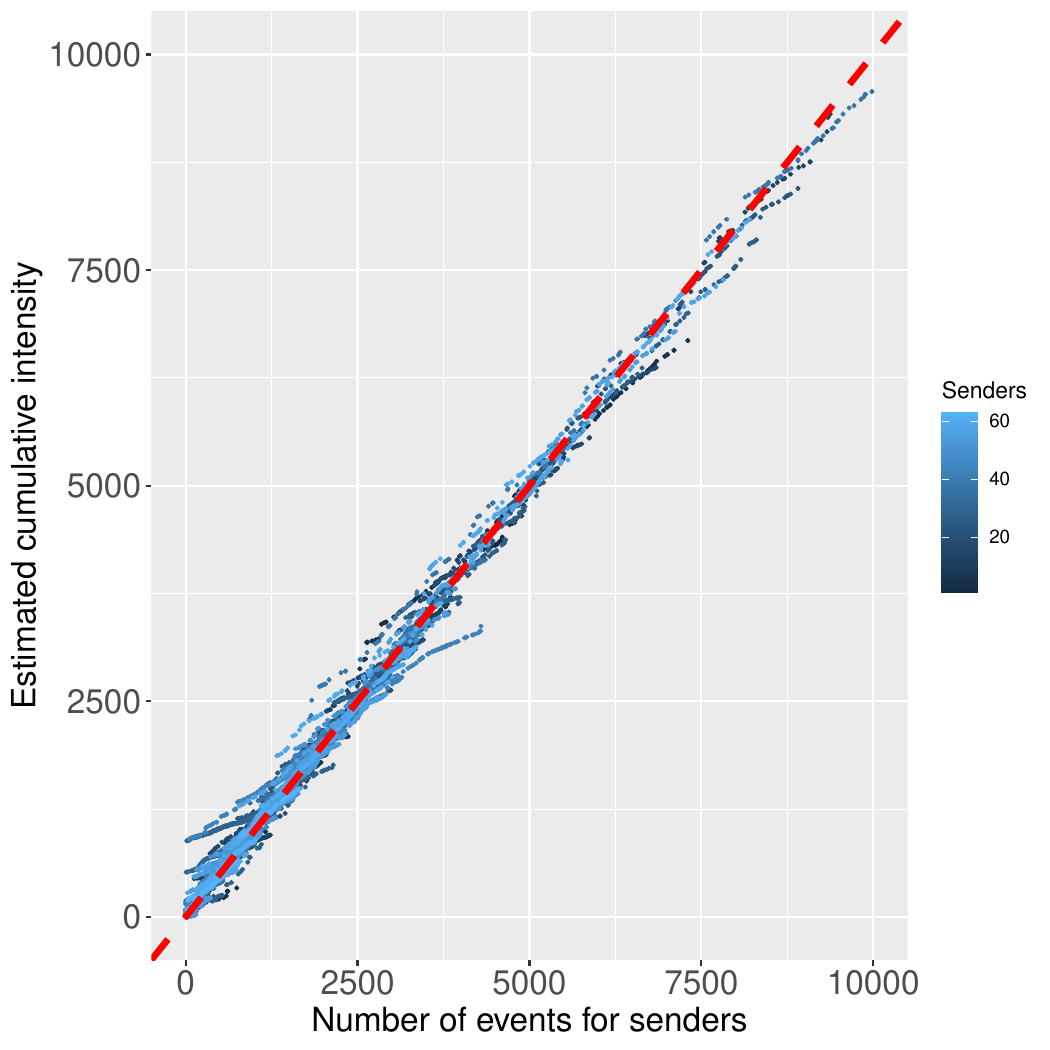}}
	\subfigure[Subfigure 2 list of figures text][Our method for in-degrees]{
		\includegraphics[width=0.33\textwidth]{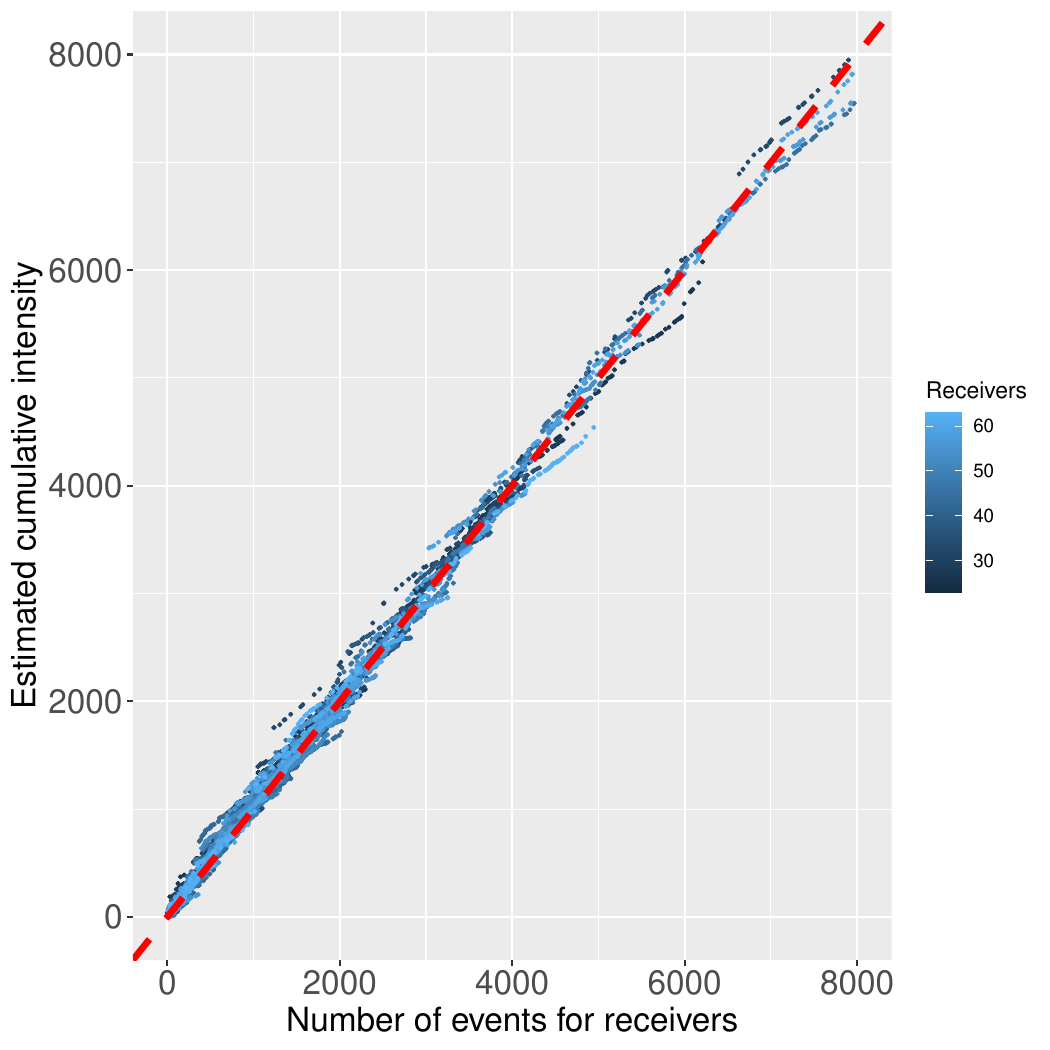}}
	\qquad
\subfigure[Subfigure 1 list of figures text][Krei\ss's method for out-degrees]{
		\includegraphics[width=0.33\textwidth]{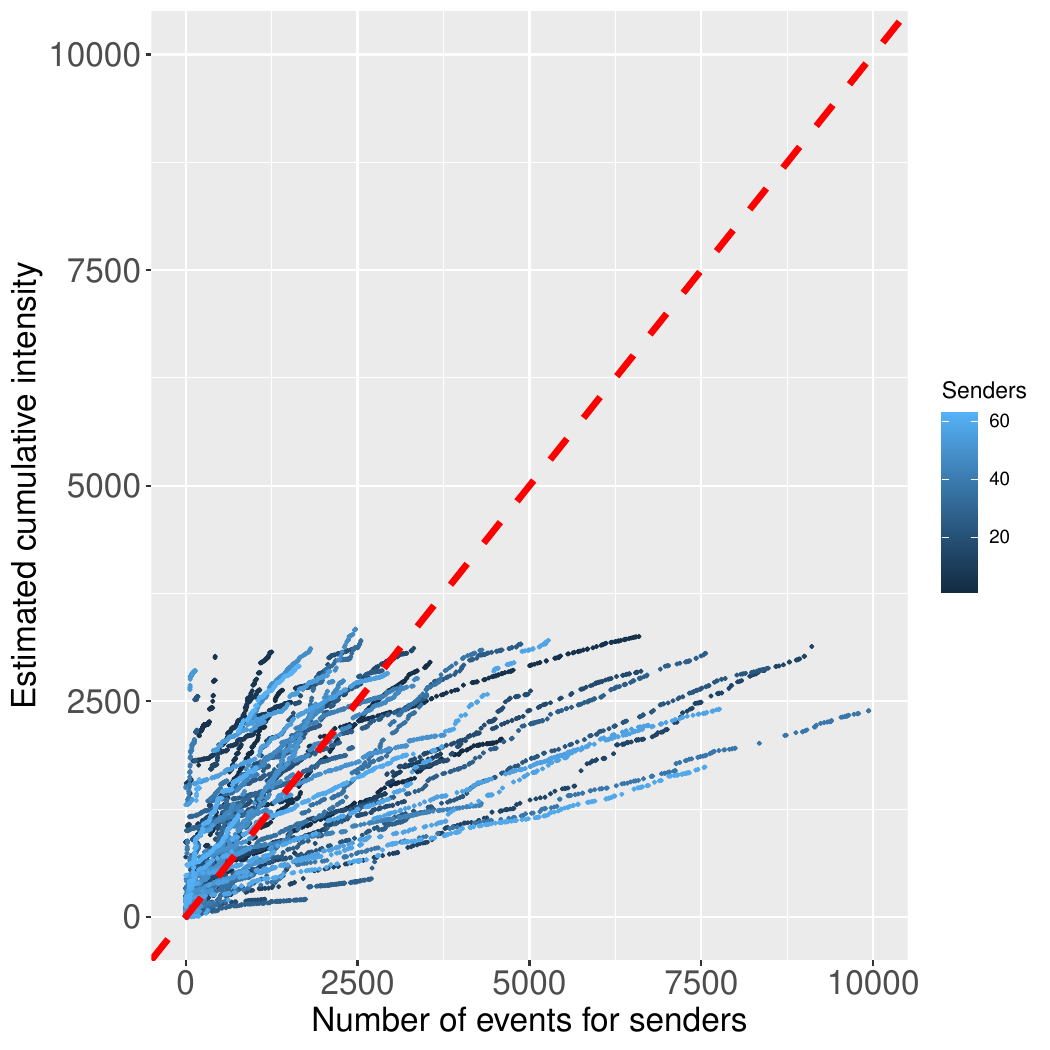}}
	\subfigure[Subfigure 2 list of figures text][Krei\ss's method for in-degrees]{
		\includegraphics[width=0.33\textwidth]{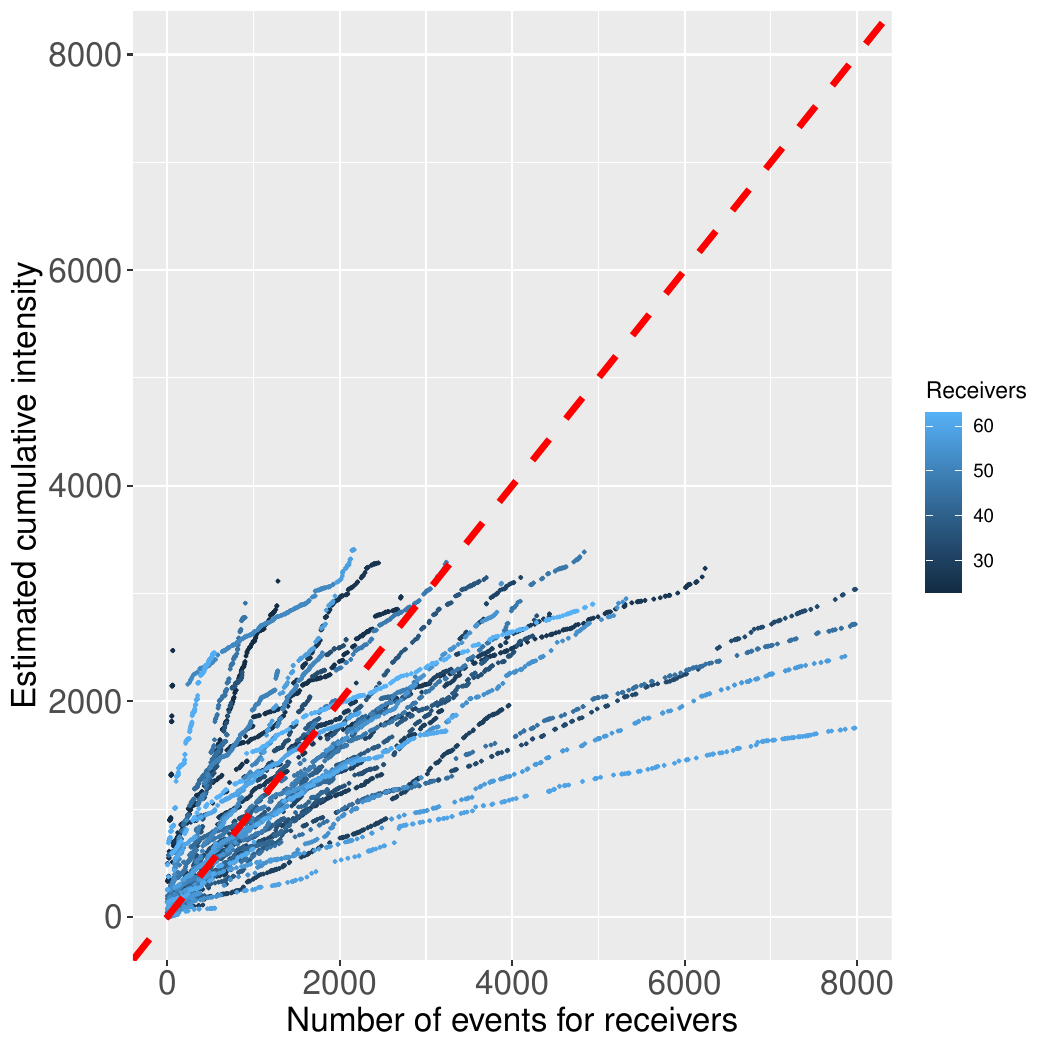}}

	\caption{MIT Social Evolution data: Goodness-of-fit comparison for two methods (red dashed line denotes the identity line)
}
	\label{fig:realGOF:MIT}
\end{figure}

Third, we investigate covariate effects and present the estimated parameter curves in Fig.~\ref{fig:realHT}. Our findings differ from those of \cite{kreib2019}. For example, their method suggests negative effects among students sharing the same admission year during certain intervals, whereas our approach reveals consistently positive effects throughout the study period. A goodness-of-fit analysis, shown in Fig.~\ref{fig:realGOF:MIT}, compares both models. Our method's curves closely align with the identity line (slope one), indicating a good fit to the data, while those of \cite{kreib2019} diverge significantly, suggesting a less accurate fit. These differences highlight the importance of accounting for subject-specific effects, as implemented in our proposed degree-corrected Cox network model, to accurately estimate covariate effects.
We conducted the testing procedure outlined in Section~\ref{sec:Trt} to assess time-varying trends in covariate effects. The resulting p-value ($< 0.001$) confirms significant time-varying trends.

Finally, to characterize students' roles and their evolution within a social network, we divide the observation period into five equally spaced time segments ($T_1$-$T_5$) to track role transitions. Each student's role in a given segment is classified into one of four social roles based on their integrated social metrics for that segment, relative to median thresholds computed across all students and time segments.

\begin{itemize}
	\item \textbf{Social receptiveness}: The receptiveness parameter is defined as $\alpha_{ik} = \frac{1}{|T_k|} \int_{T_k} \hat{\alpha}_i(t) \, dt$, and individuals are classified as having `high' or `low' receptiveness based on whether $\alpha_{ik}$ is above or below the median of $\{ \alpha_{ik} : 1 \leq i \leq n, 1 \leq k \leq 5 \}$, respectively.
	\item \textbf{Social initiative}: The social initiative parameter is defined as $\beta_{ik} = \frac{1}{|T_k|} \int_{T_k} \hat{\beta}_i(t) \, dt$, and  individuals are classified as having `high' or `low' initiative based on whether $\beta_{ik}$ is above or below the median of $\{ \beta_{ik} : 1 \leq i \leq n, 1 \leq k \leq 5 \}$,
	respectively.
\end{itemize}

\begin{figure}
	\centering
	\includegraphics[width=0.5\linewidth]{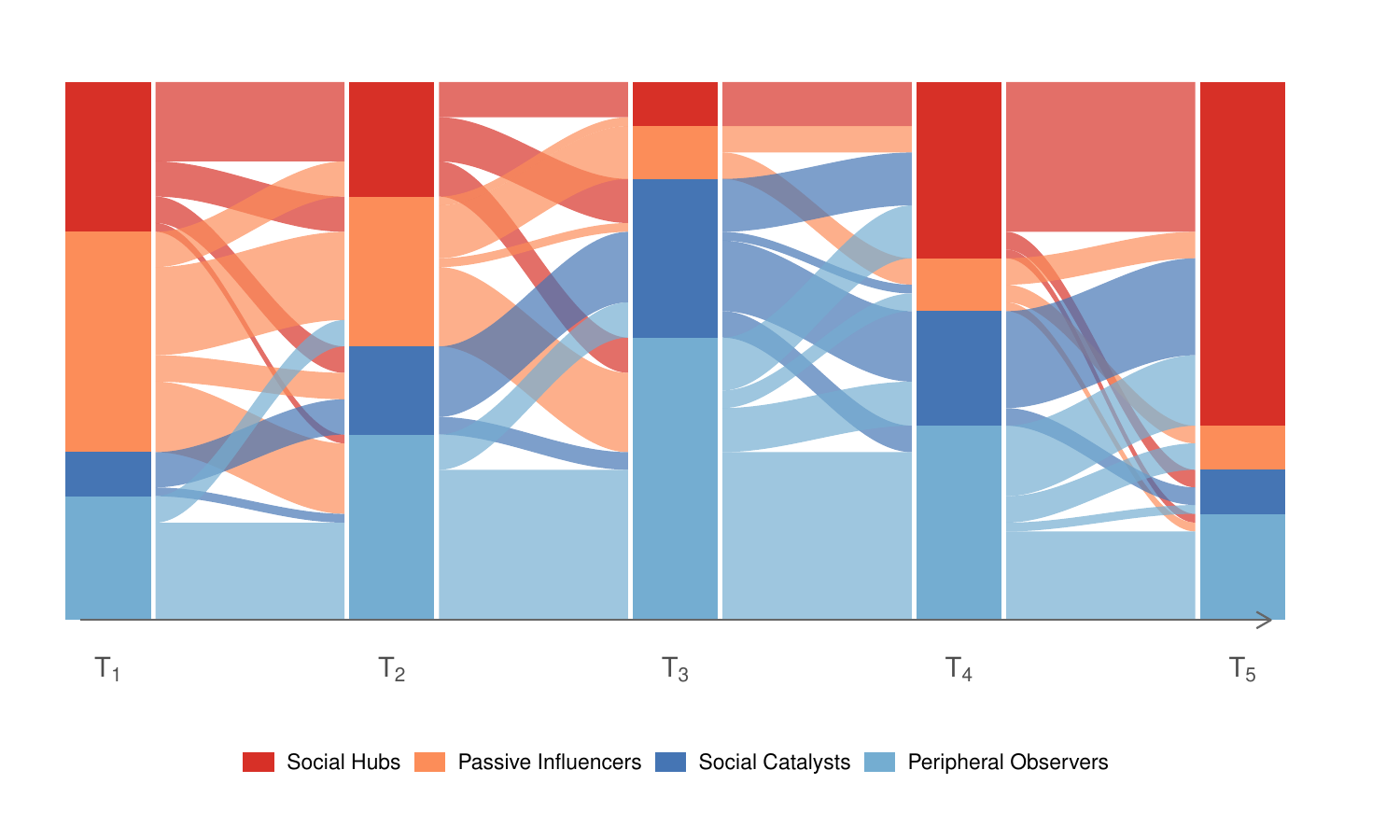}
	\caption{\label{fig:flow_mit} MIT Social Evolution data: Alluvial plot of social role transitions across four archetypes over five time segments (curves show individual shifts, thickness reflects the number of individuals, and total height represents 63 students) }
\end{figure}

\begin{figure}
	\centering
	\includegraphics[width=0.4\linewidth]{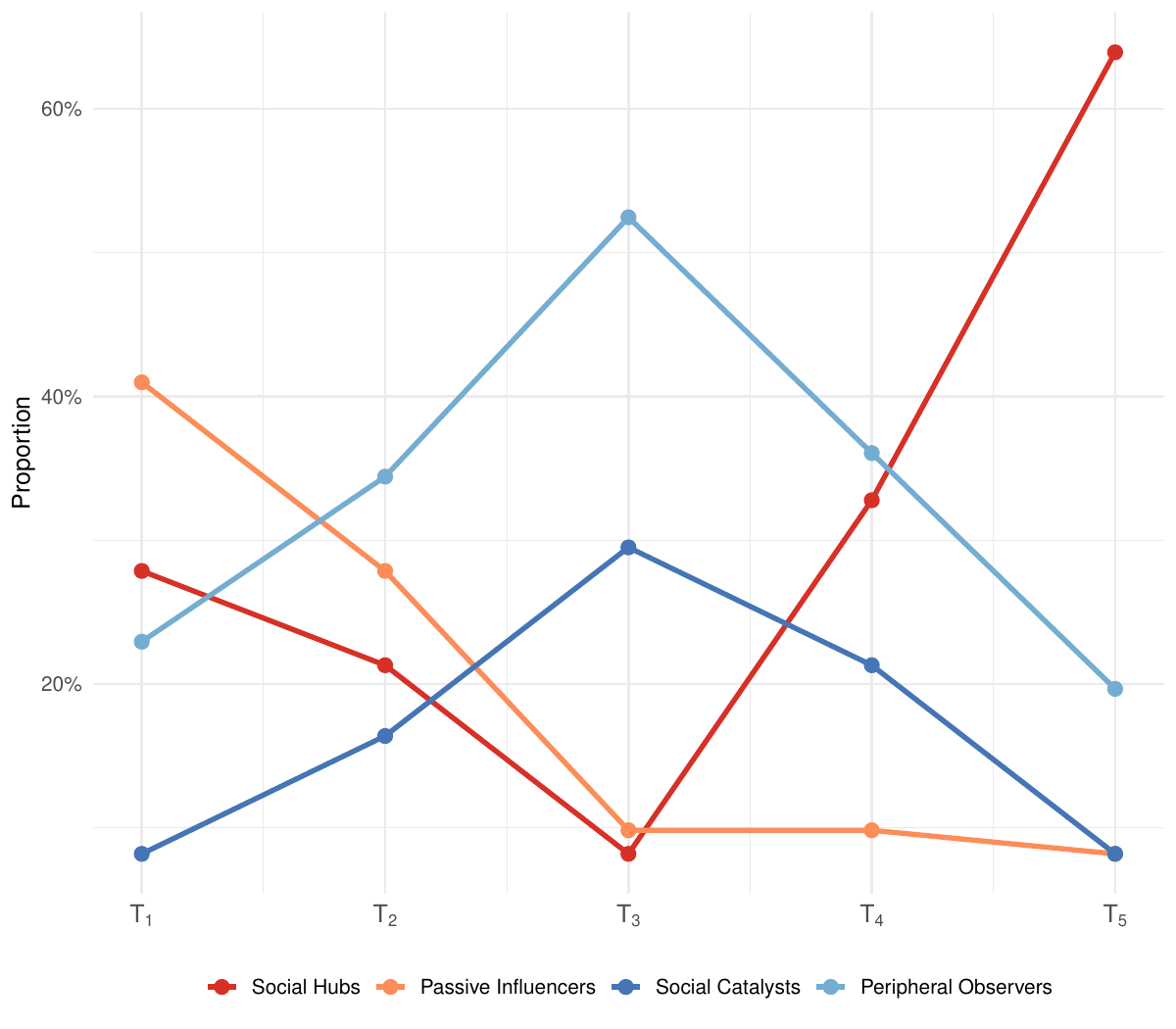}
	\caption{\label{fig:line_mit}MIT Social Evolution data: Temporal changes in the  proportions of four social roles}
\end{figure}

This classification yields four archetypes: Social Hubs (high in both receptiveness and initiative), Passive Influencers (high in receptiveness, low in initiative), Social Catalysts (low in receptiveness, high in initiative), and Peripheral Observers (low in both receptiveness and initiative). Figures~\ref{fig:flow_mit} and~\ref{fig:line_mit} illustrate a distinct three-phase social adaptation pattern. During the initial orientation phase ($T_1$), 27.9\% of students were Social Hubs, actively exploring social connections. This proportion dropped sharply by midterms ($T_3$, 8.2\% Social Hubs) due to increasing academic demands. A robust recovery ensued ($T_5$, 63.9\% Social Hubs), fueled by heightened engagement from less social students and strengthened ties among outgoing ones. This ``explore-commit" cycle aligns with social adaptation theories, with winter break ($T_4$) serving as a key turning point, deepening bonds through shared experiences. The network's eventual dominance by Social Hubs reflects a self-reinforcing system, where early engagement disparities amplify via the sociological principle of cumulative advantage.

	We summarize key observations from the MIT Social Evolution dataset:
	\begin{itemize}
		\item[(1)] The  curves of the estimates of in- and out-degree parameters for student 1 show a generally increasing trend over time,
while those for student 32 present visibly downward trends. These observations show that it is necessary to incorporate time-varying degree heterogeneity parameters to characterize the intrinsic differences of individuals
when modeling temporal networks. In addition, the out- or in-degree effects for student 1 are generally positive, while
for student 32,  out-degree effect before April 20, 2009 and in-degree effects before May 8, 2009 are positive,
but become negative soon afterwards.

		\item[(2)] Estimates for students on the same floor or from the same admission year are consistently positive, indicating stronger communication among students sharing these attributes.
		\item[(3)] The influence of same-floor students on network formation peaks by December 15, 2008, then declines until stabilizing after March 18, 2009. Similarly, the effect of same-admission-year students follows this pattern but shows a significant increase after March 18, 2009. Connections are more frequent before March 9, 2009, possibly due to students spending more time in dormitories during colder winter months.
		\item[(4)] Estimated in- and out-degree parameters enable the identification of social roles and their evolution, providing insights into the dynamic social structure.
	\end{itemize}
	These findings have implications for predicting and preventing illness spread. For example, observation (1) suggests monitoring Student 32's social interactions at the start of a flu survey \citep{Dong2012}, with increased focus on Student 1's activities over time. Observation (3) suggests that students on the same floor should strengthen flu prevention measures. 

\subsection{Capital Bikeshare data}\label{Analysis:Bike}

We analyze Capital Bikeshare data from Washington, D.C., and surrounding areas, modeling bike stations as vertices in a network. An interaction from station $i$ to station $j$ occurs when a user rents a bike at station $i$ and returns it at station $j$. This study examines biking patterns from April to December 2018, covering 542 stations and 3,542,684 connections. The data are publicly available at \url{http://www.capitalbikeshare.com/system-data}.

The covariates capture the distance and density of bike stations. The distance from station $i$ to station $j$, denoted $d_{ij}$, is the biking time from $i$ to $j$, calculated using Google Maps for a weekday afternoon. Note that $d_{ij}$ may differ from $d_{ji}$ due to one-way streets or inclines. Following \cite{kreib2019}, the density of station $i$, denoted $n(i)$, is the average number of stations reachable from $i$ or reaching $i$ within three minutes. The covariate is defined as:
\[
Z_{ij} :=
\begin{pmatrix}
	\log(d_{ij} \vee 1) \\
	(\log(d_{ij} \vee 1))^2 \\
	\log(n(i) \vee 1) \\
	\log(n(j) \vee 1)
\end{pmatrix},
\]
where $\vee$ denotes the maximum operator. The kernel and bandwidth are selected in the same way as in Section~\ref{Analysis:MIT}.

Bike-sharing patterns exhibited distinct time-varying characteristics. For instance, the following table lists four stations, including their names and geographical coordinates.\\

	\begin{tabular}{ll S[table-format=-2.5] S[table-format=2.5]}
		\toprule
		\textbf{Station ID} & \textbf{ Name} & {\textbf{Longitude}} & {\textbf{Latitude}} \\
		\midrule
		{$101$} & 20th St \& Virginia Ave NW & {$-77.04513$} & {$38.89472$} \\
		{$301$} & Georgia Ave and Fairmont St NW & {$-77.02261$} & {$38.92482$} \\
		{$401$} & New Hampshire Ave \& T St NW & {$-77.03825$} & {$38.91554$} \\
		{$501$} & Tysons West Transit Center & {$-77.23177$} & {$38.93270$} \\
		\bottomrule\\
	\end{tabular}

\begin{figure}[h]
	\centering
	\subfigure[Subfigure 1 list of figures text][$\widehat\alpha_{301}'(t)$]{
		\includegraphics[width=0.23\textwidth]{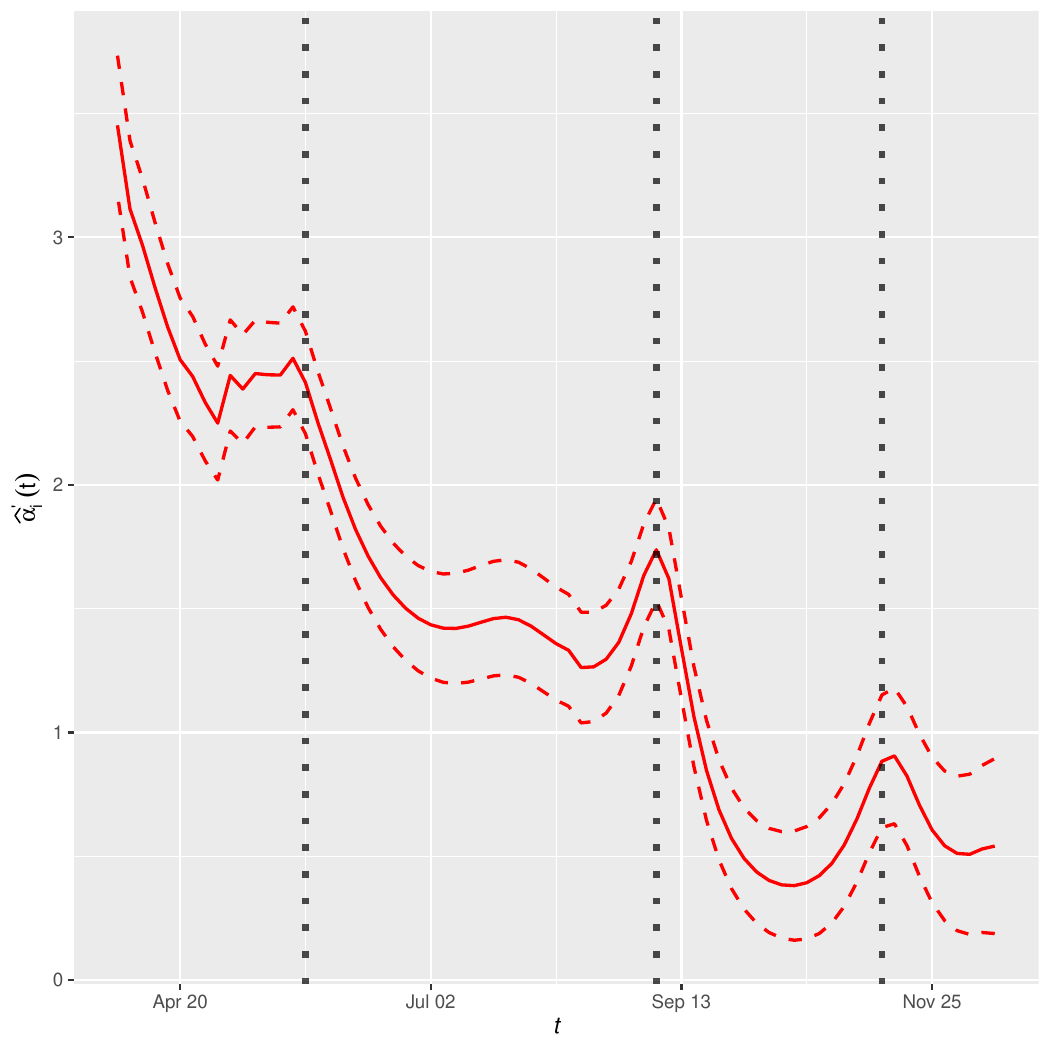}}
	\subfigure[Subfigure 1 list of figures text][$\widehat\alpha_{501}'(t)$]{
		\includegraphics[width=0.23\textwidth]{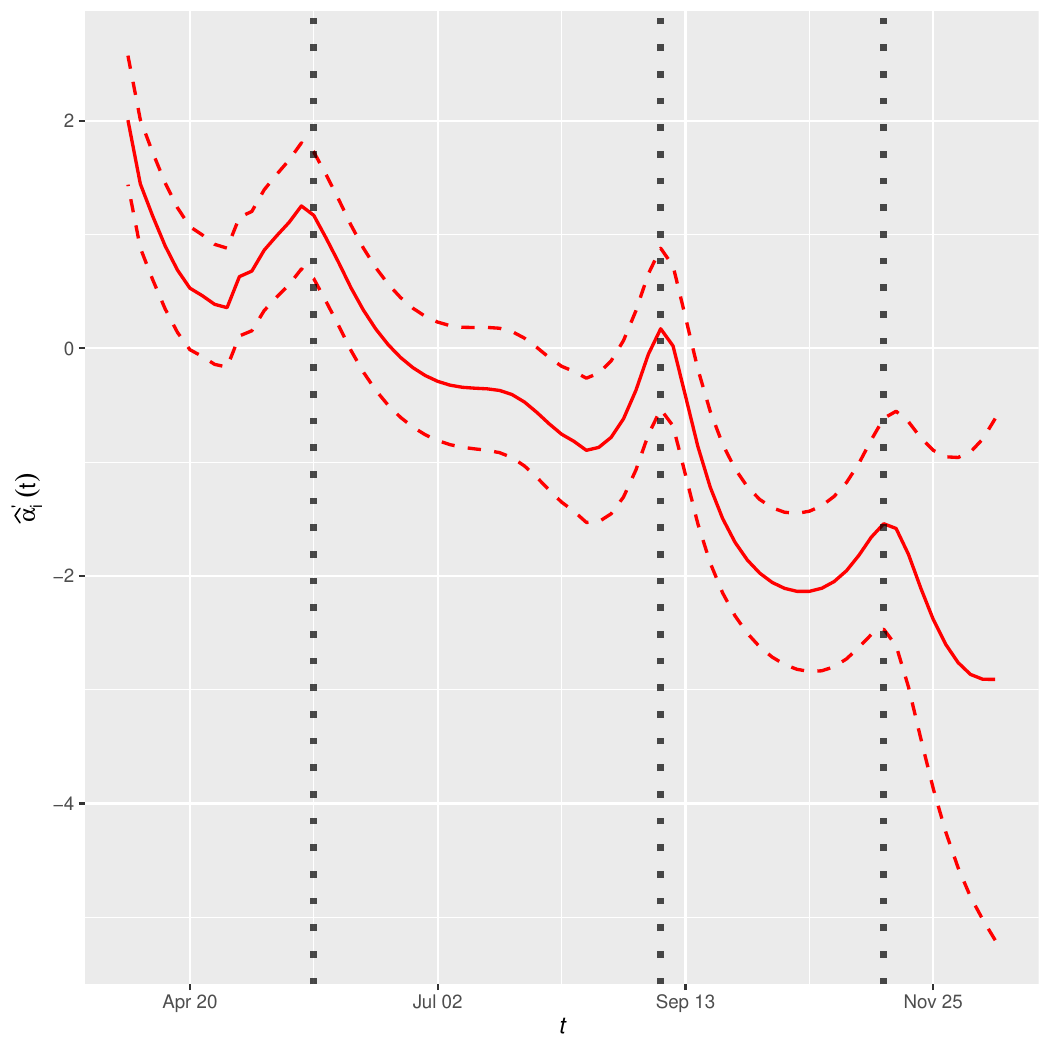}}
	\subfigure[Subfigure 1 list of figures text][$\widehat\beta_{301}'(t)$]{
		\includegraphics[width=0.23\textwidth]{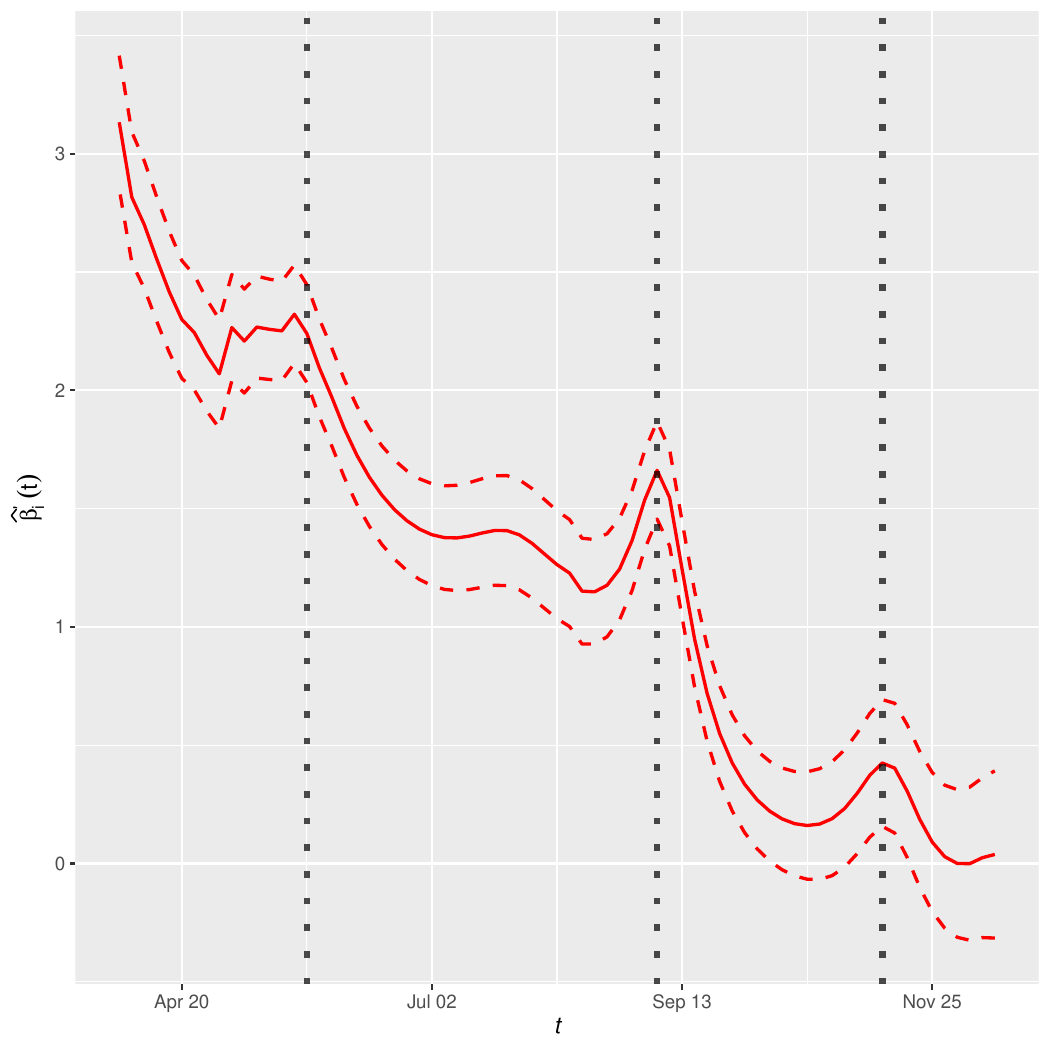}}
		\subfigure[Subfigure 1 list of figures text][$\widehat\beta_{501}'(t)$]{
			\includegraphics[width=0.23\textwidth]{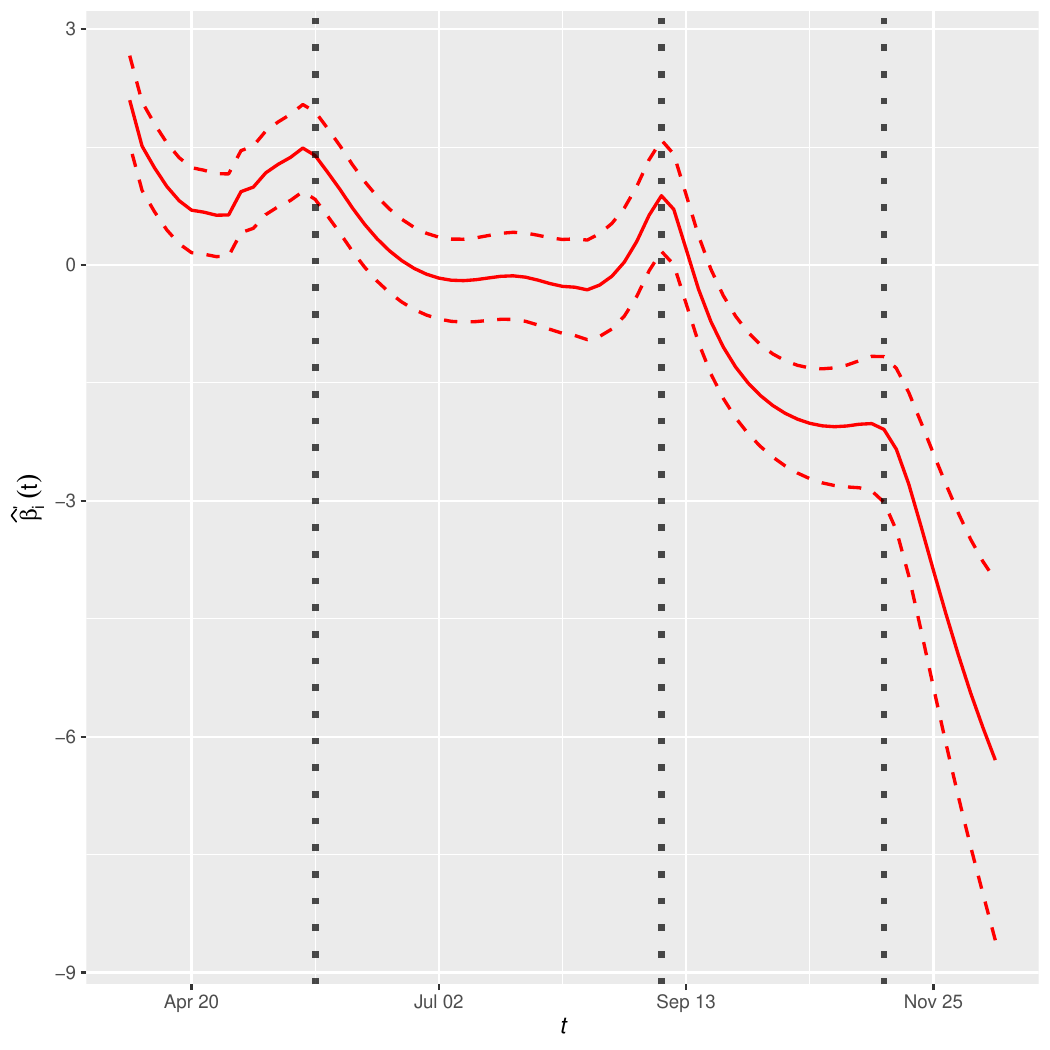}}
	\qquad
	\subfigure[Subfigure 2 list of figures text][$\widehat\alpha_{101}'(t)$]{
		\includegraphics[width=0.23\textwidth]{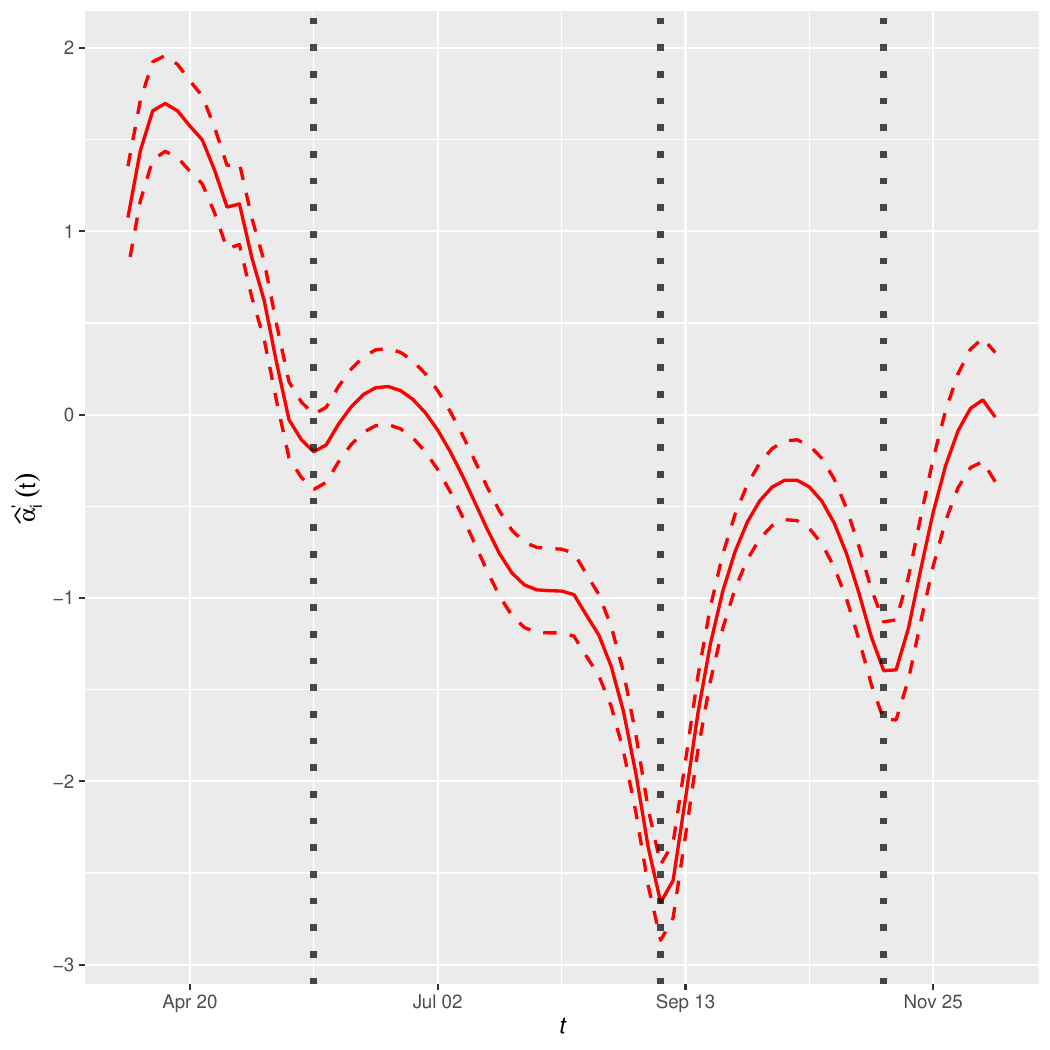}}
	\subfigure[Subfigure 2 list of figures text][$\widehat\alpha_{401}'(t)$]{
		\includegraphics[width=0.23\textwidth]{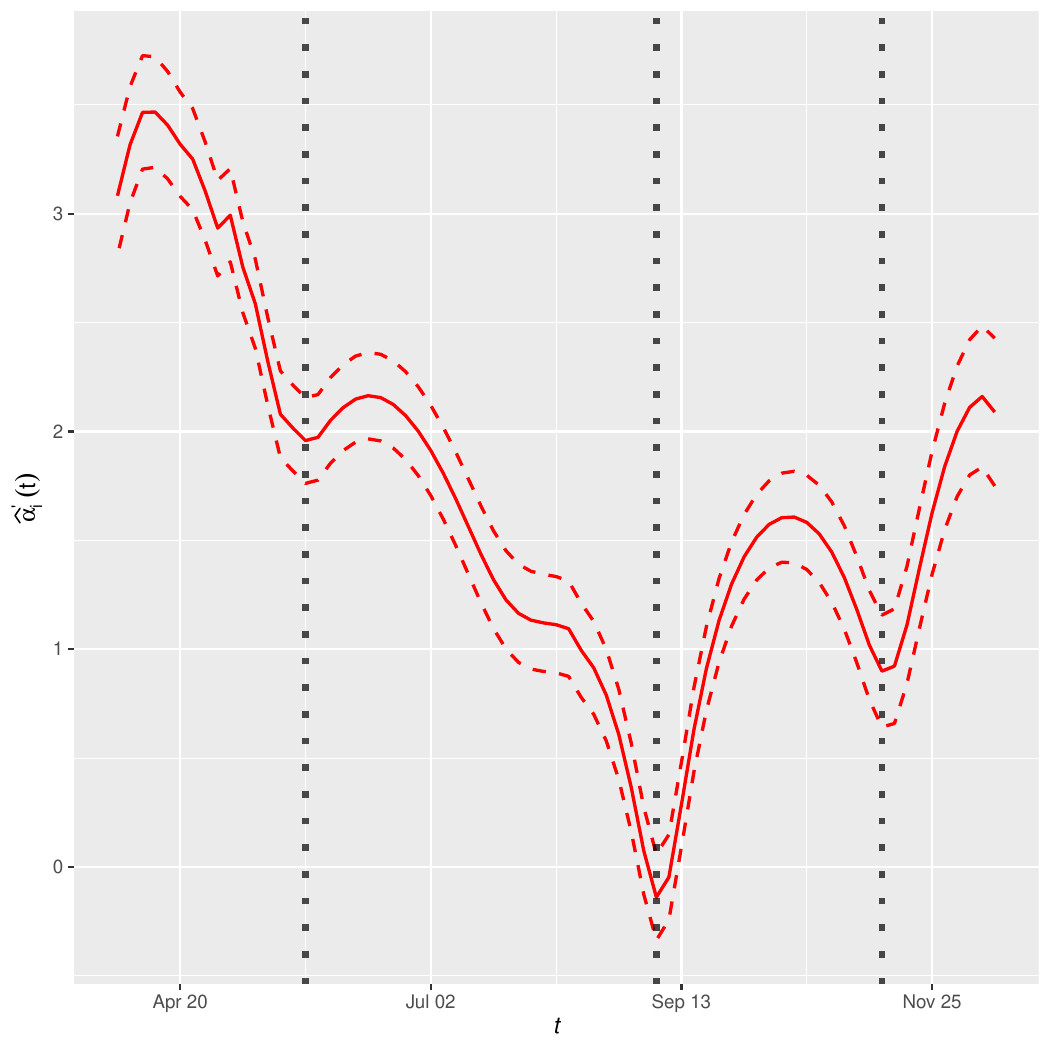}}
	\subfigure[Subfigure 2 list of figures text][$\widehat\beta_{101}'(t)$]{
		\includegraphics[width=0.23\textwidth]{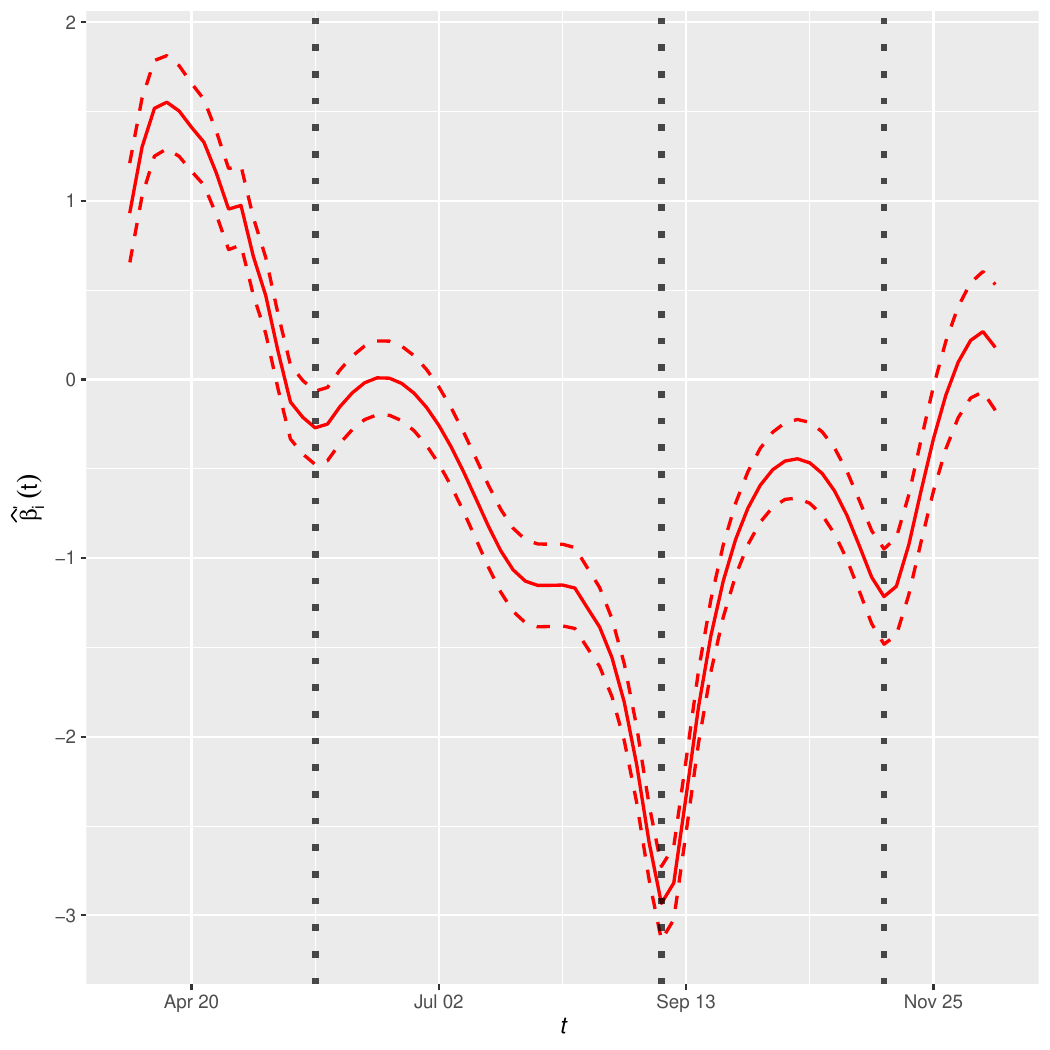}}
		\subfigure[Subfigure 2 list of figures text][$\widehat\beta_{401}'(t)$]{
			\includegraphics[width=0.23\textwidth]{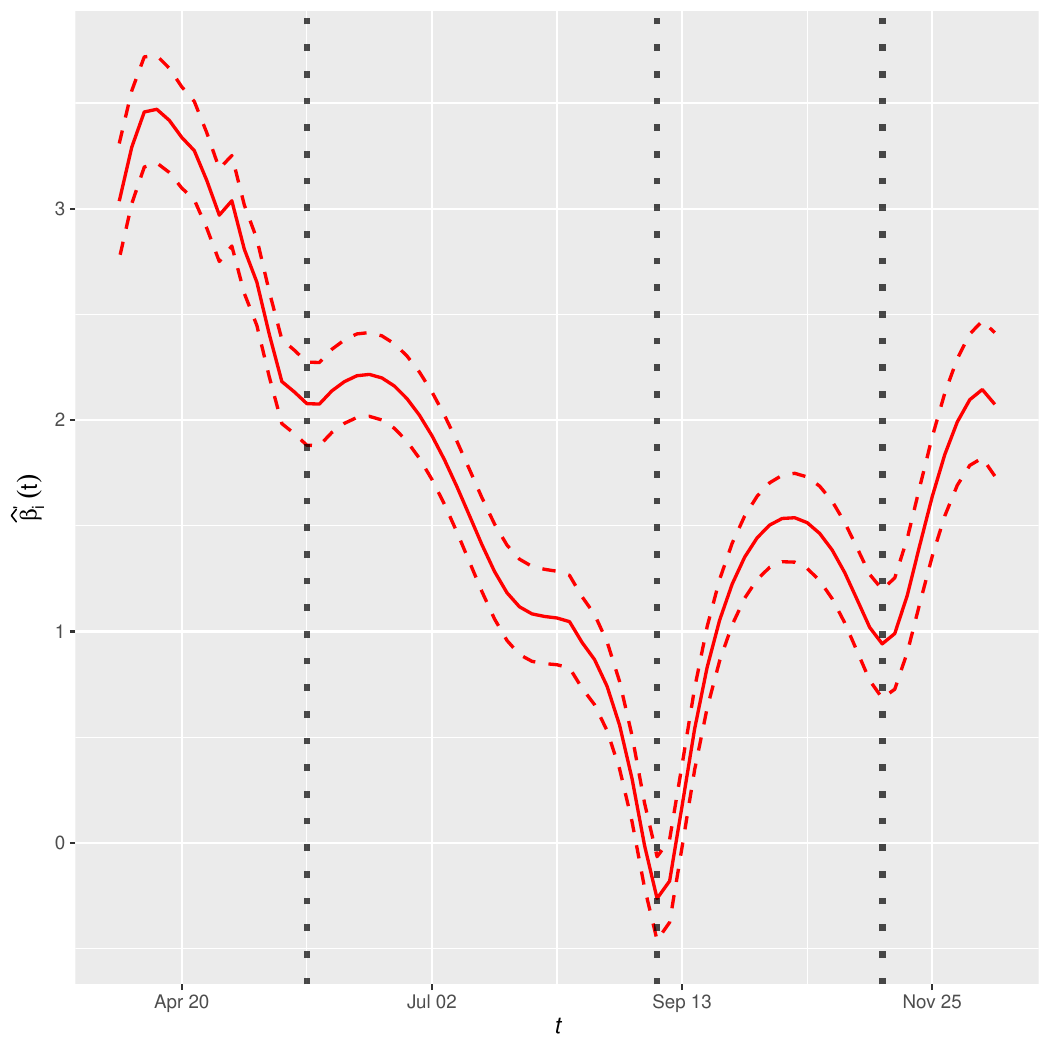}}
	\caption{Capital Bikeshare data: The solid lines represent the estimated curves of $\alpha_i^*(t)$ and $\beta_i^*(t)$ for bike stations $i=101$, $301$, $401$, and $501$. The dashed lines  represent the pointwise $95\%$ confidence intervals.
		Three vertical black dashed lines mark the dates of public holidays: May 28, September 3, and November 12.
 }
	\label{fig:realBike:DE}
\end{figure}

		\noindent Figure~\ref{fig:realBike:DE} illustrates the estimated $\alpha^*(t)$ and $\beta^*(t)$ for these stations. We observe distinct time-varying bike-sharing activity. During Memorial Day (May 28), Labor Day (September 3), and Veterans Day (November 12), central Washington, D.C. stations (20th St \& Virginia Ave NW, New Hampshire Ave \& T St NW) show reduced activity, while Northwest D.C. (Georgia Ave and Fairmont St NW) and Tysons, Virginia (Tysons West Transit Center) exhibit
        increased usage. This contrast may reflect location-specific holiday behaviour: urban core stations experience a decline due to outbound travel and reduced commercial activity, whereas Northwest D.C. benefits from recreational cycling near Howard University, and Tysons sees more activity as a transit hub. Testing procedures from Section 4, evaluating in-degree heterogeneity, out-degree heterogeneity, and time-varying degree effects, all yield $p$-values below 0.01.

\begin{figure}[h]
	\centering
	\subfigure[Subfigure 1 list of figures text][$\hat{\gamma}_{1}(t)$]{
		\includegraphics[width=0.3\textwidth]{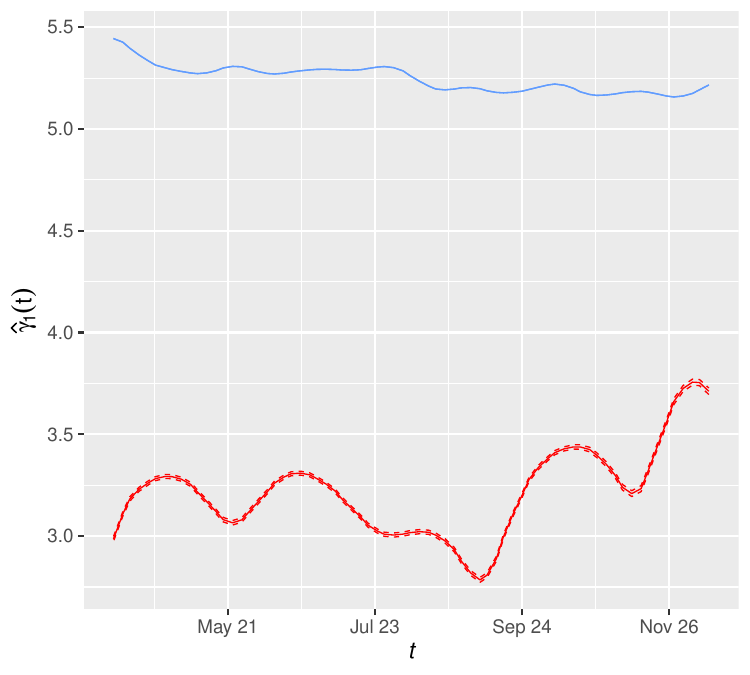}	}
	\subfigure[Subfigure 1 list of figures text][$\hat{\gamma}_{2}(t)$]{
		\includegraphics[width=0.3\textwidth]{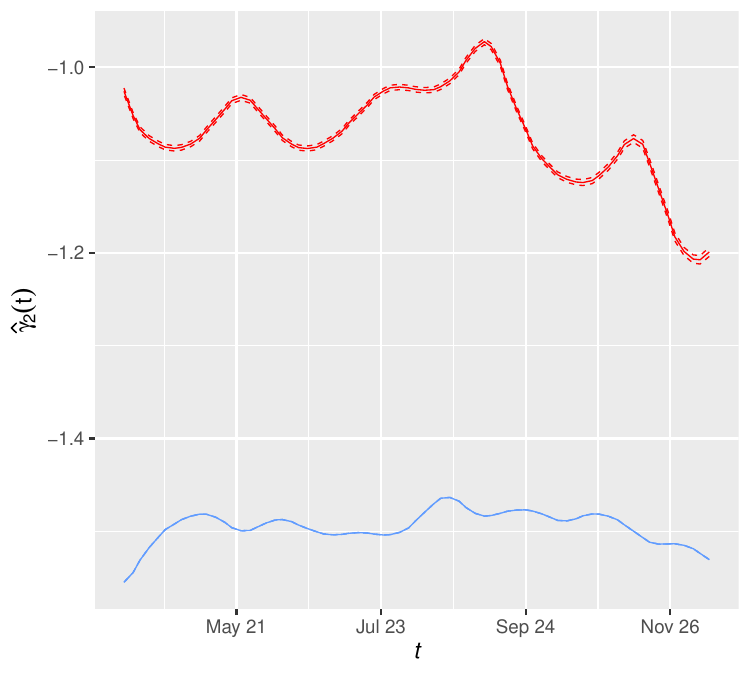}	}
		\qquad
			\subfigure[Subfigure 2 list of figures text][$\hat{\gamma}_{3}(t)$]{
			\includegraphics[width=0.3\textwidth]{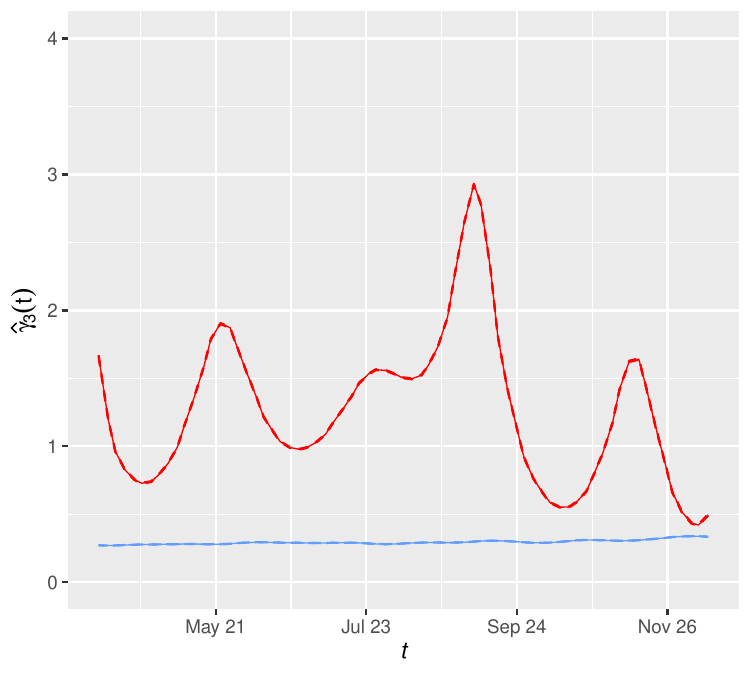}	}
	\subfigure[Subfigure 2 list of figures text][$\hat{\gamma}_{4}(t)$]{
		\includegraphics[width=0.3\textwidth]{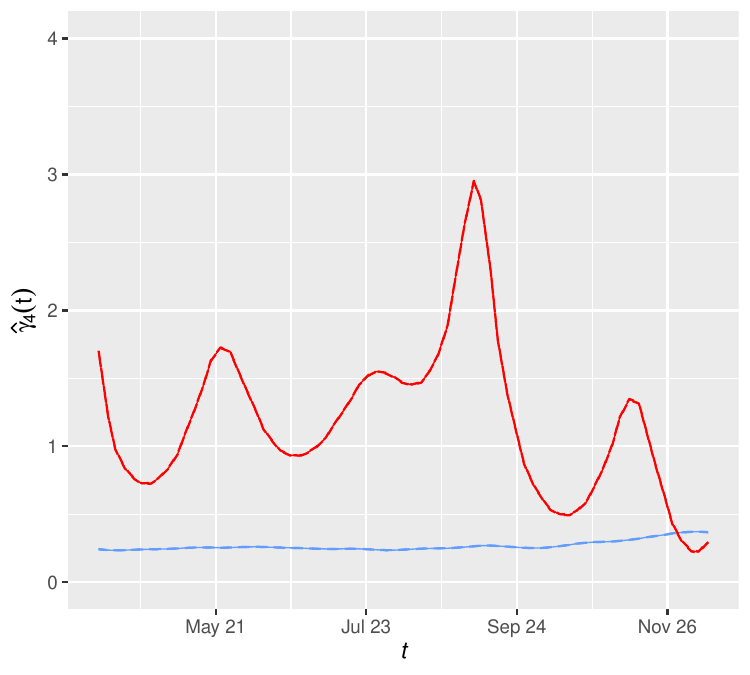}	}
		\caption{Capital Bikeshare data: The red solid curves represent the estimates of the covariate effects using our mode
while the blue curves represent the estimates obtained by the method in \cite{kreib2019}.
The red dashed lines represent the pointwise $95\%$ confidence intervals.
}
	\label{fig:realBike:CE}
\end{figure}

\begin{figure}[h]
	\centering
		\subfigure[Subfigure 1 list of figures text][Our method for out-degrees]{
		\includegraphics[width=0.38\textwidth]{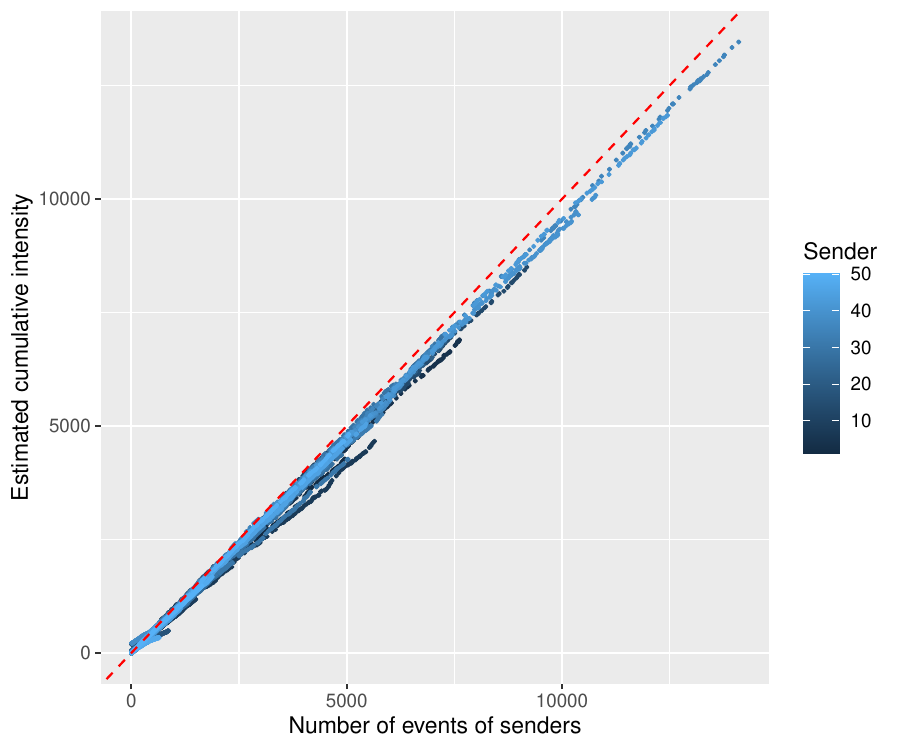}}
	\subfigure[Subfigure 2 list of figures text][Our method for in-degrees]{
		\includegraphics[width=0.38\textwidth]{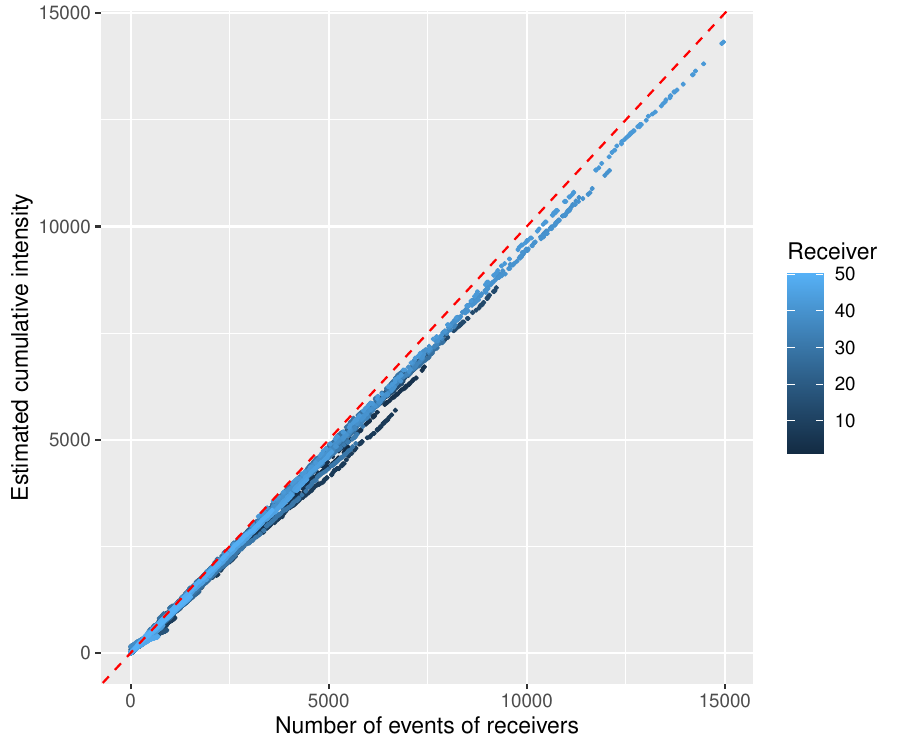}}
	\qquad
\subfigure[Subfigure 1 list of figures text][Krei\ss's method for out-degrees]{
		\includegraphics[width=0.38\textwidth]{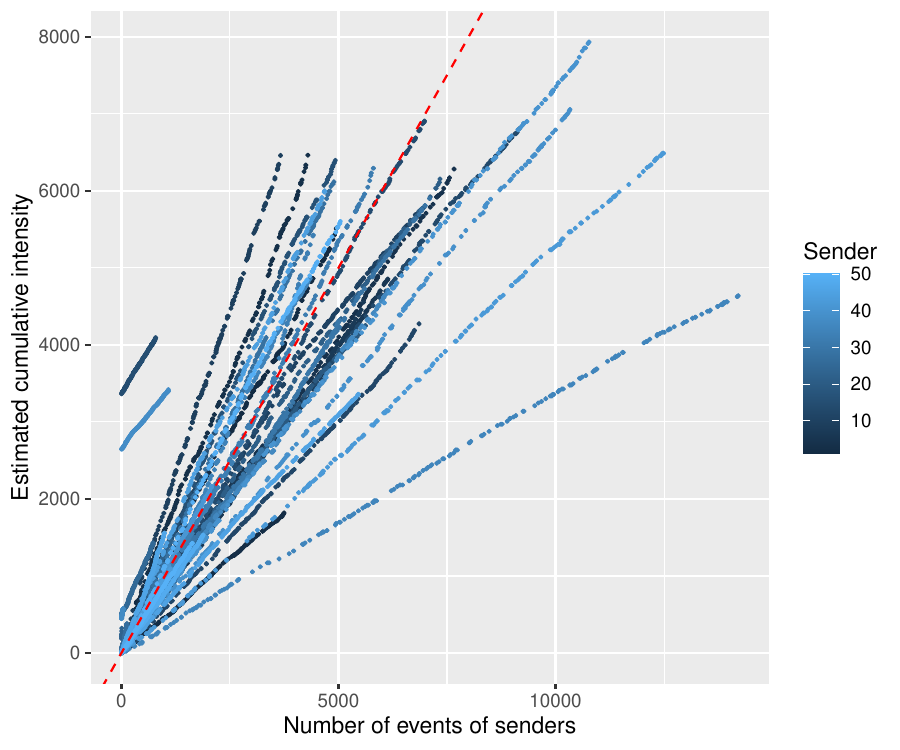}}
	\subfigure[Subfigure 2 list of figures text][Krei\ss's method for in-degrees]{
		\includegraphics[width=0.38\textwidth]{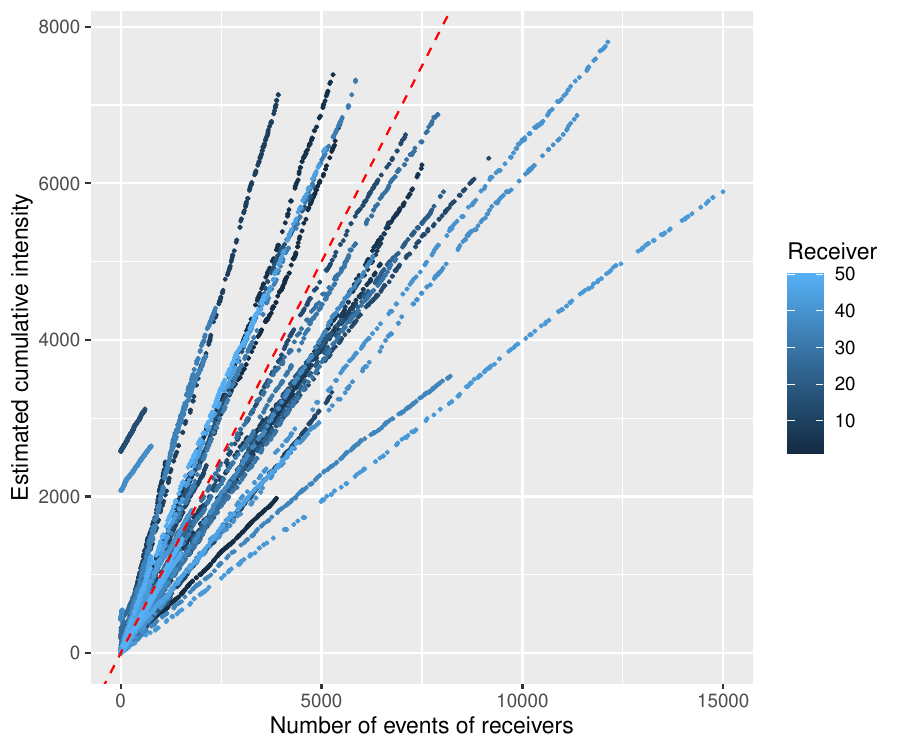}}
	\caption{Capital Bikeshare data: Goodness-of-fit comparison for two methods (red dashed line denotes the identity line)
}
	\label{fig:realGOF:Bike}
\end{figure}

Furthermore, the covariate effects exhibited
temporal variations, as evidenced by Fig.~\ref{fig:realBike:CE}.
The number of neighboring stations for both origin 
and destination stations consistently shows positive effects throughout the study period, suggesting that biking activity is significantly higher between stations with more neighbors. This phenomenon likely arises because bike stations in close proximity tend to share bicycle usage.  For instance, if a station runs out of bikes, users can easily rent from a nearby station to reach their destination.
Similarly, if a station is full, users may return bikes to nearby stations with available docks.  These dynamics increase the interaction intensity between stations with many neighbors and other stations.  Our method and that of \cite{kreib2019} both reveal time-varying covariate effects (Fig.~\ref{fig:realBike:CE}), but their reported effects differ significantly from ours. Goodness-of-fit tests (Fig.~\ref{fig:realGOF:Bike}) confirm that our model outperforms theirs.\\

\begin{figure}
	\centering
	\includegraphics[width=0.4\linewidth]{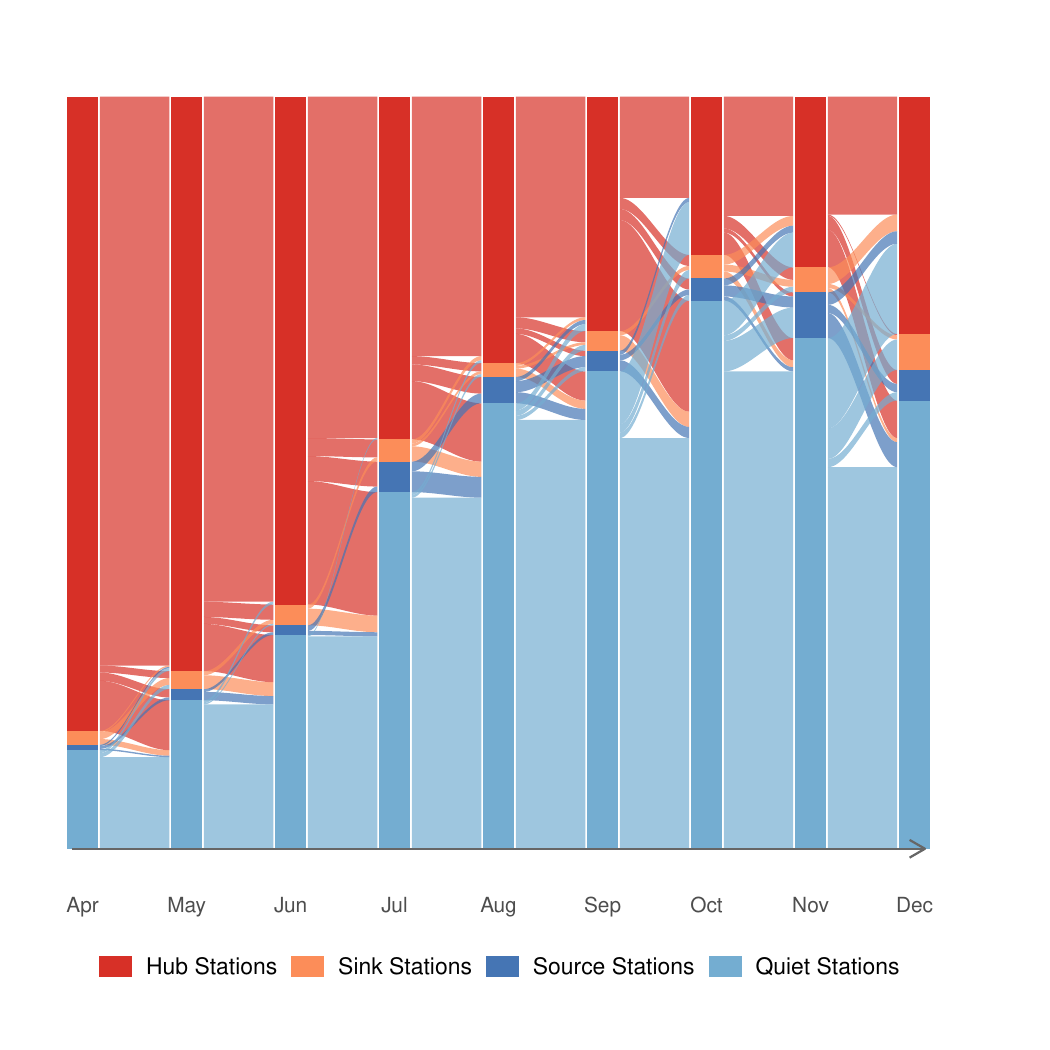}
	\caption{Capital Bikeshare data: Alluvial plot of temporal transitions in station types (curves denote transitions, and thickness reflects station count)}
	\label{fig:flow_bike}
\end{figure}
\begin{figure}
	\centering
	\includegraphics[width=0.4\linewidth]{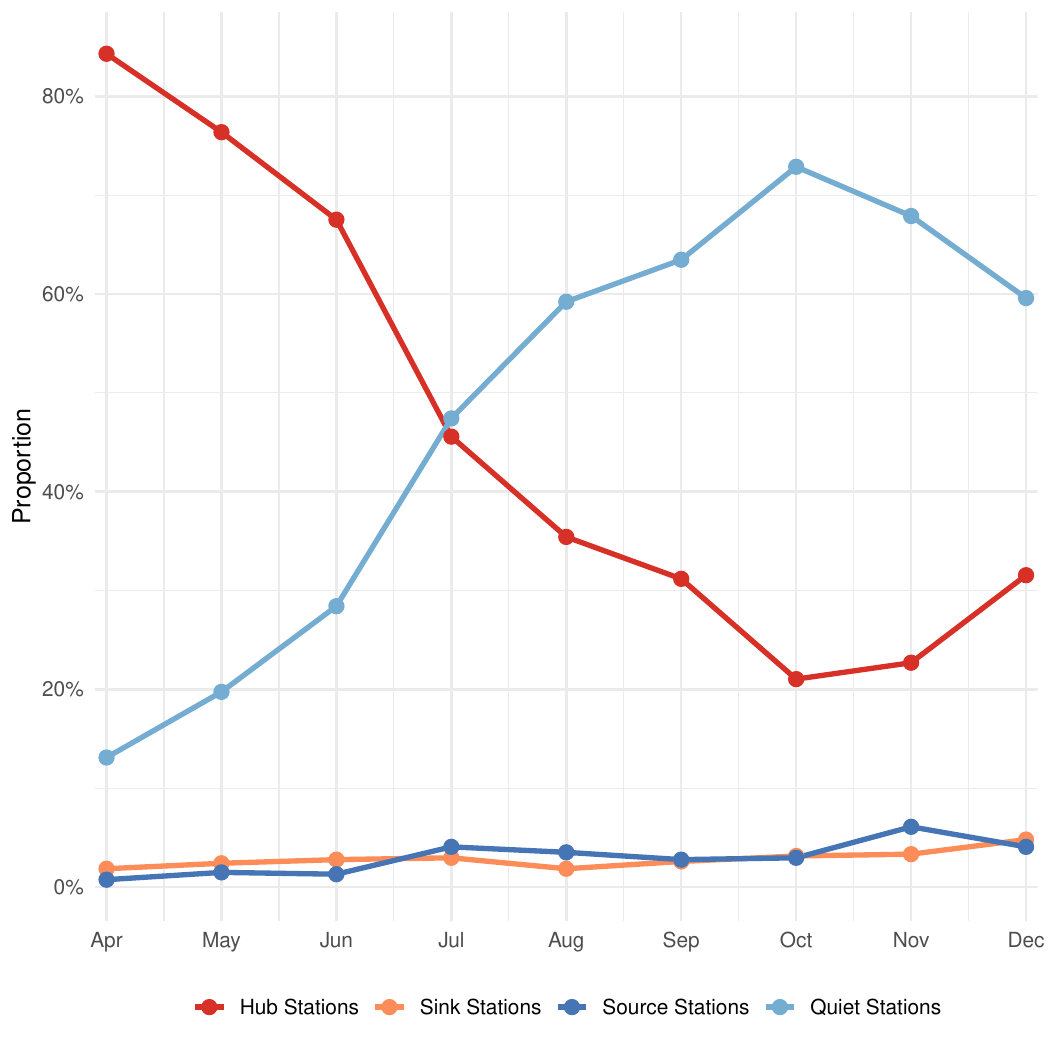}
	\caption{Capital Bikeshare data: Temporal trends in bike-sharing station classifications}
	\label{fig:line_bike}
\end{figure}
Similar to the classification of students into four social roles in Section~\ref{Analysis:MIT}, 
bike-sharing stations are classified into four types based on their rental and return activity. Stations with high rental and return rates, typically found in vibrant urban centers, are called Hub Stations. Those with high return but low rental rates, serving as primary trip destinations like workplaces, are called Sink Stations. Stations characterized by high rental but low return rates, often located in residential areas as trip origins, are called Source Stations. Lastly, Quiet Stations, with low rental and return rates, are usually situated in low-activity peripheral areas.
Temporal patterns, shown in Figures~\ref{fig:flow_bike} and~\ref{fig:line_bike}, reveal two key features of bike-sharing station dynamics:\\
\textbf{Low Proportion of Sink and Source Stations:}\\
Transport asymmetry is rare in bike-sharing networks, with Sink Stations (high-in, low-out, $\leq$5\%) and Source Stations (low-in, high-out, $\leq$6\%) far less prevalent than in the MIT social evolution data, where asymmetry is common (high-in, low-out, 8.2\% to 41.0\%; low-in, high-out, 8.2\% to 29.5\%). This contrast underscores the distinct flow dynamics between bike-sharing and social networks.\\
\textbf{Cyclical Proportion of Hub Stations:}\\
The proportion of Hub Stations (high-in, high-out) exhibits a cyclical pattern, increasing in April, decreasing through summer, and rising again in autumn, likely influenced by weather and urban activity patterns.

\begin{figure}
	\centering
	\includegraphics[width=0.5\linewidth]{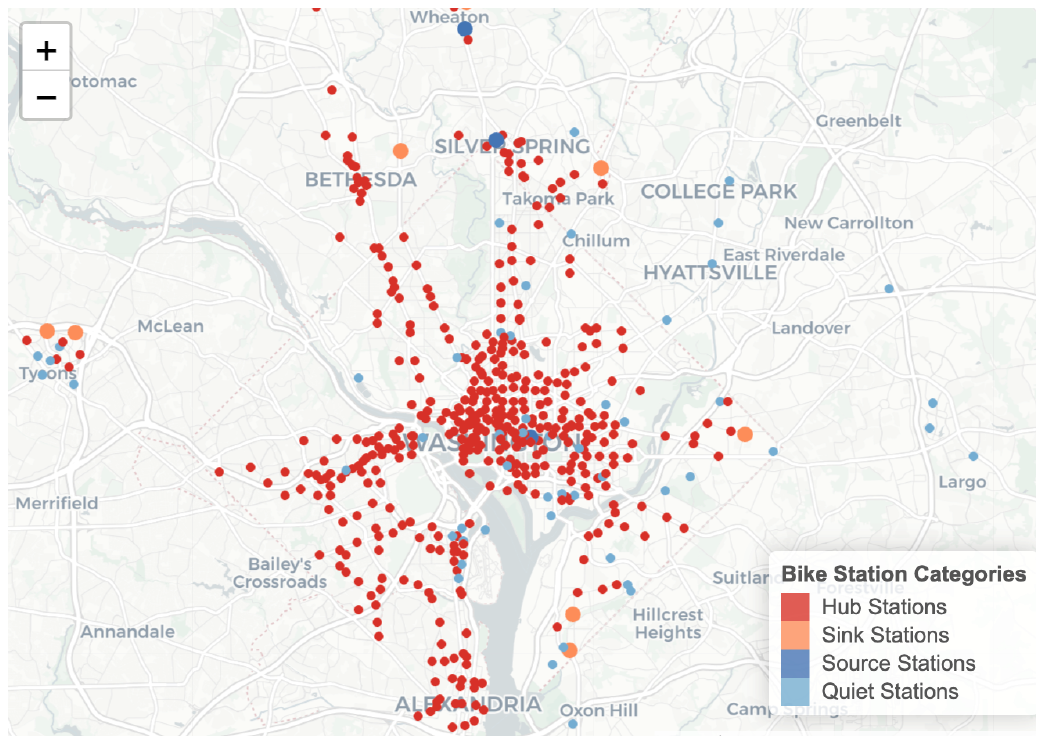}
	\caption{Capital Bikeshare data: Spatial distribution of station types, May 2018}
	\label{fig:map_may}
\end{figure}

The spatial distribution of Capital Bikeshare stations across four categories Hub, Sink, Source, and Quiet in May 2018 is shown in Fig.~\ref{fig:map_may}. Quiet stations are sparsely distributed in peripheral areas. Hub stations predominantly cluster in the central region.

\section{Discussions}
\label{section:concluding}

We proposed a degree-corrected Cox network model for characterizing both time-varying degree heterogeneity and covariate effects
in continuous-time directed network data.
We developed a local estimating equations method to estimate
all $2n+p$ unknown parameters and provided an easy-to-implement algorithm to solve the equations.
We also introduced a cross-validation method to select the optimal bandwidths.
Further, we established the uniform consistency and asymptotic normality of the proposed estimators under some mild conditions,
developed test statistics for testing whether there is time-varying degree heterogeneity,
and proposed a graphical diagnostic method to evaluate the goodness-of-fit for dynamic network models.
In addition, although the social interaction and bike-sharing datasets
are smaller in scale than  large social media or neuroscience networks, they robustly validate the model's ability
to reveal meaningful patterns. We plan to investigate the extension of our
approach to such large-scale networks
in future work. 

We characterised the temporally changing effects of degree heterogeneity using parameters $\{\alpha_i(t)\}_{i=1}^n$ and $\{\beta_i(t)\}_{i=1}^n$.
As mentioned before, we could also treat $\{\alpha_i(t)\}_{i=1}^n$ and $\{\beta_i(t)\}_{i=1}^n$ as random variables, as in frailty models \citep[e.g.,][]{A1993}.
If we impose some dependent structure on these random variables, then it models dynamic dependent structures.
However, it requires developing new methods for inference in random-effects scenarios, since
it is challenging to explore the dependent structure in network data. 

We extended the consistency result of the estimators to the case of the conditional independence assumption
(i.e., $N_{ij}(t)~(1\le i\neq j\le n)$ are independent conditional on all covariates $Z_{ij}(t)$).
We conducted simulation studies under this setting.
The details of the simulation setup are provided
in Section D of the Supplementary Material,
and the results are presented in Tables S1 and S2 of the Supplementary Material.
We observe that the MISEs decrease as the number of nodes increases.
This suggests that the proposed estimators remain consistent when $N_{ij}(t)$ are conditionally independent.
In addition, the coverage frequencies are close to the nominal level, and the lengths of the confidence intervals decrease as $n$ increases.
Figure S2 in the Supplementary Material presents the estimated curves of $\gamma^*(t)$ when $n=200$,
showing that the confidence bands tend to cover the entire true curves.
These observations indicate that the asymptotic normal approximation still appears to be valid in the conditionally independent case.
However, theoretical analysis requires a large number of non-trivial calculations to bound various remainder terms in the asymptotic representations of the estimators.
In view of the fact that the current proofs are already highly non-trivial and lengthy, it is beyond of this paper to
investigate this issue. We  defer this to future work.

\section*{Acknowledgements}
We are very grateful to the editor, the associated editor and two referees for their valuable
comments that have greatly improved the manuscript.

\section*{Supplementary material}
\label{SM}
Supplementary material includes the proofs of Theorems \ref{theorem:consistency}-\ref{theorem-central-degree},
and additional results for simulation studies, available upon on request by sending emails to tingyanty@mail.ccnu.edu.cn.

%
%
%

\end{document}